\documentclass[journal, twocolumn, 10pt, twoside]{IEEEtran}
\ifCLASSINFOpdf
\else
\fi
\usepackage[utf8]{inputenc} 
\usepackage[T1]{fontenc}
\usepackage{url}
\usepackage{ifthen}
\usepackage{cite}
\usepackage{graphicx}
\usepackage{hyperref}
\usepackage{amsmath}
\usepackage{amssymb}
\usepackage{bm}
\usepackage{bbm}
\usepackage{algorithmicx}
\usepackage{algorithm}
\usepackage{algpseudocode}
\usepackage{array}
\usepackage{booktabs}
\usepackage{multirow}
\usepackage{makecell}
\usepackage{mathtools}
\usepackage{mathrsfs}
\usepackage{amsthm}
\usepackage{xcolor}

\newtheorem{definition}{Definition}
\newtheorem{theorem}{Theorem}

\newtheorem{approximation}{Approximation}

\newcommand{\A}{\mathcal{A}}
\newcommand{\I}{\mathcal{I}}
\newcommand{\C}{\mathcal{C}}
\newcommand{\N}{\mathcal{N}}
\newcommand{\B}{\mathcal{B}}
\newcommand{\NN}{\mathbb{N}}
\newcommand{\R}{\mathbb{R}}
\newcommand{\List}{\mathcal{L}}

\newcommand{\F}{\mathbb{F}}
\newcommand{\indicator}{\mathbf{1}}
\newcommand{\E}{\mathsf{E}}
\newcommand{\Prob}{\mathsf{P}}
\newcommand{\X}{\mathcal{X}}
\newcommand{\Y}{\mathcal{Y}}
\newcommand{\Z}{\mathcal{Z}}
\newcommand{\nullset}{\mathcal{N}}

\newcommand*\diff{\mathop{}\!\mathrm{d}}
\newcommand\norm[1]{\Vert{#1}\Vert}
\newcommand{\bmx}{\bm{x}}
\newcommand{\bmc}{\bm{c}}
\newcommand{\bmy}{\bm{y}}
\newcommand{\bmr}{\bm{r}}
\newcommand{\bmz}{\bm{z}}
\newcommand{\bmX}{\bm{X}}

\newcommand{\bmv}{\bm{v}}
\newcommand{\bmP}{\bm{P}}
\newcommand{\NACK}{\textit{NACK}}

\newcommand{\SLVD}{\text{SLVD}}
\newcommand{\SSV}{\text{SSV}}
\newcommand{\trace}{\text{trace}}
\newcommand{\tlist}{\text{list}}
\newcommand{\WAVA}{\text{WAVA}}

\DeclareMathOperator*{\TBP}{TBP}
\DeclareMathOperator*{\IEE}{IEE}
\DeclareMathOperator*{\CRC}{CRC}
\DeclareMathOperator*{\Candidate}{Candi}

\DeclareMathOperator{\rcu}{rcu}
\DeclareMathOperator{\mc}{mc}

\DeclareMathOperator{\pep}{pep}
\DeclareMathOperator{\erfc}{erfc}
\DeclareMathOperator{\sign}{sign}
\DeclareMathOperator{\st}{s. t.}

\DeclareMathOperator{\cl}{cl}

\begin{document}
\title{CRC-Aided List Decoding of Convolutional Codes in the Short Blocklength Regime}

\author{Hengjie~Yang,~\IEEEmembership{Student Member,~IEEE},
        Ethan~Liang,~\IEEEmembership{Student Member,~IEEE},
        Minghao~Pan,
        and~Richard~D.~Wesel,~\IEEEmembership{Fellow,~IEEE}% <-this % stops a space
\thanks{This paper was presented in part at IEEE GLOBECOM 2018, 2019 \cite{Yang2018,Liang2019} and IEEE ISIT 2020 \cite{Yang2020}.}% <-this % stops a space
\thanks{This research is supported in part by National Science Foundation (NSF) grant CCF-2008918 and a Qualcomm Faculty Award. Any opinions, findings, and conclusions or recommendations expressed in this material are those of the author(s) and do not necessarily reflect the views of the NSF or of Qualcomm.

H.~Yang is with the Department of Electrical and Computer Engineering, University of California, Los Angeles, Los Angeles, CA, 90095 USA (e-mail: hengjie.yang@ucla.edu).

E.~Liang is with the Department of Electrical Engineering, Stanford University, Stanford, CA, 94305 USA (e-mail: emliang@stanford.edu)

M.~Pan is with the Department of Mathematics, University of California, Los Angeles, Los Angeles, CA, 90095 USA (e-mail: minghaopan@ucla.edu).

R.~D.~Wesel is with the Department of Electrical and Computer Engineering, University of California, Los Angeles, Los Angeles, CA, 90095 USA (e-mail: wesel@ucla.edu).
}}

\maketitle

\begin{abstract}
We consider the concatenation of a convolutional code (CC) with an optimized cyclic redundancy check (CRC) code as a promising paradigm for good short blocklength codes. The resulting CRC-aided convolutional code naturally permits the use of serial list Viterbi decoding (SLVD) to achieve maximum-likelihood decoding. The convolutional encoder of interest is of rate-$1/\omega$ and the convolutional code is either zero-terminated (ZT) or tail-biting (TB). The resulting CRC-aided convolutional code is called a CRC-ZTCC or a CRC-TBCC.  To design a good CRC-aided convolutional code, we propose the \emph{distance-spectrum optimal (DSO)} CRC polynomial. A DSO CRC search algorithm for the TBCC is provided.  Our analysis reveals that  the complexity of SLVD is governed by the expected list rank which converges to $1$ at high SNR. This allows a good performance to be achieved with a small increase in complexity. In this paper, we focus on transmitting $64$ information bits with a rate-$1/2$ convolutional encoder. For a target error probability $10^{-4}$, simulations show that the best CRC-ZTCC approaches the random-coding union (RCU) bound within $0.4$ dB. Several CRC-TBCCs outperform the RCU bound at moderate SNR values.
\end{abstract}

\begin{IEEEkeywords}
Convolutional code, cyclic redundancy check code, list Viterbi decoding, negative acknowledgement, undetected errors.
\end{IEEEkeywords}

% \fi
\IEEEpeerreviewmaketitle

\section{Introduction}
\label{sec: introduction}


\IEEEPARstart{R}{ecently}, the coding theory community has witnessed a growing interest in designing powerful short blocklength codes (e.g., codes with a thousand or fewer information bits). This renewed interest is mainly driven by the stringent requirement of new ultra-reliable low-latency communication in 5G \cite{Ji2018}, and advances in the finite-blocklength information theory developed by Polyanskiy, Poor and Verd\'u \cite{Polyanskiy2010}. The basic question of finite-blocklength information theory asks: what is the maximal channel coding rate achievable at a given blocklength $n$ and error probability $\epsilon$? To answer this question, Polyanskiy \emph{et al}. developed the \emph{random-coding union (RCU) bound} $\rcu(n, M)$ \cite[Theorem~16]{Polyanskiy2010} and the  \emph{meta-converse (MC) bound} $\mc(n, M)$) \cite[Theorem~27]{Polyanskiy2010} that provide, respectively,  tight upper and lower bounds on the error probability $P_e^*(n, M)$ of the best $(n, M)$ code of length $n$ and $M$ codewords. Namely,
\begin{align}
\mc(n,M)\le P_e^*(n, M)\le \rcu(n, M).
\end{align}
They also provide the \emph{normal approximation} \cite[Eq.~223]{Polyanskiy2010} that tightly approximates the performance of the best $(n, M)$ code. Thereafter, these bounds serve as benchmarks to assess the performance of a given finite-blocklength code over a broad class of channels, including the discrete memoryless channel (DMC) and the additive white Gaussian noise (AWGN) channel. Due to the prohibitive complexity of an exact computation of the RCU and MC bounds, saddlepoint approximations of these two bounds were developed that are shown to be numerically accurate \cite{Font-Segura2018}. 

For coding theorists, a central task is to construct \emph{structured} short-blocklength codes for the binary-input AWGN channel such that the probability of error falls into the region delimited by the RCU bound and the MC bound at a reasonable decoding complexity.  There are numerous approaches to achieve this goal.  As a comprehensive overview, Co{\c s}kun \emph{et al}. surveyed in detail the contemporary short-blocklength code designs developed in recent decades \cite{Coskun2019}. Important examples include extended BCH codes under ordered statistics decoding (OSD) \cite{Fossorier1995,Yue2021}, tail-biting convolutional codes under wrap-around Viterbi algorithm (WAVA)  \cite{Gaudio2017}, non-binary low-density parity-check codes \cite{Dolecek2014,Ranganathan2019}, non-binary turbo codes \cite{Liva2013, Jerkovits2016} and polar codes \cite{Arikan2009, Tal2015}. Recent advances also include the polarization-adjusted convolutional codes proposed by Ar\i kan \cite{Arikan2019,Yao2020}. It is worth noting that if no restrictions are imposed on what kind of codes should be used for the AWGN channel, Shannon \cite{Shannon1959} has ingeniously shown that the optimal $(n, M)$ code should be placed on a sphere in the $n$-dimensional Euclidean space such that the total solid angle is evenly split between the $M$ Voronoi regions and every Voronoi region is a perfect circular cone in order to achieve the minimum probability of error.

While there are many possible structures for short-blocklength coding, this paper focuses on the concatenation of a convolutional code with a cyclic redundancy check (CRC) code. The resulting concatenated code is called the \emph{CRC-aided convolutional code}. Convolutional codes were first introduced by Elias \cite{Elias1955}. Viterbi decoding of convolutional codes was developed by Viterbi \cite{Viterbi1967} and its maximum-likelihood (ML) nature was recognized by Forney \cite{Forney1973, Forney1974}. Advantages of convolutional codes include low decoding latency \cite{Hehn2009,Maiya2012} and good error correction performance at short blocklength. The term ``CRC'' stems from the use of cyclic codes for error detection \cite{Peterson1961}, where the cyclic codeword can be put into systematic form with the parity bits easily generated by a linear sequential circuit. As explained in \cite{Baicheva2019}, CRC codes are possibly shortened cyclic codes generated by a polynomial whose leading and zero coefficients are nonzero. The order of the generator polynomial defines the blocklength of the associated cyclic code. However, in practice, the CRC code is a subcode of this cyclic code whose blocklength is less than the polynomial order.

% In this paper as in many practical error detection applications, a linear sequential circuit is used to compute the parity bits, but the CRC code is not required to be cyclic. The term ``CRC'' is retained nonetheless. 

The structure of concatenating a convolutional code with a CRC code was first proposed in the context of hybrid automatic repeat request (ARQ) \cite{Rice1994} and is used in numerous practical systems where the convolutional code serves as an inner error-correcting code to combat channel errors and the CRC code serves as an error-detecting code to verify if a codeword has been correctly received. Examples include the 3GPP cellular communication standards of both 3G \cite{3GPP2006} and 4G LTE \cite{3GPP2018}.

The classical decoding approach for a CRC-aided convolutional code in a hybrid ARQ setting is Viterbi decoding with CRC verification. The input sequence identified by Viterbi decoding is checked to determine whether it is divisible by the CRC polynomial. This indicates whether a valid message has been decoded. If the decoded sequence is divisible by the CRC polynomial, the message segment of the decoded sequence is declared as the most likely message. Otherwise, a negative acknowledgement (NACK) is declared and perhaps a retransmission request is sent to the transmitter. 

Unfortunately, the classical approach of Viterbi decoding with CRC verification conceals the true potential of the CRC-aided convolutional code.  Performing a single Viterbi decoding step causes the decoder to give up too early, often before encountering a convolutional codeword whose input sequence passes the CRC verification. To unleash the power of the CRC-aided convolutional code, we consider the serial list Viterbi decoding (SLVD) pioneered by Seshadri and Sundberg \cite{Seshadri1994}.  SLVD sequentially produces a rank ordered list of codewords according to their likelihoods. Hence, CRC verification can naturally be used as a termination criterion for this list decoding.

Practical implementation of the SLVD typically assumes a \emph{constrained maximum list size} $\Psi$ to limit the peak decoding complexity. The SLVD terminates either when an input sequence passes the CRC verification or when the list rank reaches $\Psi$. The list rank at which the decoder stops is called the \emph{terminating list rank} $L$. However, it is not always possible to have $L = \Psi$. This is because $\Psi$ can be set arbitrarily large, yet only finitely many codewords exist. This implies that $L$ has an intrinsic maximum achievable value independent of $\Psi$ which is referred to as the \emph{supremum list rank} $\lambda$. Consequently, $L$ is a bounded random variable between $1$ and $\min\{\lambda, \Psi\}$. Since the decoding complexity is a function of $L$, the average decoding complexity is a function of the average list rank $\E[L]$.

Assume that $\Psi<\lambda$. In this case, there are three possible outcomes associated with the SLVD: 1) a correct decoding if SLVD identifies the transmitted message within $\Psi$ trials; 2) an undetected error (UE) if an erroneous input sequence found by SLVD passes the CRC verification within $\Psi$ trials; and 3) a NACK and the forced termination of the decoder if the SLVD fails to find an input sequence that passes CRC verification within $\Psi$ trials. In contrast, any value of $\Psi$ with $\Psi\ge \lambda$ gives the same decoder behavior where no NACK is produced. In this case, the SLVD is an implementation of ML decoding of the CRC-aided convolutional code. In the extreme case where $\Psi = 1$, the SLVD reduces to the classical Viterbi decoding with CRC verification.

A classical list decoder \cite{Elias1957} assumes a fixed list size and declares decoding success as long as the transmitted codeword is in the list. In contrast, the SLVD has a more stringent requirement for success that can lead to a higher error probability than for the classical list decoder. Several upper bounds on error probability were developed for the classical list decoder, e.g., \cite{Bocharova2008,Hof2010}. However, these results are not directly applicable to the SLVD.

This paper focuses on the concatenation of a rate-$1/\omega$ convolutional code with an optimized CRC code.  We explore both zero-terminated convolutional code (ZTCC) and tail-biting convolutional code (TBCC) \cite{Ma1986}. The resulting concatenated code is called a \emph{CRC-ZTCC} in the first case and a \emph{CRC-TBCC} in the second case. For CRC-ZTCCs, Lou \emph{et al.}\cite{Lou2015} realized that previous designs of CRC polynomials typically ignore the structure of the inner error-correcting code, which leads to suboptimal performance. Lou \emph{et al.} designed optimal CRC polynomials for a given ZTCC such that the probability of UE is minimized for a single Viterbi decoding attempt followed by CRC verification. A key point in their analysis is that when the target probability of UE is low enough, the design principle is equivalent to maximizing the minimum distance of the CRC-ZTCC. However, Lou \emph{et al.} did not address the optimal CRC design for a TBCC and did not consider SLVD.

Compared to the ZTCC, the TBCC has the advantage of avoiding the rate loss incurred by the overhead associated with the zero tail that follows the information sequence, but this overhead reduction comes with an increase in decoding complexity.   A TB codeword requires that the initial and terminating states be the same, which can be achieved, for example, by setting the initial encoder memory to be the final bits of the information sequence. However, this requirement increases the difficulty of efficiently identifying the ML path on the trellis because the common value of the initial and terminating states is unknown at the decoder. 

One approach to ML decoding of a TBCC is to perform Viterbi decoding from every possible initial state\cite{Ma1986}. 
Various \emph{approximate} algorithms are proposed for decoding the TBCC based on either ML or maximum \emph{a posteriori} probability criterion, e.g., \cite{Wang1989,Cox1994,Anderson1998,Shao2003}. Among these algorithms, the WAVA \cite{Shao2003} proves to be both efficient and near-ML. Shankar \emph{et al.} \cite{Shankar2006} introduced an efficient, iterative, two-phase algorithm for \emph{exact} ML decoding of TBCC, where an A* algorithm is applied in the second phase, using information from the first phase to compute the heuristic function. To make the exact SLVD of TBCC possible and efficient, this paper extends Shankar \emph{et al.}'s  algorithm to accommodate the CRC polynomial. Specifically, if a traceback identifies a TB path, the CRC of the corresponding input sequence is checked. If the input sequence passes the CRC verification, the algorithm terminates. Otherwise, the algorithm locates the next rank ordered path.

\subsection{Contributions}

This paper provides a design paradigm for both CRC-ZTCCs and CRC-TBCCs, a suite of  tools for performance analysis of these codes, and a complexity analysis showing that SLVD allows low-complexity decoding at low probability of UE for $\Psi\ge\lambda$, i.e., an average decoding complexity similar to standard Viterbi decoding of the convolutional code alone. These contributions combine to yield, for example, CRC-aided convolutional codes that closely approach the RCU bound while requiring decoding complexity similar to Viterbi decoding on a convolutional code trellis with $2^8$ states.

The main contributions of this paper are summarized below.
\subsubsection{CRC-Aided Convolutional Code Design} 
This paper introduces the concept of the \emph{distance-spectrum optimal (DSO) CRC polynomial}, which minimizes the theoretical union bound of the probability of UE for $\Psi\ge\lambda$. Theorem \ref{thm: theorem 1} shows that for high SNR, the DSO CRC polynomial reduces to the one that obtains the best minimum distance $d_{\min}^l$. Theorem \ref{thm: upper bound on d_crc} provides a sharp upper bound on the achievable $d_{\min}^l$  based on the distance spectrum of the convolutional code. For low target probability of UE, we present an efficient algorithm for finding DSO CRC polynomials for TBCCs of arbitrary rate, and provide these polynomials for ZTCCs and TBCCs for optimum rate-$1/2$ convolutional encoders in \cite{Lin2004} at $64$ information bits.

\subsubsection {CRC-Aided Convolutional Code Performance Analysis}
The performance of a CRC-aided convolutional code with the constrained maximum list size $\Psi$ is measured by three probabilities: probability of correct decoding $P_{c, \Psi}$, probability of UE $P_{e, \Psi}$ and probability of NACK $P_{\NACK, \Psi}$, where $P_{c, \Psi}+P_{e, \Psi}+P_{\NACK, \Psi}=1$.
This paper provides bounds, approximations, and simulation results characterizing how these probabilities vary with $\Psi$ and with SNR. Theorems \ref{theorem: probability vs. Psi} -- \ref{thm: P_UE, P_NACK vs. SNR} describe how performance evolves as $\Psi$ increases, the existence and behavior of the supremum list rank $\lambda$, and performance (in terms of $P_{c, \Psi}$, $P_{e, \Psi}$, and $P_{\NACK, \Psi}$)  as a function of SNR for extreme values of $\Psi=1$ and  $\Psi=\lambda$.

\subsubsection{CRC-Aided Convolutional Code Decoding Complexity}
This paper provides expressions for the complexity of SLVD for CRC-ZTCCs and CRC-TBCCs. These expressions reveal that complexity is a function of the expected list rank $\E[L]$. This paper characterizes $\E[L]$ including a new approach to computing $\E[L]$ in the limit of low SNR, a new analysis of conditional expected list rank given the noise magnitude, and two new approaches for approximating the conditional expected list rank. Our parametric approximation on the conditional expected list rank naturally leads to an accurate approximation of $\E[L]$ as a function of $P_{e,\lambda}$ which shows that as $P_{e,\lambda}$ converges to $0$, $\E[L]$ converges to $1$ (see Approximation \ref{approx: parametric approx} to follow). We see that for practically interesting operating points of $P_{e,\lambda}$ such as $10^{-6}$, $\E[L]\approx1$ for typical CRC lengths. This implies that for an interesting range of CRC lengths, the CRC length can be increased with negligible impact on complexity. Moreover, for these CRC lengths, the complexity of SLVD for the CRC-aided convolutional code is very similar to that of standard Viterbi decoding of the convolutional code alone.

\subsubsection{Achieving the RCU Bound with Practical Complexity}
This paper focuses on designing good CRC-aided convolutional codes for transmitting $64$ information bits. Simulation results show that the CRC-ZTCC with $8$ memory elements can approach the RCU bound within $0.4$ dB with decoding complexity similar to standard Viterbi decoding of the ZTCC.  The best CRC-TBCC with $8$ memory elements essentially achieves the RCU bound, but requires increased decoding complexity.

\subsection{Organization}

This paper is organized as follows: Section \ref{sec: preliminaries} introduces notation, the system architecture, TB trellises, Polyanskiy \emph{et al.}'s finite-blocklength bounds, and the related saddlepoint approximations. Section \ref{sec: search of DSO CRC polynomial} introduces the concept of the DSO CRC polynomial, shows that at high SNR the DSO CRC can be obtained by maximizing $d_{\min}^l$, provides an upper bound on $d_{\min}^l$, and gives a DSO CRC design algorithm for TBCCs of arbitrary rate at high SNR. Section \ref{sec: CRC-aided list decoding of CCs} presents the performance and complexity analyses of SLVD of a given CRC-aided convolutional code. Section \ref{sec: simulation results} presents simulation results of our designed CRC-aided convolutional codes and a comparison of $(128, 64)$ linear block codes. Section \ref{sec: conclusion} concludes the paper.

\begin{figure}[t]
\centering
\includegraphics[width=0.48\textwidth]{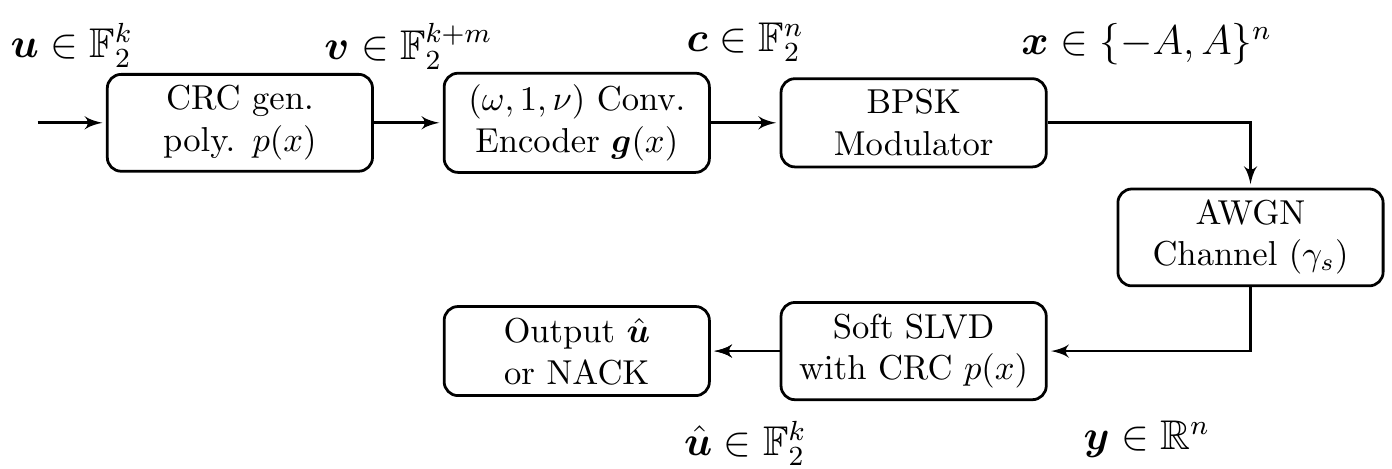}
\caption{Block diagram of the CRC-aided list decoding of convolutional codes.}
\label{fig: system model}
\end{figure}

\section{Preliminaries}
\label{sec: preliminaries}

\subsection{Notation}
 Let $\F_2 = \{0,1\}$ denote the binary field. $\F_2^n$ denotes the set of $n$-dimensional binary sequences. $\F_2[x]$ denotes the set of binary polynomials. The indicator function $\indicator_{E}$ takes the value $1$ if the event $E$ occurs, and $0$ otherwise. The polynomial $u(x) = \sum_{i=0}^{n-1}u_ix^i\in\F_2[x]$ and its row vector form $\bm{u}=[u_0, u_1,\dots, u_{n-1}]\in\F_2^n$ are used interchangeably. The CRC polynomial is represented in hexadecimal when its binary coefficients are written from the highest to lowest order. For instance, 0xD represents $x^3+x^2+1$. The convolutional generator polynomial is represented in octal when the binary coefficients of each generator polynomial are written from the lowest to highest order. For instance, $(13,17)$ represents $(1+x^2+x^3, 1+x+x^2+x^3)$. Let $w_H(\cdot)$, $d_H(\cdot,\cdot)$ and $\norm{\cdot}$ denote the Hamming weight, Hamming distance, and Euclidean norm respectively. Finally, $\cl(S)$ and $\partial(S)$ denote the closure and the boundary of a subset $S\subseteq\R^n$, respectively.

\subsection{Architecture}
This paper considers CRC-aided list decoding of convolutional codes, as depicted in Fig. \ref{fig: system model}. Let $u(x) = \sum_{i=0}^{k-1}u_ix^{i}\in\F_2[x]$ denote the $k$-bit binary information sequence, where $u_{k-1}$ is the first bit entering the CRC encoder. The information sequence $u(x)$ is first encoded with a degree-$m$ CRC generator polynomial $p(x) =1+p_1x+\cdots+p_{m-1}x^{m-1}+x^m\in\F_2[x]$ to obtain $m$ parity check bits $r(x) = x^mu(x) \mod p(x)$. Thus, we obtain $v^*(x) = x^mu(x)+r(x)$ which is divisible by the CRC polynomial $p(x)$. The final CRC-coded sequence $v(x)$ is produced by reversing $v^*(x)$, i.e., $v(x) = x^{k+m-1}v^*(x^{-1})$. This guarantees that the first bit entering the encoder, namely, $u_{k-1}$ in $u(x)$, is always the lowest degree term of $v(x)$, consistent with common representation. The concatenated codeword $\bm{c}\in\F_2^n$ of blocklength $n$ is obtained by convolutionally encoding $\bmv$ with a minimal, feedforward, $(\omega, 1, \nu)$ encoder $\bm{g}(x) = [g_1(x), g_2(x),\dots, g_{\omega}(x)]$, $g_{i}(x)=\sum_{j=0}^{\nu}g_{i,j}x^j$, with $\nu$ memory elements. To terminate a convolutional code into a linear block code, we consider either the ZT or TB method. 

This paper focuses on  CRC-aided convolutional codes, but our analysis also involves the higher-rate convolutional code for which the CRC codeword $\bmv$ is the input message.  To describe the two codes of interest as concisely as possible, define the higher-rate code $\C_{h}$ and the lower-rate code  $\C_{l}$, where the latter is the CRC-aided convolutional code, as follows:
\begin{align}
  &\C_{h} \triangleq \big\{\bm{c}\in\F_2^n: \bm{c} = \bm{v}\bm{G}, \forall \bm{v}\in\F_2^{k+m} \big\},\\
  &\C_{l} \triangleq \big\{\bm{c}\in\F_2^n: \bm{c} = \bm{v}\bm{G}, \forall \bm{v}\in\F_2^{k+m} \text{ s.t. } p(x)|v^*(x) \big\},
\end{align}
where $\bm{G}\in\F_2^{(k+m)\times n}$ is the matrix representation of the convolutional encoder. Intuitively, the effect of $p(x)$ is to obtain a subcode $\C_{l}$ from the given higher-rate code $\C_{h}$. 
%Hence, $\C_l$ is the CRC-CC concatenated code. 
The exact definition of $\C_{h}$ and $\C_{l}$ require the specification of the ZTCC or TBCC. For a ZTCC, $n = \omega(k+m+\nu)$ and
\begin{align*}
\bm{G} = \begin{bmatrix}
G_0 & G_1 & \cdots & G_\nu    &        & & \\
    & G_0 & G_1    & \cdots & G_\nu    & & \\
    &     & \ddots & \ddots & \ddots & \ddots & \\
    &     &        & G_0    & G_1    & \cdots & G_{\nu}
\end{bmatrix},
\end{align*}
where 
\begin{align*}
G_i = \begin{bmatrix} g_{1,i} & g_{2,i} & \cdots & g_{\omega, i}  \end{bmatrix}, \quad i = 1,2,\dots, \nu.
\end{align*}
Similarly, for a TBCC, $n = \omega(k+m)$ and 
\begin{align*}
\bm{G} = \begin{bmatrix}
G_0 & G_1 & \cdots & \cdots & G_\nu & & &\\
    & G_0 & G_1 & \cdots & \cdots & G_\nu & &\\
    &     & \ddots & \ddots & \ddots & & \ddots & \\
    &     &  & G_0 & G_1    & \cdots & \cdots & G_\nu\\
G_\nu &     &  &     & G_0    & G_1 & \cdots & G_{\nu-1}\\
G_{\nu-1} & G_{\nu} & &  & & \ddots & \ddots & \vdots\\
\vdots & & \ddots & & & & \ddots & G_1\\
G_1 & G_{2} & \cdots & G_{\nu}  & & & & G_0
\end{bmatrix}.
\end{align*}
Clearly, $\C_{l}\subseteq \C_{h}$, $|\C_{h}|=2^{k+m}$ and $|\C_{l}|=2^k$. The rate of the CRC-aided convolutional code (i.e., the lower-rate code) $R = k/n$. A fundamental quantity associated with a linear block code is its minimum distance. To aid our discussion, we define
\begin{align}
  d_{\min}^{h} &\triangleq \min\{w_H(\bmc): \bmc\in\C_{h}\setminus\{\bm{0}\} \},\\
  d_{\min}^{l} &\triangleq \min\{w_H(\bmc): \bmc\in\C_{l} \setminus\{\bm{0}\} \}.
\end{align}
As a corollary, $0<d_{\min}^{h}\le d_{\min}^{l}$. Note that for a ZTCC, $d_{\min}^h$ is in fact an order-$(k+m-1)$ row distance and is thus no less than the free distance of the convolutional code \cite{Johannesson1999}. 

The binary phase shift keying (BPSK) modulated sequence $\bmx=[x_0, x_1, \dots, x_{n-1}]$ for codeword $\bmc$ is obtained via $x_i = (1-2c_i)A$, where $A$ is the BPSK amplitude, and is then transmitted over the AWGN channel with channel SNR $\gamma_s$. Therefore, the channel model is
\begin{align}
y_i = x_i + z_i,\quad i = 0, 1,\dots, n-1,
\end{align}
where $z_i$'s are independent and identically distributed (i.i.d.) according to the standard normal distribution. Thus, $\gamma_s = A^2$ or $A = \sqrt{\gamma_s}$.

Upon receiving the channel observations $\bm{y}$, the (soft) SLVD with a constrained maximum list size $\Psi$ using CRC polynomial $p(x)$ is employed to determine the most likely information sequences $\hat{u}(x)$ from the trellis of the higher-rate code $\C_h$ based on $\bm{y}$ in a sequential manner using a maximum of  $\Psi$ trials. We assume that the SLVD sequentially produces rank ordered codewords\footnote{The input sequence that generates this higher-rate codeword is also known simultaneously.} that are also higher-rate codewords in $\C_h$. This is true when $\C_h$ is a ZTCC and may not be true when it is a TBCC in practice. If an input sequence $\hat{v}^*(x)$ associated with a higher-rate codeword passes the CRC verification, decoding terminates and the list stops growing. The corresponding list rank is marked as the terminating list rank $L$ and the most likely information sequence $\hat{u}(x)$ is recovered from the last $k$ bits of $\hat{v}^*(x)$. If an input sequence divisible by $p(x)$ is not found after $\Psi$ attempts, the decoder terminates at list rank $\Psi$ with a NACK as the output. As mentioned earlier, there exists a supremum list rank $\lambda$ (whose formal definition will be given in \eqref{eq: supremum list rank}) which is independent of $\Psi$. If $\Psi\ge \lambda$, no NACK will occur. Consequently, $L$ is always bounded between $1$ and $\min\{\lambda, \Psi\}$.

A UE occurs if the SLVD erroneously identifies an input sequence $\hat{v}^*(x)$ that is divisible by $p(x)$ and $\hat{v}^*(x)\ne v^*(x)$. This is equivalent to the case where the UE polynomial $\hat{v}^*(x) - v^*(x)\in\F_2[x]$ is nonzero and is divisible by $p(x)$. Hence, an \emph{error event} is given by the input-output pair $(\hat{v}(x) - v(x), \hat{c}(x) - c(x))$, where $\hat{v}(x)\ne v(x)$ and $\hat{c}(x)$ is a higher-rate codeword associated with $\hat{v}(x)$. By linearity, each error event corresponds to a pair of a nonzero input sequence $v(x)$ and its corresponding codeword $c(x)$. When restricted to convolutional codes, we can also use a trellis path to represent an error event.

The performance of the CRC-aided convolutional code is measured by three probabilities: probability of correct decoding $P_{c, \Psi}$, probability of UE $P_{e, \Psi}$, and probability of NACK $P_{\NACK, \Psi}$, where $P_{c, \Psi}+P_{e, \Psi}+P_{\NACK, \Psi}=1$. In the special case where $\Psi\ge \lambda$, $P_{c, \Psi}+P_{e, \Psi}=1$. For ease of reference, we use $P_{e,\lambda}$ to represent $P_{e,\Psi}$ for which $\Psi\ge \lambda$.

%For the SLVD, the peak decoding complexity is controlled by the constrained maximum list size $\Psi$ and its average decoding complexity is controlled by the expected list rank $\E[L]$.  

\subsection{Tail-Biting Trellises}

We follow \cite{Koetter2003} in describing a TB trellis. Let $V$ be a set of vertices (or states).   The set $\A$  is the output alphabet, and $E$ is the set of edges described as ordered triples  $(v, a, v')$ with $v, v'\in V$, and $a\in\A$. In words, $(v, a, v')\in E$ denotes an edge that starts at $v$, ends at $v'$ and has output $a$.

\begin{definition}[Tail-biting trellises]
A tail-biting trellis $T=(V, E, \A)$ of depth $N$ is an edge-labeled directed graph with the following property: the vertex set $V$ can be partitioned as
\begin{align}
V = V_0\cup V_1\cup\cdots \cup V_{N-1} \label{eq: TB trellis}
\end{align}
such that every edge in $T$ either begins at a vertex of $V_i$ and ends at a vertex of $V_{i+1}$ for some $i=0,1,\dots,N-2$, or begins at a vertex of $V_{N-1}$ and ends at a vertex of $V_0$.
\end{definition}

Geometrically, a TB trellis can be viewed as a cylinder of $N$ sections defined on some circular time axis. Alternatively, we can also define a TB trellis on a sequential time axis $\I=\{0,1,\dots,N\}$ with the restriction that $V_0 = V_N$ so that we obtain a conventional trellis. 

For a trellis $T$ of depth $N$, a trellis section connecting time $i$ and $i+1$ is a subset $T_i\subseteq V_i\times \A \times V_{i+1}\subseteq E$ that specifies the allowed combination $(s_i, a_i, s_{i+1})$ of state $s_i\in V_i$, output symbol $a_i\in \A$, and state $s_{i+1}\in V_{i+1}$, $i=0,1,\dots,N-1$. Such allowed combinations are called trellis branches. A trellis path $(\bm{s},\bm{a})\in T$ is a state/output sequence pair, where $\bm{s}\in V_0\times V_1\times\cdots\times V_N$, $\bm{a}\in \A^N$. Since $\bm{s}$ equivalently specifies the input sequence, an error event can also be described by its corresponding trellis path $(\bm{s}, \bm{a})$. 

For a TB trellis $T$ of depth $N$, a TB path $(\bm{s},\bm{a})$ of length $N$ on $T$ is a \emph{closed} path through $N$ vertices. If $T$ is defined on a sequential time axis $\I=\{0,1,\dots,N\}$, then any TB path $(\bm{s},\bm{a})$ of length $N$ satisfies $s_0=s_N$.

\subsection{Finite-Blocklength Bounds and Approximations}
In \cite{Polyanskiy2010}, Polyanskiy \emph{et al.} derived the RCU bound and the MC bound that upper and lower bound the probability of error of the best $(n, M)$ code. These two bounds serve as benchmarks to assess the performance of a given finite-blocklength code. 

We follow the notation in \cite{Font-Segura2018} to introduce the RCU bound and the MC bound. Let $W^n(\cdot|\cdot)$ denote a length-$n$ channel. Let $\alpha_{\beta}(P, Q)$ denote the smallest type-I error probability among all tests discriminating between distributions $P$ and $Q$, with a type-II error probability at most $\beta$ \cite[Chapter 11.7]{Cover2006}. For a random-coding ensemble defined over distribution $P^n$, the RCU bound is given by
\begin{align}
\rcu(n,M) &\triangleq \E[\min\{1, (M-1)\pep(X^n, Y^n)],
\end{align}
where $(X^n, Y^n)\sim P^n\times W^n$ and the pairwise error probability $\pep(x^n, y^n)$ is defined as
\begin{align*}
\pep(x^n, y^n)&\triangleq \Prob\big(W^n(y^n|\bar{X}^n)\ge W^n(y^n|x^n)\big),
\end{align*}
with $\bar{X}^n\sim P^n$.
The MC bound is a minimax of a particular smallest type-I error probability
\begin{align}
\mc(n,M) &\triangleq \min_{P^n}\max_{Q^n}\left\{\alpha_{\frac{1}{M}}(P^n\times W^n, P^n\times Q^n) \right\},
\end{align}
where the minimization is over all input distributions $P^n$, and the maximization is over a set of auxiliary, independent of the input, output distributions $Q^n$.

An exact evaluation of the RCU bound and the MC bound involves integrating tail probabilities of $n$-dimensional random variables, which is computationally difficult even for simple channels and moderate values of $n$. In \cite{Font-Segura2018}, the authors provided saddlepoint approximations of these two bounds for memoryless symmetric channels, including the binary-input AWGN channel. These approximations are shown to be tight for a wide range of rates and blocklengths. Section \ref{sec: simulation results}  uses saddlepoint approximations to evaluate the RCU bound and the MC bound for the binary-input AWGN channel.

\begin{approximation}[MC bound, \cite{Font-Segura2018}] \label{approx: MC bound}
For memoryless symmetric channels for which $Y\sim W(\cdot|x)$ is independent of $x$,
\begin{align}
&\mc(n, M)\approx \max_{\rho\ge0}\Big\{e^{-n(E_0(\rho)-\rho E_0'(\rho))}\cdot\notag\\
  &\phantom{=}\Big(\psi\big(\sqrt{nU(\rho)}\big)+\psi\big(\rho\sqrt{nU(\rho)}\big)-e^{-n(R-E_0'(\rho))} \Big) \Big\},
\end{align}
where
\begin{align}
E_0(\rho, P) &= -\log\int_{\Y}\Big(\sum_{x\in\X}P(x)W(y|x)^{\frac{1}{1+\rho}} \Big)^{1+\rho}\diff y, \\
E_0(\rho) &= \max_{P} E_0(\rho, P), \label{eq: 12}\\
\psi(x)&= \frac12\erfc\left(\frac{|x|}{\sqrt{2}}\right)e^{\frac{x^2}{2}}\sign(x), \\
U(\rho) &= -(1+\rho)E_0''(\rho),
\end{align}
where $\X$ and $\Y$ denote the input and output alphabets of the channel, and the maximization in \eqref{eq: 12} is over all possible probability distributions on $\X$.
\end{approximation}

\begin{approximation}[RCU bound, \cite{Font-Segura2018}] \label{approx: RCU bound}
For memoryless symmetric channels for which $Y\sim W(\cdot|x)$ is independent of $x$,
\begin{align}
\rcu(n, M)\approx \tilde{\xi}_n(\hat{\rho}) + \varphi_n(\hat{\rho})e^{-n(E_0(\hat{\rho}, P) - \hat{\rho} R)},
\end{align}
where $\hat{\rho}$ is the value for which $E_0'(\rho, P)=R$, and
\begin{align}
Q_{\rho}(y)&=\frac{1}{e^{-E_0(\rho, P)}}\Big(\sum_{x\in\X}P(x)W(y|x)^{\frac{1}{1+\rho}}\Big)^{1+\rho}, \\
\bar{\omega}''(\hat{\rho}) &= \int_{\Y} Q_{\hat{\rho}}(y)\bigg[\frac{\partial^2}{\partial \tau^2}\Big(\log\sum_{x\in\X}P(x)W(y|x)^{\tau} \Big)\Big|_{\tau=\hat{\tau}} \bigg]\diff y, \\
\theta_n(\hat{\rho})&=\frac{1}{\sqrt{1+\hat{\rho}}}\left(\frac{1+\hat{\rho}}{\sqrt{2\pi n \bar{\omega}''(\hat{\rho})}} \right)^{\hat{\rho}}, \\
\tilde{\xi}_n(\hat{\rho})&=\begin{cases}
1, & \hat{\rho} < 0\\
0, & 0\le \hat{\rho}\le 1\\
e^{-n(E_0(1, P)-R)}\theta_n(1), & \hat{\rho} > 1,
\end{cases}\\
V(\hat{\rho}) &= -E_0''(\hat{\rho}, P), \\
\varphi_n(\hat{\rho})&=\theta_n(\hat{\rho})\Big(\psi\big(\hat{\rho}\sqrt{nV(\hat{\rho})}\big)+\psi\big((1-\hat{\rho})\sqrt{nV(\hat{\rho})}\big) \Big).
\end{align}
\end{approximation}

\section{The search for the DSO CRC polynomial}
\label{sec: search of DSO CRC polynomial}
In this section, we seek to design good CRC-aided convolutional codes that provide the lowest possible probability of UE $P_{e,\lambda}$. To this end, for a given convolutional code, we design CRC polynomials that minimize the union bound on the probability of undetected error   $P_{e,\lambda}$. The resulting CRC polynomial is known as the DSO CRC polynomial.

\subsection{General Theory}
\label{sec:GeneralTheory}
For a given convolutional code and a desired CRC degree $m$, we wish to identify the degree-$m$ CRC polynomial 
\begin{align}
p(x) = 1 + p_1x+\dots+p_{m-1}x^{m-1}+x^m \in\F_2[x]
\end{align}
that minimizes the probability of UE $P_{e, \lambda}$. Since the exact probability $P_{e, \lambda}$ has no closed-form expression that can facilitate a design procedure, we use the union bound as an objective function that only involves the \emph{distance spectrum}, $C_{d_{\min}^l}, \dots, C_n$, of the lower-rate code $\C_l$, where $C_d$ denotes the number of codewords in $\C_l$ of Hamming weight $d$, $d_{\min}^l\le d\le n$.  The distance spectrum of the lower-rate code $\C_l$ is a function of both the CRC polynomial $p(x)$ and the higher-rate code $\C_h$.  For any candidate polynomial $p(x)$, the union bound on $P_{e, \lambda}$ is given by
\begin{align}
P_{e, \lambda} &\le \sum_{\bmc\in\C_l\setminus\{\bar{\bmc}\}}\Prob\Big(Z>\frac12\norm{\bmx(\bmc)-\bmx(\bar{\bmc})}\big| \bmX = \bmx(\bar{\bmc})\Big)\notag  \\
  &=\sum_{d = d_{\min}^l}^nC_dQ\big(A\sqrt{d}\big), \label{eq: union bound}
\end{align}
where $\bar{\bmc}\in\C_l$ is the transmitted codeword, $\bmx(\bmc)\in\{-A, A\}^n$ is the BPSK-modulated point for codeword $\bmc$, $Z\sim\N(0, 1)$, and 
\begin{align}
Q(x) &\triangleq \int_x^\infty \frac{1}{\sqrt{2\pi}}e^{-u^2/2}\diff u
\end{align}
is the complementary Gaussian cumulative distribution function. $Q\big(A\sqrt{d}\big)$ computes the pairwise error probability of two codewords at distance $d$. For a given higher-rate code $\C_h$, a given SNR $\gamma_s$ (i.e., $A = \sqrt{\gamma_s}$), and a CRC degree $m$, we define the degree-$m$ \emph{DSO CRC polynomial} as the one that minimizes the union bound on $P_{e, \lambda}$. Namely, the degree-$m$ DSO CRC polynomial is the solution to the following optimization problem:
\begin{align}
  &\min_{p(x)}\quad \sum_{d=d_{\min}^l}^nC_{d}Q \big(A\sqrt{d} \big).
\end{align}

Theoretically, the distance spectrum $C_{d_{\min}^l}, \dots, C_n$ of $\C_l$ can be found through Viterbi search of the trellis of the higher-rate code $\C_h$, retaining only codewords whose input sequences are divisible by the candidate CRC polynomial $p(x)$. However, this approach requires the calculation of distance spectra for $2^{m-1}$ candidate CRC polynomials and quickly becomes computationally expensive as the information length $k$ gets large. The degree-$m$ DSO CRC polynomial depends on the specific higher-rate code and the SNR at which $P_{e, \lambda}$ is being minimized. Note that the DSO CRC polynomial can be different for different values of $k$. In \cite{Lou2015}, Lou \emph{et al.} investigated how DSO CRC polynomials vary with information length $k$. Their essential finding is that a DSO CRC polynomial for a large $k$ is usually ``good'' for shorter $k$. If the SNR is not sufficiently high, the CRC polynomial that minimizes the union bound in \eqref{eq: union bound} may not minimize the actual  $P_{e, \lambda}$.

Nevertheless, when SNR is sufficiently high or equivalently if the target probability of UE $P_{e, \lambda}$ is sufficiently low (typically less than $10^{-6}$), the union bound \eqref{eq: union bound} will be dominated by its first term $C_{d^l_{\min}} Q\Big(A\sqrt{d_{\min}^l}\Big)$ which becomes asymptotically tight to $P_{e, \lambda}$. Furthermore, in most cases at high SNR  where the operating $A$ is large enough, the first term in \eqref{eq: union bound} is only dominated by $d_{\min}^l$. The following theorem justifies this statement.
\begin{theorem}\label{thm: theorem 1}
For a given higher-rate code $\C_h$, let $C_{d_{\min, 1}^l},\dots, C_n$ and $C'_{d_{\min, 2}^l},\dots, C'_n$ be two distance spectra associated with lower-rate codes generated by CRC polynomials $p_1(x)$ and $p_2(x)$, respectively. If $d_{\min, 1}^l<d_{\min, 2}^l$, there exists a positive threshold $A^*$ such that if $A>A^*$,
\begin{align}
  \sum_{d=d_{\min, 1}^l}^nC_dQ\big(A\sqrt{d}\big)> \sum_{d=d_{\min, 2}^l}^nC_d'Q\big(A\sqrt{d}\big).
\end{align}
In the special case where $d_{\min, 1}^l=d_{\min, 2}^l$ and $C_{d_{\min, 1}^l}> C_{d_{\min, 2}^l}'$, the above conclusion still holds.
\end{theorem}

\begin{IEEEproof}
Assume that $d_{\min, 1}^l<d_{\min, 2}^l$. Since coefficients $C_{d_{\min,1}^l}, C_{d_{\min,2}^l}'$ are positive and bounded,
\begin{align}
\phantom{=}&\lim_{A\to\infty}\frac{\sum_{d=d_{\min, 1}^l}^nC_dQ(A\sqrt{d})}{\sum_{d=d_{\min, 2}^l}^nC_d'Q(A\sqrt{d})}\\
=&\lim_{A\to\infty}\frac{C_{d_{\min, 1}^l}\exp\Big(-\frac{A^2d_{\min, 1}^l}{2}\Big)}{C_{d_{\min, 2}^l}'\exp\Big(-\frac{A^2d_{\min, 2}^l}{2}\Big)}\notag\\
\phantom{=}&\cdot\frac{\Big[1+\sum_{d=d_{\min, 1}^l+1}^n\frac{C_d}{C_{d_{\min, 1}^l}}\exp\Big(-\frac{A^2(d-d_{\min, 1}^l)}{2}\Big)\Big]}{\Big[1+\sum_{d=d_{\min, 2}^l+1}^n\frac{C_d'}{C_{d_{\min, 2}^l}'}\exp\Big(-\frac{A^2(d-d_{\min, 2}^l)}{2}\Big)\Big]}\\
=&\lim_{A\to\infty}\frac{C_{d_{\min, 1}^l}}{C_{d_{\min, 2}^l}'}\exp\left(\frac{A^2}{2}(d_{\min, 2}^l-d_{\min, 1}^l)\right)\label{eq: 28}\\
=&\infty.\notag
\end{align}
Hence, there exists a threshold $A^*$ such that when $A>A^*$, $\sum_{d=d_{\min, 1}^l}^nC_dQ(A\sqrt{d})> \sum_{d=d_{\min, 2}^l}^nC_d'Q(A\sqrt{d})$. In the special case where $d_{\min, 1}^l=d_{\min, 2}^l$ and $C_{d_{\min, 1}^l}> C_{d_{\min, 2}^l}'$, the limit in \eqref{eq: 28} is still greater than $1$. Thus, the same conclusion follows.
\end{IEEEproof}
For sufficiently low target $P_{e, \lambda}$, the operating amplitude $A$ is typically large enough such that $A>A^*$ is easily met in practice. In these common situations, the DSO CRC design principle reduces to maximizing the minimum distance $d_{\min}^l$ of the  lower-rate code. 

\begin{figure}[t]
\centering
\includegraphics[width=0.45\textwidth]{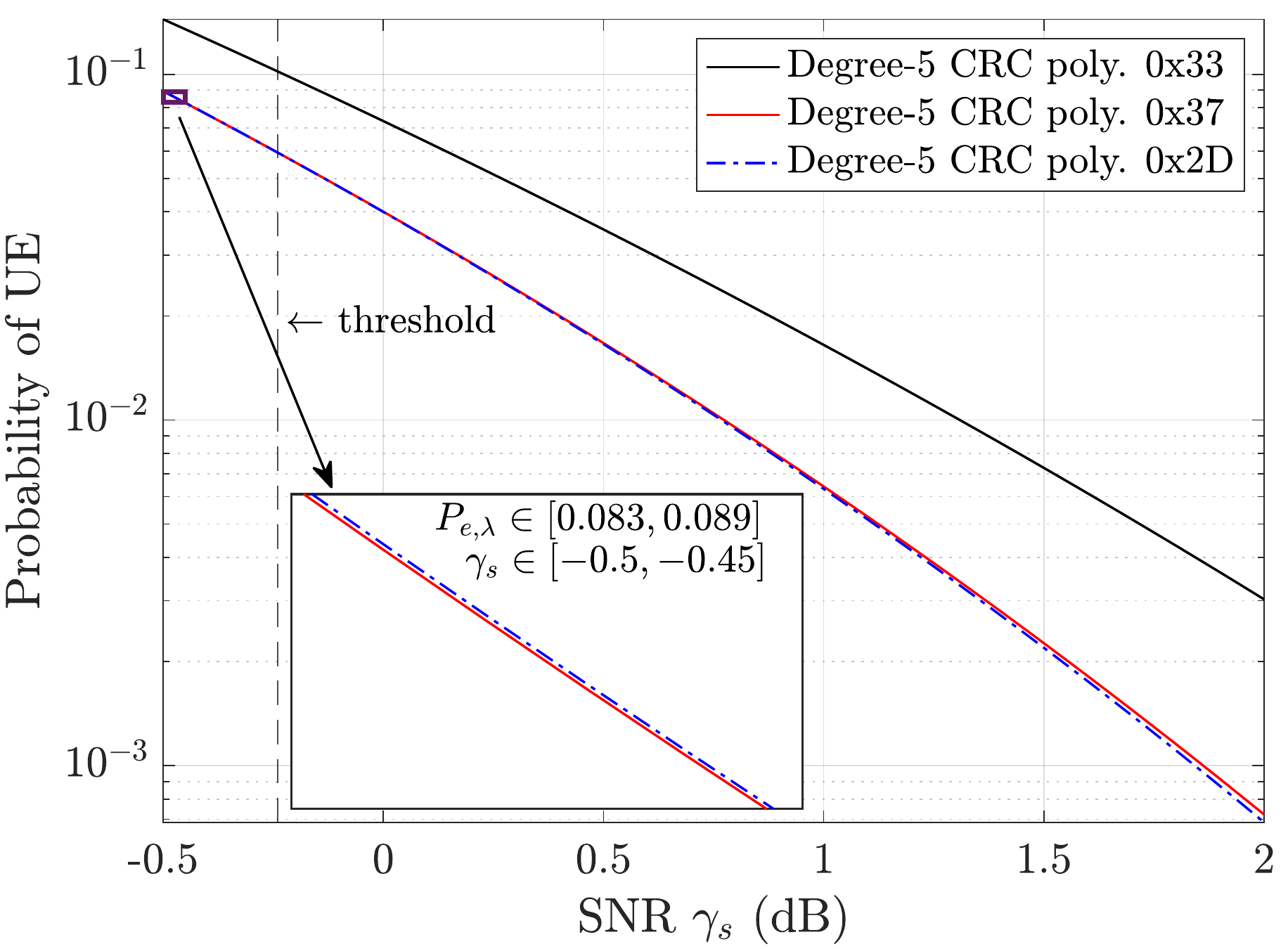}
\caption{Comparison of the DSO CRC polynomials for $k = 10$, $m = 5$ and ZTCC $(13, 17)$. The blocklength of the CRC-ZTCC $n = 36$. The threshold value is $-0.2398$ dB.}
\label{fig: comparison of DSO CRCs}
\end{figure}

As an illustrative example, Fig. \ref{fig: comparison of DSO CRCs} shows the union bounds \eqref{eq: union bound} for three degree-$5$ CRC polynomials among the $16$ candidates for $k = 10$ and ZTCC $(13, 17)$. The CRC 0x37 minimizes the union bound at low SNR, whereas the CRC 0x2D minimizes the union bound at high SNR. On the contrary, the CRC 0x33 yields the worst possible union bound among all candidates. A detailed computation reveals that $d_{\min}^l=11$, $C_{d_{\min}^l}=17$ for 0x37, $d_{\min}^l=12$, $C_{d_{\min}^l}=76$ for 0x2D. Thus, the DSO CRC may not necessarily have the best minimum distance. The worst CRC polynomial 0x33 has $d_{\min}^l=8$, $C_{d_{\min}^l}=10$. In this example, the threshold at which the DSO CRC polynomial switches from  0x37 to 0x2D is $-0.2398$ dB. However, the gap between the performance of the two CRC polynomials is minimal, especially at low SNR. Nevertheless, both 0x37 and 0x2D achieve a gain of $0.5$ dB compared to 0x33 at $10^{-2}$, showing that the optimal CRC polynomial is crucial to achieving good performance.

For a given convolutional code and a specified CRC degree $m$, one may ask: how large can $d_{\min}^l$ be? The next theorem gives a tight upper bound on $d_{\min}^l$ in terms of the distance spectrum of the higher-rate code $\C_h$.

\begin{theorem}\label{thm: upper bound on d_crc}
Given  a specified CRC degree $m$ and a higher-rate code $\C_h$ with distance spectrum $B_{d_{\min}^h}, \dots, B_n$, define $w^*$ as the minimum $w$ for which $\sum_{d=d_{\min}^h}^w B_d\ge 2^m$. For any degree-$m$ CRC polynomial, we have $d_{\min}^l\le 2w^*$.
\end{theorem}

\begin{IEEEproof}
Define the set $V(\bmc)$ to be the set of codewords from the higher-rate code $\C_h$ that unambiguously decode to codeword $\bmc$ of the lower-rate code $\C_l$. Specifically, for each $\bmc\in\C_l$, define
\begin{align}
V(\bmc) \triangleq \{\bmr\in\C_h: d_H(\bmr, \bmc)< d_H(\bmr, \bmc'),\ \forall \bmc'\in\C_l\}.
\end{align}
Hence, by linearity of the higher-rate code, the cardinality of $V(\bmc)$ for every $\bmc\in\C_l$ is exactly the same. Hence, 
\begin{align}
|V(\bmc)|\le \frac{|\C_h|}{|\C_l|} = 2^m, \label{eq:inequality}
\end{align}
where \eqref{eq:inequality} is an inequality because some codewords $\bmr\in\C_h$ may be equidistant from two or more lower-rate codewords. 

Next, we show that for a given $\bmc\in\C_l$, $d_H(\bmr, \bmc)<\frac12d_{\min}^l$ implies that $\bmr\in V(\bmc)$. By definition of the minimum distance, for two arbitrary distinct codewords $\bmc, \bmc'\in\C_l$, $d_H(\bmc, \bmc')\ge d_{\min}^l$. Hence, for any $\bmr\in\C_h$, by triangle inequality,
\begin{align}
d_H(\bmr, \bmc) + d_H(\bmr, \bmc') \ge d_H(\bmc, \bmc')\ge d_{\min}^l.
\end{align}
Thus, if $d_H(\bmr, \bmc)<\frac12d_{\min}^l$, this implies that $d_H(\bmr, \bmc')> \frac12 d_{\min}^l$ for any other $\bmc'\in\C_l$, i.e., $d_H(\bmr, \bmc)<d_H(\bmr, \bmc')$ for all $\bmc'\in\C_l$. By definition of $V(\bmc)$, we conclude that $\bmr\in V(\bmc)$.

By law of contraposition, if $\bmr\notin V(\bmc)$, then $d_H(\bmr, \bmc)\ge \frac{1}{2}d_{\min}^l$. Indeed, when $\sum_{d=d_{\min}^h}^wB_d\ge 2^m$ (i.e., $\sum_{d=0}^wB_d\ge 2^m+1$), by pigeonhole principle, there exists a codeword $\bmr\in\C_h$ that is outside of $V(\bmc)$ and whose distance from $\bmc$ satisfies $d_H(\bmr, \bmc)\le w$. Therefore, for this codeword $\bmr$, $w\ge d_H(\bmr, \bmc)\ge\frac{1}{2}d_{\min}^l$ or equivalently, $d_{\min}^l\le 2d_H(\bmr, \bmc)\le2w$. Since this holds for any $w$ satisfying $\sum_{d=d_{\min}^h}^wB_d\ge 2^m$, the minimum such value $w^*$ yields the tightest upper bound.
\end{IEEEproof}

Table \ref{table: upper bound on d_crc} shows the comparison between $d_{\min}^l$ and the upper bound $2w^*$ in Theorem \ref{thm: upper bound on d_crc} for both ZTCC and TBCC generated with the rate-$1/2$ convolutional encoder $(13, 17)$ at $k = 64$. We see that the upper bound is sharp as there exist DSO CRC polynomials that achieve this bound.

% create a table of d_min vs. d^* here
\begin{table}[t]
\caption{Comparison Between $d_{\min}^l$ Associated With the DSO CRC Polynomial and $2w^*$ Computed From Theorem \ref{thm: upper bound on d_crc} for $k=64$}
\centering
\begin{tabular}{r|l|c|c|l|c|c}
\hline
\multirow{2}{*}{$m$} & \multicolumn{3}{c|}{ZTCC $(13, 17)$} &\multicolumn{3}{c}{TBCC $(13, 17)$} \\\cline{2-7}
 & $p(x)$ & $d_{\min}^l$ & $2w^*$ & $p(x)$ & $d_{\min}^l$ & $2w^*$\\\hline\hline
0 & 0x1  & 6  & 12 & 0x1 & 6 & 12 \\
3 & 0x9  & 10 & 12 & 0xF & 8 & 12\\
4 & 0x1B & 10 & 12 & 0x1F & 9 & 12\\
5 & 0x2D & 12 & 12 & 0x2D & 10 & 12\\
6 & 0x43 & 12 & 12 & 0x63 & 12 & 12\\
7 & 0xB5 & 13 & 14 & 0xED & 12 & 14\\
8 & 0x107& 14 & 14 & 0x107& 12 & 14\\
9 & 0x313& 14 & 16 & 0x349& 14 & 16\\
10& 0x50B& 15 & 18 & 0x49D& 14 & 18\\\hline
\end{tabular}
\label{table: upper bound on d_crc}
\end{table}

%\subsection{Collection/Search Approach to TBCC DSO CRC Design}
\subsection{A Two-Phase DSO CRC Design Algorithm for TBCCs}
\label{sec:2Phase}
We focus on finding the DSO CRC polynomial for low target $P_{e, \lambda}$.  As discussed earlier, the design principle under this circumstance conveniently reduces to maximizing the $d_{\min}^l$ of the lower-rate code. Thus, the optimal CRC polynomial depends on the convolutional code but not the SNR.  

In principle, the DSO CRC design algorithm for low target $P_{e,\lambda}$ comprises a \emph{collection phase} that gathers error events of the higher-rate code $\C_h$ up to a certain distance $\tilde{d}$, and a \emph{search phase} that identifies the degree-$m$ DSO CRC polynomial using the error events gathered in the collection phase. In this section, we propose a two-phase DSO CRC design algorithm particularized to TBCCs of arbitrary rate (including rate $1/\omega$). Later, we point out that our algorithm is also applicable to ZTCCs of arbitrary rate with a few distinctions.

The difficulty of designing DSO CRC polynomials for a TB trellis lies in the fact that a TB trellis is a union of $2^{\nu}$ subtrellises that share trellis branches in the middle. Thus, to collect error events that meet the TB condition, a straightforward collection method is to perform Viterbi search separately at each possible start state to identify the \emph{irreducible error event} (IEE) that leaves the start state once and rejoins it once, and then use them to reconstruct length-$N$ TB paths with distance less than $\tilde{d}$. These IEEs constitute the error events of interest. However, this scheme will be \emph{inefficient} in that for each nonzero start state, there exists a catastrophic IEE that spends a majority of time in the self-loop of the zero state. Such an IEE has the catastrophic property that its length grows unbounded with a finite weight. As a consequence, they are rarely used during reconstruction yet occupy a significant portion of total IEEs.

The algorithm we are about to propose follows the straightforward algorithm with the distinction in collecting IEEs. To circumvent the aforementioned catastrophic IEEs, we wish to identify IEEs whose weight is proportional to its length. To this end, we first partition the TB trellis into several sets that are closed under cyclic shifts. Next, all elements in each set are reconstructed via the concatenation of the corresponding IEEs and circular shifts of the resulting path.

For a given length-$N$ TB trellis associated with a minimal convolutional encoder $\bm{g}(x)$, let  $V_0 = \{0,1,\dots, 2^{\nu}-1\}$ be the set of possible encoder states. We seek a partition of the TB trellis, i.e., mutually exclusive sets that, together, contain all length-$N$ TB paths. To do this, we define $\TBP(0)$ as the set that contains all TB paths that traverse state $0$; $\TBP(1)$ contains the TB paths that traverse state $1$ but not state $0$; and so on. In general, the set $\TBP(\sigma)$ for $\sigma\in V_0$ is defined as follows:
\begin{align}
\TBP(\sigma)&\triangleq \big\{(\bm{s},\bm{a})\in V_0^{N+1}\times \A^N: s_0=s_N;  \notag\\
  \exists i\in\I& \text{ s.t. }  s_i=\sigma;\ \forall i\in\I,\ s_i\notin\{0,1,\dots,\sigma-1\} \big\}. 
\end{align}

An important property of the above decomposition is that each set $\TBP(\sigma)$ is closed under cyclic shifts, as circularly shifting a TB path preserves the sequence of states that it traverses. Furthermore, such a partition of the TB trellis motivates the following IEE.

\begin{definition}[Irreducible error events]
For a TB trellis $T$ on sequential time axis $\I=\{0,1,\dots,N\}$, the set of irreducible error events $(\bm{s},\bm{a})$ at state $\sigma\in V_0$ is defined as
\begin{align}
\IEE(\sigma)\triangleq\bigcup_{i=1,2,\dots,N}\overline{\IEE}(\sigma,i),
\end{align}
where
\begin{align}
    \overline\IEE(\sigma,i)\triangleq&\{(\bm{s},\bm{a})\in V_0^{i+1}\times \A^{i}: s_0=s_i=\sigma;\notag\\
  &\forall j, 0<j<i,\ s_{j}\notin\{0, 1,\dots,\sigma\}\}.
\end{align}
\end{definition}

For ZTCCs, Lou \emph{et al.}\cite{Lou2015} considered finding IEEs that start and end at the zero state and counting the allowed combinations. Hence, The IEE defined above generalizes Lou \emph{et al.}'s IEEs. Since for a nonzero start state, no IEE can traverse the zero state, this guarantees that the weight of the IEE grows proportionally with its length, thus avoiding the catastrophic IEEs incurred in the straightforward algorithm. 

With the sets $\TBP(\sigma)$ defined as above, the following theorem describes how to efficiently find all elements in each $\TBP(\sigma)$ via the corresponding IEEs.

\begin{theorem}\label{theorem: irreducible error events}
Every TB path $(\bm{s},\bm{a})\in\TBP(\sigma)$ can be constructed from the IEEs in $\IEE(\sigma)$ via concatenation and subsequent cyclic shifts.
\end{theorem}

\begin{algorithm}[t]
\caption{The Collection Procedure}
\label{algorithm: collection}
\algrenewcommand\algorithmicrequire{\textbf{Input:}}
\algrenewcommand\algorithmicensure{\textbf{Output:}}
\begin{algorithmic}[1]
\Require The TB trellis $T$, threshold $\tilde{d}$
\Ensure The list of IEEs $\List_{\IEE}(\tilde{d})=\{(\bm{s},\bm{a},\bm{v})\}$
\State Initialize lists $\List_{\sigma}$ to be empty for all $\sigma\in V_0$;
\For{$\sigma\gets0,1,\dots,|V_0|-1$}
\State Perform Viterbi search at $\sigma$ on $T$ to collect list $\List_{\sigma}(\tilde{d})$ of all IEEs of distances less than $\tilde{d}$;
\EndFor
\State \Return $\List_{\IEE}(\tilde{d})\gets\bigcup_{\sigma\in V_0}\List_{\sigma}(\tilde{d})$;
\end{algorithmic}
\end{algorithm}

\begin{algorithm}[t]
\caption{The Search Procedure}
\label{algorithm: search}
\algrenewcommand\algorithmicrequire{\textbf{Input:}}
\algrenewcommand\algorithmicensure{\textbf{Output:}}
\begin{algorithmic}[1]
\Require The trellis length $N$, degree $m$, list of IEEs $\List_{\IEE}(\tilde{d})$
\Ensure The degree-$m$ DSO CRC polynomial $p(x)$
\State Initialize the list $\List_{\CRC}$ of $2^{m-1}$ CRC candidates and empty lists $\List_{\TBP}(d)$ of TBPs, $d=1,\dots,\tilde{d}-1$;
\For{$d\gets1,2\dots,\tilde{d}-1$}
\State Construct all TBPs $(\bm{s},\bm{a},\bm{v})$ from $\List_{\IEE}(\tilde{d})$ s.t. $w_H(\bm{a})=d$, $|\bm{v}|=N$, via concatenation and cyclic shifts;
\State For each TBP, $\List_{\TBP}(d)\gets\List_{\TBP}(d)\cup\{(\bm{s},\bm{a},\bm{v})\}$;
\EndFor
\State $\Candidate(1)\gets\List_{\CRC}$;
\For{$d\gets1,\dots,\tilde{d}-1$}
\For{$p_i(x)\in \Candidate(d)$}
\State Pass all $\bm{v}(x)\in\List_{\TBP}(d)$ to $p_i(x)$;
\State $C^{(i)}\gets$ the number of divisible $\bm{v}(x)$ of dist. $d$;
\EndFor
\State $C^*\gets\min_{i\in\Candidate(d)}C^{(i)}$
\State $\Candidate(d+1)\gets\{p_i(x)\in\Candidate(d): C^{(i)}=C^*\}$;
\If{$|\Candidate(d+1)|=1$}
\State \Return $\Candidate(d+1)$;
\EndIf
\EndFor
\end{algorithmic}
\end{algorithm}

\begin{IEEEproof}
Let us consider $T$ as a TB trellis defined on a sequential time axis $\I=\{0,1,\dots,N\}$. For any TB path $(\bm{s},\bm{a})\in\TBP(\sigma)$ of length $N$ on $T$, we can first circularly shift it to some other TB path $(\bm{s}^{(0)},\bm{a}^{(0)})\in\TBP(\sigma)$ on $T$ such that $s_0^{(0)}=s_N^{(0)}=\sigma$.

Now, we examine $\bm{s}^{(0)}$ over $\I$. If $\bm{s}^{(0)}$ is already an element of $\IEE(\sigma)$, then there is nothing to prove. Otherwise, there exists a time index $j$, $0<j<N$, such that $s_j=\sigma$. In this case, we break the TB path $(\bm{s}^{(0)},\bm{a}^{(0)})$ at time $j$ into two sub-paths $(\bm{s}^{(1)},\bm{a}^{(1)})$ and $(\bm{s}^{(2)},\bm{a}^{(2)})$, where
\begin{align*} 
\bm{s}^{(1)}=&(s_0,s_1,\dots,s_j),\ \bm{a}^{(1)}=(a_0,a_1,\dots,a_{j-1}),\\
\bm{s}^{(2)}=&(s_j,s_{j+1},\dots,s_N),\ \bm{a}^{(2)}=(a_{j},a_{j+1},\dots,a_{N-1}).
\end{align*}
Note that after segmentation of $(\bm{s}^{(0)},\bm{a}^{(0)})$, the resultant two sub-paths, $(\bm{s}^{(1)},\bm{a}^{(1)})$ and $(\bm{s}^{(2)},\bm{a}^{(2)})$, still meet the TB condition. Repeat the above procedure on $(\bm{s}^{(1)},\bm{a}^{(1)})$ and $(\bm{s}^{(2)},\bm{a}^{(2)})$. Since the length of a new sub-path is strictly decreasing after each segmentation, the boundary case is the atomic sub-path $(\bm{s},\bm{a})$ of some length $j^*$ satisfying $s_0=s_{j*}=\sigma$, $s_{j'}\ne\sigma$, $\forall j'\in(0,j^*)$. Clearly, this atomic path is an element of $\IEE(\sigma)$. Thus, we successfully decompose a length-$N$ TB path into elements of $\IEE(\sigma)$. Hence, reversing the above procedure will turn elements of $\IEE(\sigma)$ into a length-$N$ TB path.
\end{IEEEproof}

We now present our two-phase DSO CRC polynomial design algorithm for TBCCs of arbitrary rate (including rate $1/\omega$) at low target $P_{e,\lambda}$ that consists of the collection procedure as described in Algorithm \ref{algorithm: collection} and the search procedure as described in Algorithm \ref{algorithm: search}. In the collection procedure, $(\bm{s},\bm{a},\bm{v})$ denotes the triple of states $\bm{s}$, outputs $\bm{a}$ and inputs $\bm{v}$, where the inputs $\bm{v}$ are uniquely determined by state transitions $s_i\to s_{i+1}$, $i=0,1,\dots,N-1$. The TB trellis considered in the collection procedure should set a sufficiently large trellis length so that IEEs with bounded distance less than $\tilde{d}$ are fully collected. Once the collection procedure is done, one can reuse the collected IEEs in the search procedure for various trellis lengths. For a given higher-rate code $\C_h$ and a specified CRC degree $m$, according to Theorem \ref{thm: upper bound on d_crc}, it suffices to consider distance threshold $\tilde{d}\le 2w^*+1$, where $w^*$ is the minimum weight determined in the theorem, to identify the degree-$m$ DSO CRC polynomial.

In the search procedure, let $|\bm{v}|$ denote the length of $\bm{v}$. Steps from lines 2 to 5 use the IEEs to build all length-$N$ trellis paths with distance less than $\tilde{d}$. In practice, this can be accomplished using dynamic programming. Specifically, for a given state $\sigma\in V_0$, let $\List_{\sigma}(w, l)$ denote the list of TB paths of weight $w$, of length $l$, and with initial state $\sigma$, $0\le w< \tilde{d}$, $1\le l\le N$. Then, the update rule of $\List_{\sigma}(w, l)$ is as follows: given an IEE $(\bm{s}, \bm{a}, \bm{v})\in\IEE(\sigma)$ with $w_H(\bm{a})\le w$ and $|\bm{v}|<l$,
\begin{align}
\List_{\sigma}(w, l)\,{\gets}\,\List_{\sigma}(w, l)\cup\{\List_{\sigma}(w - w_H(\bm{a}), l - |\bm{v}|)\oplus(\bm{s}, \bm{a}, \bm{v}) \},\notag
\end{align}
where $\List_{\sigma}(w,l)\oplus(\bm{s}, \bm{a}, \bm{v})$ denotes appending $(\bm{s}, \bm{a}, \bm{v})$ to the rear of each element in $\List_{\sigma}(w, l)$. The update rule inherently requires that $w, l$ be enumerated in ascending order and $w_H(\bm{a}), |\bm{v}|$ in descending order. Finally, the set of length-$N$ TB paths of distance less than $\tilde{d}$ via direct concatenation are given by $\bigcup_{\sigma\in V_0}\List_{\sigma}(\tilde{d}-1, N)$. The rest of the TB paths are obtained by circularly shifting elements in $\bigcup_{\sigma\in V_0}\List_{\sigma}(\tilde{d}-1, N)$.

\begin{table}[t]
\caption{Optimum Rate-$1/2$ ZTCCs and Their DSO CRC Polynomials for $k=64$ at Sufficiently Low Probability of UE $P_{e, \lambda}$}
\centering
\begin{tabular}{r|c|cm{.15cm}m{.15cm}m{.15cm}m{.2cm}m{.3cm}m{.3cm}m{.4cm}}
\hline
\multirow{2}{*}{$\nu$} & \multirow{2}{*}{ZTCC $\bm{g}(x)$} &\multicolumn{8}{c}{DSO CRC Polynomials} \\
 &  & $m=3$ & 4 & 5 & 6 & 7 & 8 & 9 & 10\\\hline\hline
3 & $(13,17)$ & 9 & 1B & 2D & 43 & B5 & 107 & 313 & 50B\\
4 & $(27,31)$ & F & 15 & 33 & 4F & D3 & 13F & 2AD & 709\\
5 & $(53,75)$ & 9 & 11 & 25 & 49 & EF & 131 & 23F & 73D \\
6 & $(133,171)$ & F & 1B & 23 & 41 & 8F & 113 & 2EF & 629\\
7 & $(247,371)$ & 9 & 13 & 3F & 5B & E9 & 17F & 2A5 & 61D\\
8 & $(561,753)$ & F & 11 & 33 & 49 & 8B & 19D & 27B & 4CF\\
9 & $(1131,1537)$ & D & 15 & 21 & 51 & B7 & 1D5 & 20F & 50D\\
10 & $(2473,3217)$ & F & 13 & 3D & 5B & BB & 105 & 20D & 6BB\\
\hline
\end{tabular}
\label{table: CRC-ZTCC codes}
\end{table}

\begin{table}[t]
\caption{Optimum Rate-$1/2$ TBCCs and Their DSO CRC Polynomials for $k=64$ at Sufficiently Low Probability of UE $P_{e, \lambda}$}
\centering
\begin{tabular}{r|c|cm{.15cm}m{.15cm}m{.15cm}m{.2cm}m{.3cm}m{.3cm}m{.4cm}}
\hline
\multirow{2}{*}{$\nu$} & \multirow{2}{*}{TBCC $\bm{g}(x)$} &\multicolumn{8}{c}{DSO CRC Polynomials} \\
 &  & $m=3$ & 4 & 5 & 6 & 7 & 8 & 9 & 10\\\hline\hline
3 & $(13,17)$ & F & 1F & 2D & 63 & ED & 107 & 349 & 49D \\
4 & $(27,31)$ & F & 11 & 33 & 4F & B5 & 1AB & 265 & 4D1\\
5 & $(53,75)$ & 9 & 11 & 3F & 63 & BD & 16D & 349 & 41B \\
6 & $(133,171)$ & F & 1B & 3D & 7F & FF & 145 & 2BD & 571\\
7 & $(247,371)$ & F & 11 & 33 & 63 & EF & 145 & 3A1 & 5D7\\
8 & $(561,753)$ & F & 11 & 33 & 7F & FF & 1AB & 301 & 4F5\\
9 & $(1131,1537)$ & D & 15 & 33 & 51 & C5 & 1FF & 349 & 583 \\
10 & $(2473,3217)$ & F & 1B & 33 & 79 & BB & 199 & 217 & 4DD\\
\hline
\end{tabular}
\label{table: CRC-TBCC codes}
\end{table}

We remark that our algorithm can be generalized to ZTCCs of arbitrary rate yet comes with the following distinctions: the collection procedure only collects IEEs that start and terminate at the zero state; the search procedure only performs dynamic programming to reconstruct all ZT paths with the target trellis length $N$ and distances less than $\tilde{d}$; termination tails of each ZT path should be removed before CRC verification. For interested readers, the MATLAB routines are available for ZTCCs \cite{CRC_ZTCC_link} and for TBCCs \cite{CRC_TBCC_link}.

Table \ref{table: CRC-ZTCC codes} presents the DSO CRC polynomials of degree $m$ from $3$ to $10$ that maximize $d_{\min}^l$ of CRC-ZTCCs based on a family of optimum rate-$1/2$ convolutional encoders in \cite[Table 12.1(c)]{Lin2004} with constraint length $v$ from $3$ to $10$ for $k = 64$.  These DSO CRC polynomials are for a sufficiently low $P_{e, \lambda}$. Table \ref{table: CRC-TBCC codes} presents the TBCC counterpart in the same setting. The code generated by the DSO CRC polynomial and convolutional encoder in the above tables is our designed CRC-aided convolutional code. In Section \ref{sec: simulation results}, we will present the performance and complexity trade-off of these codes.

\section{Performance and Complexity of SLVD}
\label{sec: CRC-aided list decoding of CCs}
This section explores the performance and complexity of SLVD.  For a specified CRC-aided convolutional code, performance under SLVD is characterized by three probabilities: $P_{c, \Psi}$, $P_{e, \Psi}$ and $P_{\NACK, \Psi}$. The average decoding complexity of SLVD is a function of expected list rank $\E[L]$. In order to understand the performance-complexity trade-off, we investigate how these quantities vary with system parameters including the SNR $\gamma_s$ and the constrained maximum list size $\Psi$.

Geometrically speaking, the process of SLVD is to draw a list decoding sphere around the received sequence $\bmy$ with an increasing radius until the sphere touches the closest lower-rate codeword. To formalize this procedure, let us consider the set of received sequences $\bmy\in \R^n\setminus{\nullset}$ where $\nullset$ is the probability-zero set defined by $\nullset\triangleq\{\bmy\in\R^n: \exists\, \bmc_1,\bmc_2\in\C_h\,\st\,\norm{\bmy-\bmx(\bmc_1)}=\norm{\bmy-\bmx(\bmc_2)}\}$. For every $\bmy\in\R^n\setminus\nullset$, let
\begin{align}
  \bmc_1(\bmy), \bmc_2(\bmy), \dots, \bmc_{|\C_h|}(\bmy)
\end{align}
be an enumeration of $\C_h$ such that
\begin{align}
\norm{\bmy\,{-}\,\bmx(\bmc_1(\bmy))}< \norm{\bmy\,{-}\,\bmx(\bmc_2(\bmy))}\,{<}\cdots{<}\,\norm{\bmy\,{-}\,\bmx(\bmc_{|\C_h|}(\bmy))}.\notag
\end{align}

Using the above enumeration, we formally define the terminating list rank $L(\bmy)$ and the terminating Euclidean distance $d_t(\bmy)$ for $\bmy$ as follows:
\begin{align}
  L(\bmy) &\triangleq \min\{s\in\{1,2,\dots, |\C_h|\}: \bmc_s(\bmy)\in\C_l \},\\
  d_t(\bmy)&\triangleq \min_{\bmc\in\C_l}\norm{\bmy - \bmx(\bmc)}.
\end{align}
Thus, the list decoding sphere of $\bmy$ can be expressed as
\begin{align}
\B_{\SLVD}(\bmy) = \{\bmc\in\C_h: \norm{\bmy - \bmx(\bmc)}\le d_t(\bmy)\}.
\end{align}
Clearly, $L(\bmy) = |\B_{\SLVD}(\bmy)|$.

The concepts above are defined for each individual received point $\bmy\in\R^n\setminus\nullset$. Alternatively, we can also consider the decoding region $\Y(\bmc)$ (i.e., the Voronoi region) of each lower-rate codeword $\bmc\in\C_l$:
\begin{align}
\Y(\bmc)\triangleq& \big\{\bmy\in\R^n\setminus\nullset: \norm{\bmy-\bmx(\bmc)}<\norm{\bmy-\bmx(\bmc')},\notag\\
  &\phantom{==}\forall \bmc'\in\C_l\setminus\{\bmc\} \big\}.
\end{align}
For SLVD, the decoding region $\Y(\bmc)$ can be further decomposed into finer subsets according to the list rank. Namely, for each $\bmc\in\C_l$ and a particular list rank $s\in\{1,2,\dots, |\C_h|-|\C_l|+1\}$,
\begin{align}
  \Z_s(\bmc) &\triangleq\Big\{\bmy\in\R^n\setminus\nullset:\exists\bmc_1,\dots,\bmc_{s-1}\in\C_h\setminus\C_l\,\st\, \notag\\
    &\phantom{==}\norm{\bmy-\bmx(\bmc)}>\max_{j=1,2,\dots,s-1}\norm{\bmy-\bmx(\bmc_j)}\,\text{and}\,\notag\\
    &\phantom{==} \norm{\bmy-\bmx(\bmc)}<\min_{\bmc'\notin\C_h\setminus\{\bmc, \bmc_1,\dots,\bmc_{s-1}\}}\norm{\bmy-\bmx(\bmc')} \Big\}.
\end{align}
Here, each $\Z_s(\bmc)$ is referred to as the \emph{order-$s$ decoding region of $\bmc$}. Obviously, for each $\bmc\in\C_l$, we have
\begin{align}
&\Z_{s_1}(\bmc)\cap \Z_{s_2}(\bmc) = \varnothing,\quad \text{if } s_1\ne s_2\\
&\Y(\bmc) = \bigcup_{s=1,2,\dots,|\C_h|-|\C_l|+1}\Z_s(\bmc).
\end{align}
By linearity of the code, the order-$s$ decoding regions of all lower-rate codewords are isomorphic. With BPSK modulation, the bisection hyperplane of any two codewords passes through the origin of $\R^n$, making each order-$s$ decoding region a polyhedron. Note that there exists a \emph{supremum list rank} $\lambda$ 
\begin{align}
\lambda \triangleq \max\{s: \Z_s(\bmc)\ne\varnothing, \forall \bmc\in\C_l\}. \label{eq: supremum list rank}
\end{align}
Here, the supremum list rank $\lambda$ only depends on $\C_l$ and $\C_h$ and is independent of $\Psi$. Hence, if $\Psi\ge \lambda$, the possible outcomes of SLVD include only correct decoding or UE. Namely, NACKs are not possible.

\subsection{Performance Analysis}
\label{subsec: performance analysis}

We first give our results on how $P_{c,\Psi}, P_{e, \Psi}$ and $P_{\NACK, \Psi}$ vary with $\Psi$ for a fixed SNR. Each of these probabilities may be understood as the probability of an event defined as a set of received sequences $\bmy$.  For example, with $\bar{\bmc}\in\C_l$ as the transmitted codeword, by linearity, we have
\begin{align}
  P_{c,\Psi} &= \Prob\left(\bigcup_{s=1,2\dots,\lambda\wedge\Psi }\Z_s(\bar{\bmc})\bigg|\bmX=\bmx(\bar{\bmc})\right)\notag\\
    &= \sum_{s=1}^{\lambda\wedge\Psi}\Prob\big(\Z_s(\bar{\bmc})|\bmX=\bmx(\bar{\bmc})\big), \label{eq: 44}\\
  P_{e, \Psi} &= \sum_{\bmc\in\C_l\setminus{\{\bar{\bmc}\}}}\Prob\left(\bigcup_{s=1,2,\dots\lambda\wedge\Psi}\Z_s(\bmc)\bigg|\bmX=\bmx(\bar{\bmc})\right)\notag\\
    &= \sum_{s=1}^{\lambda\wedge\Psi}\sum_{\bmc\in\C_l\setminus{\{\bar{\bmc}\}}}\Prob\big(\Z_s(\bmc)|\bmX=\bmx(\bar{\bmc})\big), \label{eq: 45}
\end{align}
where $\lambda\wedge \Psi \triangleq\min\{\lambda,\Psi\}$.

\begin{theorem}\label{theorem: probability vs. Psi}
For a given CRC-aided convolutional code decoded with SLVD at a fixed SNR, $P_{c, \Psi}$ and $P_{e, \Psi}$ are both strictly increasing in $\Psi$ and will converge to $P_{c, \lambda}$ and $P_{e, \lambda}$ respectively, where $P_{c, \lambda}+P_{e, \lambda} = 1$.
\end{theorem}

\begin{IEEEproof}
According to \eqref{eq: 44} and \eqref{eq: 45}, $P_{c,\Psi}$ and $P_{e,\Psi}$ are summations of the order-$s$ decoding regions $\Prob(\Z_s(\bmc)|\bmX=\bmx(\bar{\bmc}))$, thus are non-decreasing in $\Psi$. For each $\bmc\in\C_l$ and  $s=1,2,\dots,\lambda$, $\Prob(\Z_s(\bmc)|\bmX=\bmx(\bar{\bmc}))$ is solely determined by the SNR value and is independent of $\Psi$. Since every order-$s$ decoding region $\Z_{s}(\bmc)$ is the intersection of halfplanes, it follows that each $\Z_{s}(\bmc)$ is an open set. Hence, it suffices to show that each  $\Z_{s}(\bmc)$ is nonempty. To this end, we use induction to show that all $\Z_{s}(\bmc)$, $s=1,2,\dots,\lambda$, are open and nonempty.

By definition, $\Z_{\lambda}(\bmc)$ is open and nonempty. Assume $\Z_{s}(\bmc)$ is open and nonempty for some fixed $s\le\lambda$. Hence, there exists $\bmy\in\Z_{s}(\bmc)$ with $\bmc_1, \bmc_2, \dots, \bmc_s\in\B_{\SLVD}(\bmy)$, where $\bmc_1,\dots,\bmc_{s-1}\in\C_h\setminus\C_l$ and $\bmc_s\in\C_l$. Next, we show that with probability 1, a point $\bmy'$ can be constructed from $\bmy$ such that $\bmc_1, \bmc_2,\dots,\bmc_{j-1}, \bmc_{j+1}, \dots,\bmc_{s-1}, \bmc_s\in\B_{\SLVD}(\bmy')$ for some $j\in\{2,3,\dots, s-2\}$. 

The new point  $\bmy'$ is constructed as $\bmy' = \bmy+t(\bmx(\bmc_s)-\bmy)$, where $t\in[0, 1]$. Hence, 
\begin{align}
\norm{\bmx(\bmc_s)-\bmy'} &=(1-t)\norm{\bmy-\bmx(\bmc_s)}.
\end{align}
Therefore, it is equivalent to showing that there exists $t\in(0, 1)$ such that for some $j\in\{1,2,\dots, s-1\}$,
\begin{align}
&\norm{\bmy'-\bmx(\bmc_{j})}>(1-t)\norm{\bmy-\bmx(\bmc_s)}\\
\max_{ \substack{i\in\{1,\dots,s-1\}\setminus\{j\} } }&\norm{\bmy'-\bmx(\bmc_{i})}<(1-t)\norm{\bmy-\bmx(\bmc_s)}.
\end{align}
To this end, we show that the set of $\bmy$ for which no such $t$ exists has a probability of zero. First, consider function
\begin{align}
F(t)\triangleq \max_{i=1,2,\dots,s-1}\norm{\bmy'-\bmx(\bmc_{i})}-(1-t)\norm{\bmy-\bmx(\bmc_s)}.\notag
\end{align}
Since each $\norm{\bmy'-\bmx(\bmc_{i})}$, $i=1,2,\dots,s-1$, is a continuous function in $t$, $F(t)$ is also a continuous function in $t\in[0, 1]$. Note that
\begin{align}
F(0) &= \max_{i=1,2,\dots,s-1}\norm{\bmy-\bmx(\bmc_{i})}-\norm{\bmy-\bmx(\bmc_s)}<0\\
F(1) &= \max_{i=1,2,\dots,s-1}\norm{\bmx(\bmc_s)-\bmx(\bmc_{i})}>0.
\end{align}
By the intermediate value theorem, there exists a $t^*\in(0, 1)$ such that
\begin{align}
\max_{i=1,2,\dots,s-1}\norm{\bmy'-\bmx(\bmc_{i})}=(1-t^*)\norm{\bmy-\bmx(\bmc_s)}.
\end{align}
Hence, the converse case can only occur if there exist two codewords $\bmc_{j_1}$ and $\bmc_{j_2}$, $j_1\ne j_2$, such that 
\begin{align}
\norm{\bmy'-\bmx(\bmc_{j_1})} = \norm{\bmy'-\bmx(\bmc_{j_2})}=(1-t^*)\norm{\bmy-\bmx(\bmc_s)}.\label{eq: 55}
\end{align}
If \eqref{eq: 55} holds, this implies that $\bmy'$ lies on the intersection of two hyperplanes: one that bisects $\bmx(\bmc_{j_1})\bmx(\bmc_{s})$ and the other that bisects $\bmx(\bmc_{j_2})\bmx(\bmc_{s})$. Namely, $\bmy'$ lies on an $(n-2)$-dimensional hyperplane that crosses the origin. Hence, such $\bmy'$ only occurs if line segment $\bmy\bmx(\bmc_s)$ intersects with any of these  $(n-2)$-dimensional hyperplanes. Therefore, the set of $\bmy$ for which the converse case occurs is the union of finitely many $(n-1)$-dimensional hyperplanes, and thus has probability of zero. Hence, we can construct a $\bmy'$ from $\bmy\in\Z_s(\bmc)$ such that $L(\bmy')=s-1$ with probability 1. Namely, $\Z_{s-1}(\bmc)$ is open and nonempty.

By induction, every order-$s$ decoding region $\Z_s(\bmc)$, $s=1,2,\dots,\lambda$, is open and nonempty. Thus, $P_{c,\Psi}$ and $P_{e, \Psi}$ are both strictly increasing in $\Psi$ and will converge to $P_{c, \lambda}$ and $P_{e, \lambda}$ respectively provided that $\Psi\ge \lambda$.
\end{IEEEproof}

\begin{figure}[t]
\centering
\includegraphics[width=0.45\textwidth]{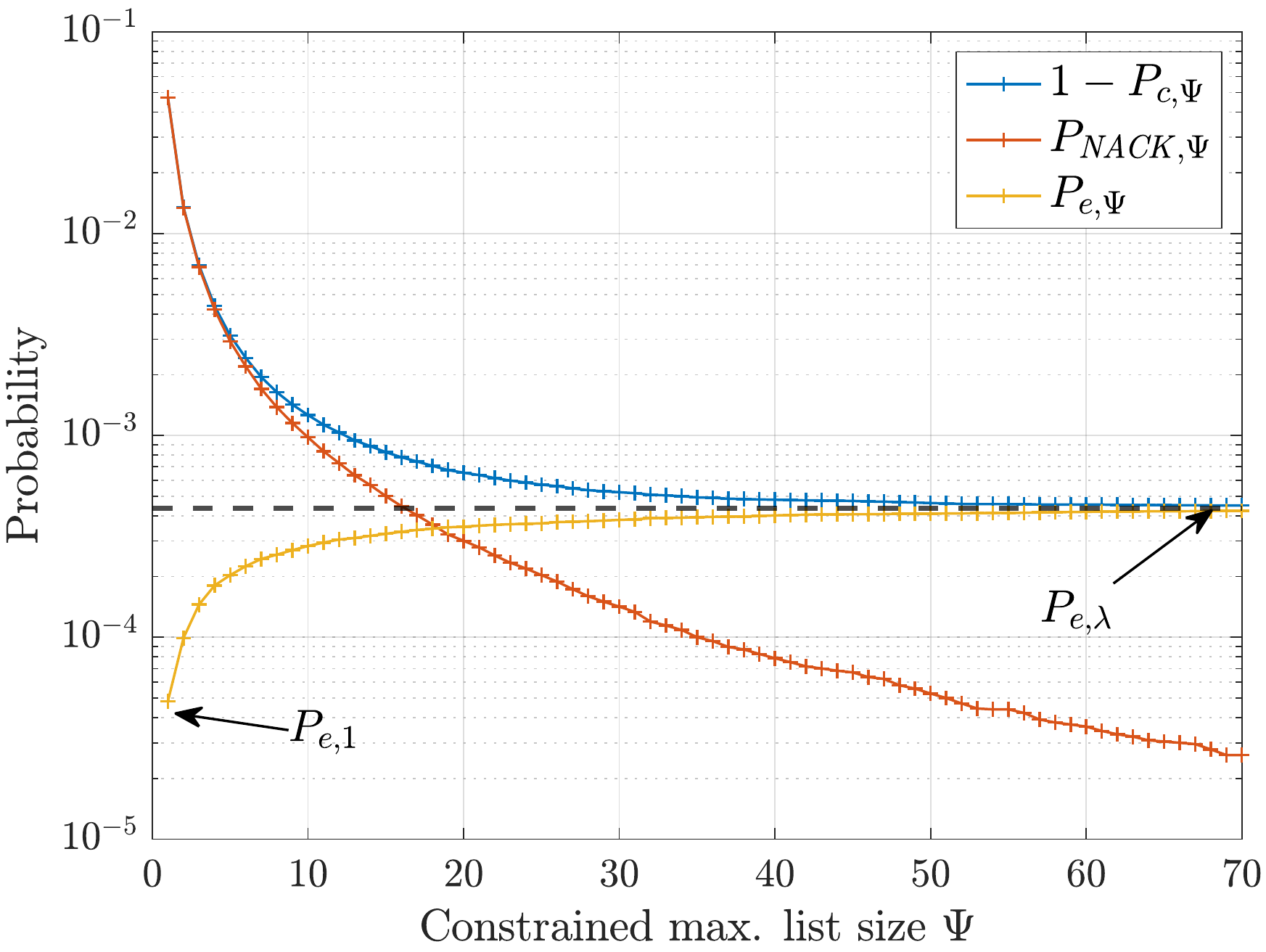}
\caption{$1-P_{c,\Psi}, P_{\NACK, \Psi}, P_{e, \Psi}$ vs. the constraint maximum list size $\Psi$ at SNR $\gamma_s = 3$ dB for ZTCC $(13, 17)$, degree-$6$ DSO CRC polynomial 0x43 and $k=64$ in Table \ref{table: CRC-ZTCC codes}. The black, dashed line represents $P_{e,\lambda}$.}
\label{fig: prob_vs_Psi}
\end{figure}

As an example, Fig. \ref{fig: prob_vs_Psi} shows the probability of UE $P_{e, \Psi}$ and probability of NACK $P_{\NACK, \Psi}$ vs. the constrained maximum list size $\Psi$ for $k = 64$, degree-$6$ DSO CRC polynomial 0x43 and ZTCC $(13, 17)$. It can be seen that $P_{e, \Psi}$ quickly increases and converges to $P_{e, \lambda}$ when $\Psi$ is relatively small.

The monotone property of $P_{e, \Psi}$ with $\Psi$ in Theorem \ref{theorem: probability vs. Psi} indicates that for a fixed SNR value, 
\begin{align}
P_{e, 1}\le P_{e, \Psi}\le P_{e,\lambda},\quad \forall \Psi\in\NN^+.
\end{align}
The proof of Theorem \ref{theorem: probability vs. Psi} also implies that the closure of the order-$\lambda$ decoding region must intersect with the boundary of $\Y(\bmc)$, $\bmc\in\C_l$.  We formalize this notion in Theorem \ref{theorem: supremum list rank region}.

\begin{theorem}\label{theorem: supremum list rank region}
For any lower-rate codeword $\bmc\in\C_l$, $\cl(\Z_{\lambda}(\bmc))\cap\partial \Y(\bmc)\ne\varnothing$. 
\end{theorem}

\begin{IEEEproof}
Fix a lower-rate codeword $\bmc\in\C_l$. Let $\bmy\in\Z_{\lambda}(\bmc)$. Consider $\bmy'=\bmy+t(\bmy - \bmx(\bmc))$, $t\ge0$. By the proof in Theorem \ref{theorem: probability vs. Psi}, if $\bmy'\in\Y(\bmc)$, $L(\bmy')\ge L(\bmy)=\lambda$. Since $\lambda$ is the maximum list rank, $L(\bmy')=\lambda$ for all $0\le t<t^*$, where $t^*$ is the threshold at which $\bmy'\in\partial\Y(\bmc)$. This implies that $\cl(\Z_{\lambda}(\bmc))\cap\partial \Y(\bmc)\ne\varnothing$.
\end{IEEEproof}

Theorem \ref{theorem: supremum list rank region} indicates that one can find $\lambda$ by following along the boundary of $\Y(\bmc)$ and making a slight deviation towards the decoding region $\Y(\bmc)$. This approach is computationally challenging in $\R^n$ for interesting values of $n$. While  $\lambda\le |\C_h| - |\C_l|+1$ provides an initial upper bound on $\lambda$, it remains an open problem to identify a tighter bound on $\lambda$ and to develop an efficient algorithm to compute $\lambda$.

We next direct our attention to quantifying $P_{e, 1}$, $P_{e, \lambda}$ in terms of the SNR (or equivalently in terms of amplitude $A$) and the distance spectra of both the lower-rate code $\C_l$ and the higher-rate code $\C_h$.

\begin{theorem} \label{thm: P_UE, P_NACK vs. SNR}
Under SLVD of a CRC-aided convolutional code with higher-rate distance spectrum $B_{d_{\min}^h},\dots, B_n$ and lower-rate distance spectrum $C_{d_{\min}^l},\dots, C_n$,
\begin{align}
P_{e, 1}&\le \min\Bigg\{2^{-m},\ \sum_{d=d_{\min}^l}^nC_{d}Q\big(A\sqrt{d}\big)\Bigg\}\label{eq: 57}\\
  &\approx \min\Bigg\{2^{-m},\ C_{d_{\min}^l}Q\Big(A\sqrt{d_{\min}^l}\Big)\Bigg\},\label{eq: 58} \\
P_{e, \lambda}&\le \min\Bigg\{1,\ \sum_{d=d_{\min}^l}^nC_dQ\big(A\sqrt{d}\big)\Bigg\}\\
  &\approx \min\Bigg\{1,\ \sum_{d=d_{\min}^l}^{\tilde{d}}C_dQ\big(A\sqrt{d}\big)\Bigg\}, \label{eq: 60} \\
P_{\NACK, 1}&\approx \min\Bigg\{1 - 2^{-m},\notag\\
  &\quad\quad\quad \sum_{d=d_{\min}^h}^{\tilde{d}}B_dQ\big(A\sqrt{d}\big)-C_{d_{\min}^l}Q\Big(A\sqrt{d_{\min}^l}\Big)\Bigg\},\label{eq: 61}
\end{align}
where the second approximation in braces in \eqref{eq: 58} is called the nearest neighbor approximation, and the second approximation in \eqref{eq: 60} is called the truncated union bound (TUB) at distance $\tilde{d}$.
\end{theorem}

\begin{IEEEproof}
First, note that $P_{e,\Psi}$ is a monotonically decreasing function of $A$ for any $\Psi$. This can be seen from \eqref{eq: 45} where as $A$ increases, the center of the Gaussian density is moving away from every $\bmx(\bmc)$ for $\bmc \in\C_l\setminus\{\bar{\bmc}\}$. Hence, the corresponding probability $\Prob(\Z_s(\bmc)|\bmX = \bmx(\bar{\bmc}))$ decreases with $A$, causing $P_{e,\Psi}$ to decrease with $A$.

Now we focus on the $\Psi=1$ case. The previous paragraph reveals that $P_{e,1}$ has its maximum value at $A=0$.  As $A\to0$, the transmitted point converges to the origin $\bm{O}$ in $\R^n$. At the limit where $\bmx(\bar{\bmc})=\bm{O}$, the symmetry of the Gaussian density and linearity of the code ensures that each order-$1$ decoding region has a probability of $2^{-(k+m)}$. Hence,
\begin{align}
P_{e,1}&=\sum_{\bmc\in\C_l\setminus\{\bar{\bmc}\}}\Prob(\Z_1(\bmc)|\bmX=\bmx(\bar{\bmc}))\\
  &\le \lim_{A\to0}\sum_{\bmc\in\C_l\setminus\{\bar{\bmc}\}}\Prob(\Z_1(\bmc)|\bmX=\bmx(\bar{\bmc}))\\
  &= \sum_{\bmc\in\C_l\setminus\{\bar{\bmc}\}}\Prob(\Z_1(\bmc)|\bmX=\bm{O}))\\
  &= (2^k-1)2^{-(k+m)}\le 2^{-m}.
\end{align}

For any SNR value, $P_{e,1}<P_{e,\lambda}$ so that the union bound \eqref{eq: union bound} is also an upper bound for $P_{e,1}$. Hence, the minimum between the two is an upper bound on $P_{e, 1}$. As SNR increases, the majority of probability will concentrate on the nearest neighbors of $\bar{\bmc}$, hence, we can approximate $P_{e, 1}$ only using the nearest neighbors.

For $P_{e, \lambda}$, we upper bound it by the union bound \eqref{eq: union bound}. For ease of computation, we can consider the TUB up to a sufficient distance $\tilde{d}$ to approximate the original union bound.

For $P_{\NACK, 1}$, in the extremely low SNR regime (i.e., when $A$ is close to $0$), $P_{c, 1}\approx 2^{-(k+m)}$ and $P_{e, 1}\approx 2^{-m}(1-2^{-k})$. It follows that
\begin{align}
P_{\NACK, 1} &= 1 - P_{e, 1} - P_{c, 1} \approx 1 - 2^{-m}.
\end{align}
For an arbitrary SNR, invoking the union bound on $P_{\NACK, 1}+P_{e, 1}$ yields
\begin{align}
  P_{\NACK, 1}+P_{e, 1}&\le \sum_{d=d_{\min}^h}^nB_dQ\big(A\sqrt{d}\big)\approx \sum_{d=d_{\min}^h}^{\tilde{d}}B_dQ\big(A\sqrt{d}\big).\notag
\end{align}
Hence,
\begin{align}
  P_{\NACK, 1}&\approx \sum_{d=d_{\min}^h}^{\tilde{d}}B_dQ\big(A\sqrt{d}\big) - C_{d_{\min}^l}Q\Big(A\sqrt{d_{\min}^l}\Big).
\end{align}
This concludes the proof of Theorem \ref{thm: P_UE, P_NACK vs. SNR}.
\end{IEEEproof}

\begin{figure}[t]
\centering
\includegraphics[width=0.45\textwidth]{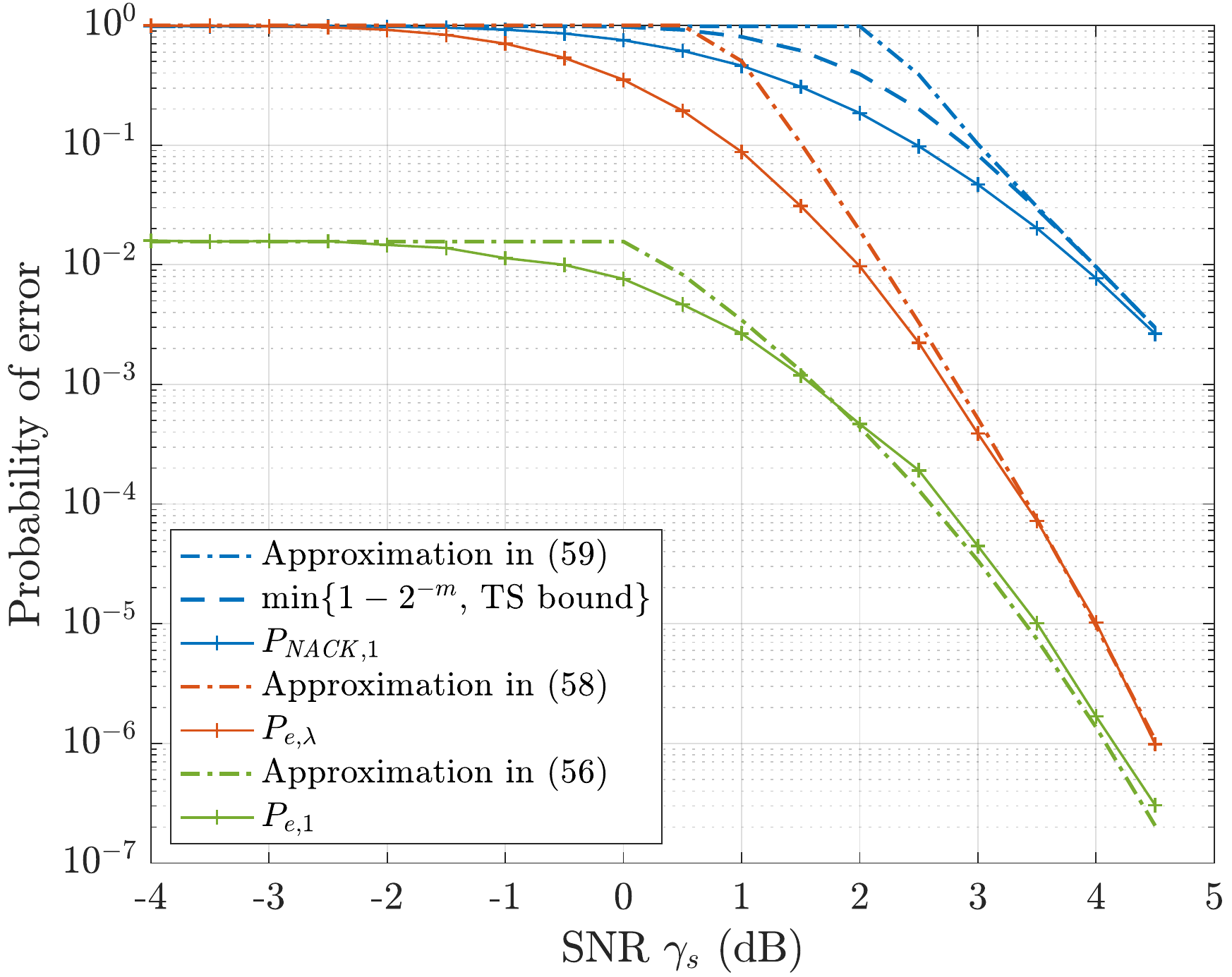}
\caption{$P_{\NACK, 1}$, $P_{e, \lambda}$ and $P_{e, 1}$ vs. SNR $\gamma_s$ for ZTCC $(13, 17)$, degree-$6$ DSO CRC polynomial 0x43 and $k=64$ in Table \ref{table: CRC-ZTCC codes}. The TUBs in \eqref{eq: 60} and \eqref{eq: 61} are obtained at $\tilde{d} = 24$. The TS bound on $P_{\NACK, 1}$ is plotted using \cite[Eq. (14)]{Yousefi2004_J2}. }
\label{fig: prob of UE vs SNR}
\end{figure}

Fig. \ref{fig: prob of UE vs SNR} shows simulation results and  approximations for the three probabilities addressed in Theorem \ref{thm: P_UE, P_NACK vs. SNR}:  $P_{\NACK, 1}$, $P_{e, 1}$, and $P_{e, \lambda}$. As SNR increases, all three approximations become asymptotically tight to the respective $P_{e, 1}$, $P_{\NACK, 1}$, and $P_{e, \lambda}$.  The nearest neighbor approximation of the union bound on $P_{e, \lambda}$ eventually will become asymptotically tight for $P_{e, \lambda}$, but is a tight approximation for $P_{e, 1}$ at a much lower SNR.

We remark that improved upper bounds on $P_{\NACK, 1}$ and $P_{e,\lambda}$ can be derived using Gallager's first bounding technique \cite{Gallager1963}, provided that the full distance spectra of $\C_h$ and $\C_l$ are known, respectively. Some classical examples include the tangential bound \cite{Berlekamp1980}, the tangential sphere (TS) bound \cite{Poltyrev1994,Yousefi2004_J2}, and the added-hyperplane bound \cite{Yousefi2004}. These bounds provide a tight estimation at high noise levels and converge to the union bound at low noise levels. As an example, in Fig. \ref{fig: prob of UE vs SNR}, we plot the minimum between $(1-2^{-m})$ and the TS bound for $P_{\NACK, 1}$ following \cite[Eq. (14)]{Yousefi2004_J2}. It can be seen that the TS bound quickly converges to the TUB as SNR increases. Since this paper mainly focuses on low target error probability, we only consider the TUB for estimating $P_{\NACK, 1}$ and $P_{e,\lambda}$.

\subsection{Analysis of the Expected List Rank}
\label{subsec: analysis of E[L]}

For a fixed transmitted point $\bar{\bmx}$, observe that $\Prob(L=s|\bmX=\bar{\bmx})=\sum_{\bmc\in\C_l}\Prob(Z_s(\bmc)|\bmX=\bar{\bmx})$ is independent of $\Psi$. Combining with the linearity $\E[L] = \E[L|\bmX = \bar{\bmx}]$, it follows that $\E[L]$ is a strictly increasing function in $\Psi$. In the subsequent analysis, we assume that $\Psi\ge\lambda$ and the terminating list rank $L$ ranges from $1$ to $\lambda$ unless otherwise specified.

\begin{theorem}
For a given CRC-aided convolutional code decoded with SLVD, $\lim_{\gamma_s\to0}\E[L] = \E[L|\bmX=\bm{O}]$.
\end{theorem} 
\begin{IEEEproof}
We use the projection method to show the convergence of $\E[L]$ in the low SNR regime.

For ease of discussion, let $\B(\bm{a}, r)$ denote the \emph{spherical surface} of center $\bm{a}\in\R^n$ and radius $r$ in $\R^n$. With BPSK modulation, all codewords sit on the \emph{codeword sphere} $\B(\bm{O}, A\sqrt{n})$, whereas the received point $\bmy$ lies on the \emph{noise sphere} $\B(\bar{\bmx}, w)$ for some noise vector with Euclidean norm $w$ added to the transmitted point $\bar{\bmx}$. The projection method projects the received point $\bmy$ onto the codeword sphere. Namely, the projected point $\bmy_p$ of $\bmy$ is given by $\bmy_p = (A\sqrt{n}/\norm{\bmy})\bmy$. Fig. \ref{fig: projection method} illustrates the geometry of the projection method.

\begin{figure}[t]
\centering
\includegraphics[width=0.23\textwidth]{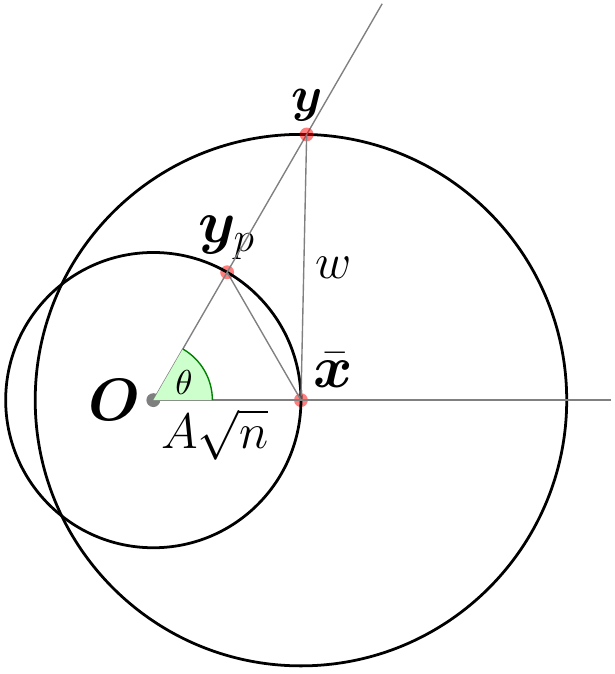}
\caption{An illustration of the projection method. }
\label{fig: projection method}
\end{figure}%

The significance of the projection method introduced above lies in the fact that it preserves the order of list decoded codewords. By law of cosines at angle $\theta$ in Fig. \ref{fig: projection method}, we obtain
\begin{align}
\norm{\bmy_p - \bar{\bmx}}=\begin{cases}
\sqrt{\frac{\norm{\bmy-\bar{\bmx}}^2 - \norm{\bmy-\bmy_p}^2}{1+\frac{\norm{\bmy-\bmy_p}}{A\sqrt{n}}}},& \text{if $\bmy_p$ in between $\bm{O}$, $\bmy$}\\
\sqrt{\frac{\norm{\bmy-\bar{\bmx}}^2 - \norm{\bmy-\bmy_p}^2}{1-\frac{\norm{\bmy-\bmy_p}}{A\sqrt{n}}}},& \text{otherwise}.
\end{cases}
\end{align}
Hence, the monotone relation between $\norm{\bmy_p - \bar{\bmx}}$ and $\norm{\bmy-\bar{\bmx}}$ ensures that performing SLVD using $\bmy$ is equivalent to that using $\bmy_p$. The essential motivation of projecting points onto the codeword sphere is to transfer the computation on the noise sphere to the codeword sphere.

To see how the projection method helps to show the convergence of $\E[L]$, we first decompose the expected list rank $\E[L]$ according to the noise vector norm $W = w$. By linearity of the code,
\begin{align}
\E[L]&=\E[L|\bmX=\bar{\bmx}] \notag\\
  &= \int_{0}^{\infty}f_W(w)\E[L|W = w, \bmX=\bar{\bmx}]\diff w, \label{eqL: E[L] decomposition}
\end{align}
where $f_W(w)$ denotes the density function of norm $W = w$. To find $f_W(w)$, let
\begin{align}
\phi_n(w)&\triangleq\frac{1}{(\sqrt{2\pi})^n}\exp\left(-\frac{w^2}{2}\right),\\
S_{n-1}(w)&\triangleq \frac{2\pi^{\frac{n}{2}}}{\Gamma(\frac{n}{2})}w^{n-1}
\end{align}
be the $n$-dimensional standard normal density function and the spherical area of $\B(\bar{\bmx}, w)$ in $\R^n$, respectively. Then,
\begin{align}
f_W(w) &=\phi_n(w)S_{n-1}(w)=\frac{w^{n-1}}{2^{\frac{n-2}{2}}\Gamma(\frac{n}{2})}\exp\left(-\frac{w^2}{2}\right).
\end{align}
For a given norm $W = w$, it follows that
\begin{align}
\E[L|W = w, \bmX=\bar{\bmx}]&=\frac{1}{S_{n-1}(w)}\int_{\bmy\in\B(\bar{\bmx},w)\setminus\nullset}L(\bmy)\diff\bm{\sigma}, \label{eq: 69}
\end{align}
where $\bm{\sigma}$ denotes the spherical measure on $\B(\bar{\bmx}, w)$. Using the projection method, the integral in \eqref{eq: 69} can be transformed to the codeword sphere at the cost of introducing an induced density function $g_w(\bmy_p)$. Namely,
\begin{align}
\E[L|W = w,\bmX=\bar{\bmx}]&=\int_{\bmy_p\in\B(\bm{O}, A\sqrt{n})\setminus\nullset }L(\bmy_p)g_w(\bmy_p)\diff\bm{\sigma}.
\end{align}
In Appendix \ref{appendix: induced density function}, the induced density function, for $w\ge A\sqrt{n}$, is given by
\begin{align}
g_w(\bmy_p)&=\left(\frac{\norm{\bmy(\bmy_p)}}{w}\right)^{n-1}\frac{1}{\cos\angle\bar{\bmx}\bmy(\bmy_p)\bm{O}}\frac{1}{S_{n-1}(A\sqrt{n})},
\end{align}
where $\bmy(\bmy_p)$ is the pre-image of $\bmy_p$ on the noise sphere $\B(\bar{\bmx}, w)$. Note that $g_w(\bmy_p)$ is rotationally symmetric with respect to axis $\bm{O}\bar{\bmx}$. Appendix \ref{appendix: induced density function} also shows that
\begin{align}
g_w(\bmy_p)&\ge\frac{1}{S_{n-1}(A\sqrt{n})}\left(1-\frac{A\sqrt{n}}{w}\right)^{n-1},\\
g_w(\bmy_p)&\le\frac{1}{S_{n-1}(A\sqrt{n})}\left(1+\frac{A\sqrt{n}}{w}\right)^{n-1}.
\end{align}
This implies that for a fixed norm $w$,
\begin{align}
\lim_{A\to0}\frac{g_w(\bmy_p)}{(S_{n-1}(A\sqrt{n}))^{-1}}=1.
\end{align}
Hence, for a fixed norm $w$, it follows that
\begin{align}
\phantom{=}&\lim_{A\to0}\E[L|W = w,\bmX=\bar{\bmx}]\notag\\
=& \lim_{A\to0}\int_{\bmy_p\in\B(\bm{O}, A\sqrt{n})\setminus\nullset }L(\bmy_p)g_w(\bmy_p)\diff\bm{\sigma}\\
  =& \lim_{A\to0}\int_{\bmy_p\in\B(\bm{O}, A\sqrt{n})\setminus\nullset }L(\bmy_p)\frac{1}{S_{n-1}(A\sqrt{n})}\diff\bm{\sigma}\\
  =& \lim_{A\to0}\E[L|W = A\sqrt{n}, \bmX=\bm{O}]\\
  =& \E[L|\bmX=\bm{O}],
\end{align} 
where we have used the fact that $\E[L|W = w, \bmX=\bm{O}]=\E[L|\bmX=\bm{O}]$ for all $w>0$. Similarly, we can also show that, for a fixed amplitude $A$,
\begin{align}
\lim_{w\to\infty}\E[L|W = w,\bmX=\bar{\bmx}]=\E[L|\bmX=\bm{O}].
\end{align}
As a consequence,
\begin{align}
\lim_{\gamma_s\to0}\E[L]&=\lim_{A\to0}\int_0^{\infty}f_W(w)\E[L|W = w, \bmX=\bar{\bmx}]\diff w\notag\\
  &=\int_0^{\infty}f(w)\lim_{A\to0}\E[L|W = w, \bmX=\bar{\bmx}]\diff w\notag\\
  &=\int_0^{\infty}f(w)\E[L|\bmX=\bm{O}]\diff w\notag\\
  &=\E[L|\bmX=\bm{O}].
\end{align}
This completes the proof.
\end{IEEEproof}

The proof above implies that in the low SNR regime, most of the probability will concentrate on the limit of $\E[L|W = w, \bmX=\bar{\bmx}]$ as $w\to\infty$, i.e., $\E[L|\bmX=\bm{O}]$. In general, $\E[L|\bmX=\bm{O}]$ depends on the geometric structure of the lower-rate code $\C_l$ and the higher-rate code $\C_h$ on $\B(\bm{O}, A\sqrt{n})$ and it is not easy to obtain an analytic expression. Still, using a simple random coding argument, we show that a good concatenated code could achieve $\E[L|\bmX=\bm{O}]\le 2^m$.

\begin{theorem}\label{theorem: random coding argument}
For a given higher-rate code $\C_h$ with $|\C_h|=2^{k+m}$, let $\A_l\triangleq\{\C'\subset\C_h: |\C'|=2^k\}$. Let $\Prob(\C')=\frac{1}{|\A_l|}$ be the uniform distribution defined over $\A_l$. Assume $\C'$ is drawn according to $\Prob(\C')$. Then,
\begin{align}
\E_{\C'}\big[\E[L|\bmX=\bm{O}, \C']\big]\le 2^m.
\end{align}
This implies that there exists a lower-rate code $\C'$ (which may not be a linear code) such that $\E[L|\bmX=\bm{O}, \C']\le 2^m$.
\end{theorem}

\begin{IEEEproof}
Let $L(\bmy, \C')$ be the terminating list rank for received point $\bmy\in\R^n$ when a lower-rate code is selected as $\C'\in\A_l$\footnote{If there exist two codewords $\bmc_{j_1}$ and $\bmc_{j_2}$ that are equidistant from $\bmy$, the decoder adopts a pre-determined order relation between $\bmc_{j_1}$ and $\bmc_{j_2}$. }. Hence, we obtain
\begin{align}
\phantom{=}&\E_{\C'}\big[\E[L|\bmX=\bm{O}, \C']\big]\notag\\
=&\sum_{\C'\in\A_l}\Prob(\C')\frac{1}{S_{n-1}(A\sqrt{n})}\int_{\bmy\in\B(\bm{O},A\sqrt{n})}L(\bmy, \C')\diff \bm{\sigma}\notag\\
=&\frac{1}{S_{n-1}(A\sqrt{n})}\int_{\bmy\in\B(\bm{O},A\sqrt{n})}\sum_{\C'\in\A_l}\Prob(\C')L(\bmy, \C')\diff \bm{\sigma}\notag\\
=&\frac{1}{S_{n-1}(A\sqrt{n})}\int_{\bmy\in\B(\bm{O},A\sqrt{n})}\E_{\C'}[L(\bmy, \C')|\bmy] \diff \bm{\sigma}. \label{eq: p84}
\end{align}
Next, we show that for any $\bmy\in\B(\bm{O}, A\sqrt{n})$, 
\begin{align}
\E_{\C'}[L(\bmy, \C')|\bmy]\le 2^m \label{eq: p85}
\end{align}
for $\C'$ uniformly drawn from $\A_l$. Fix a $\bmy\in\B(\bm{O}, A\sqrt{n})$ and let $\bmc_1(\bmy), \bmc_2(\bmy), \dots, \bmc_{|\C_h|}(\bmy)$ be an enumeration of $\C_h$ such that
\begin{align}
\norm{\bmy-\bmx(\bmc_1(\bmy))}\le\dots\le\norm{\bmy-\bmx(\bmc_{|\C_h|}(\bmy))}.\notag
\end{align}
Hence, the terminating list rank $L(\bmy, \C')$ of $\bmy$ is given by
\begin{align}
L(\bmy, \C') = \min\{s: \bmc_s(\bmy)\in\C'\}.
\end{align}
For $\C'$ uniformly drawn in $\A_l$, computing $\E_{\C'}[L(\bmy, \C')|\bmy]$ is equivalent to solving the following problem: there are $|\C_h|$ balls in a basket, among which $|\C'|$ of them are red and the rest are white. Balls are picked up $|\C_h|$ times without replacement and the time at which the first red ball emerges is marked as the terminating list rank. Since every ordering of ball picking is equiprobable and is bijective with $\A_l$, the expected list rank in ball picking problem is equal to $\E_{\C'}[L(\bmy, \C')|\bmy]$. Hence,
\begin{align}
\E_{\C'}[L(\bmy, \C')|\bmy]&= \sum_{s=1}^{|\C_h|-|\C'|+1}s\frac{\binom{|\C_h|-s}{|\C'|-1}}{\binom{|\C_h|}{|\C'|}}\label{eq: 90}\\
  &=\frac{|\C_h|+1}{|\C'|+1}\label{eq: 91}\\
  &\le2^m,\notag
\end{align}
where \eqref{eq: 91} follows from a variant of the Chu-Vandermonde identity.

Finally, substituting \eqref{eq: p85} into \eqref{eq: p84} proves Theorem \ref{theorem: random coding argument}.
\end{IEEEproof}

\begin{figure}[t]
\centering
\includegraphics[width=0.45\textwidth]{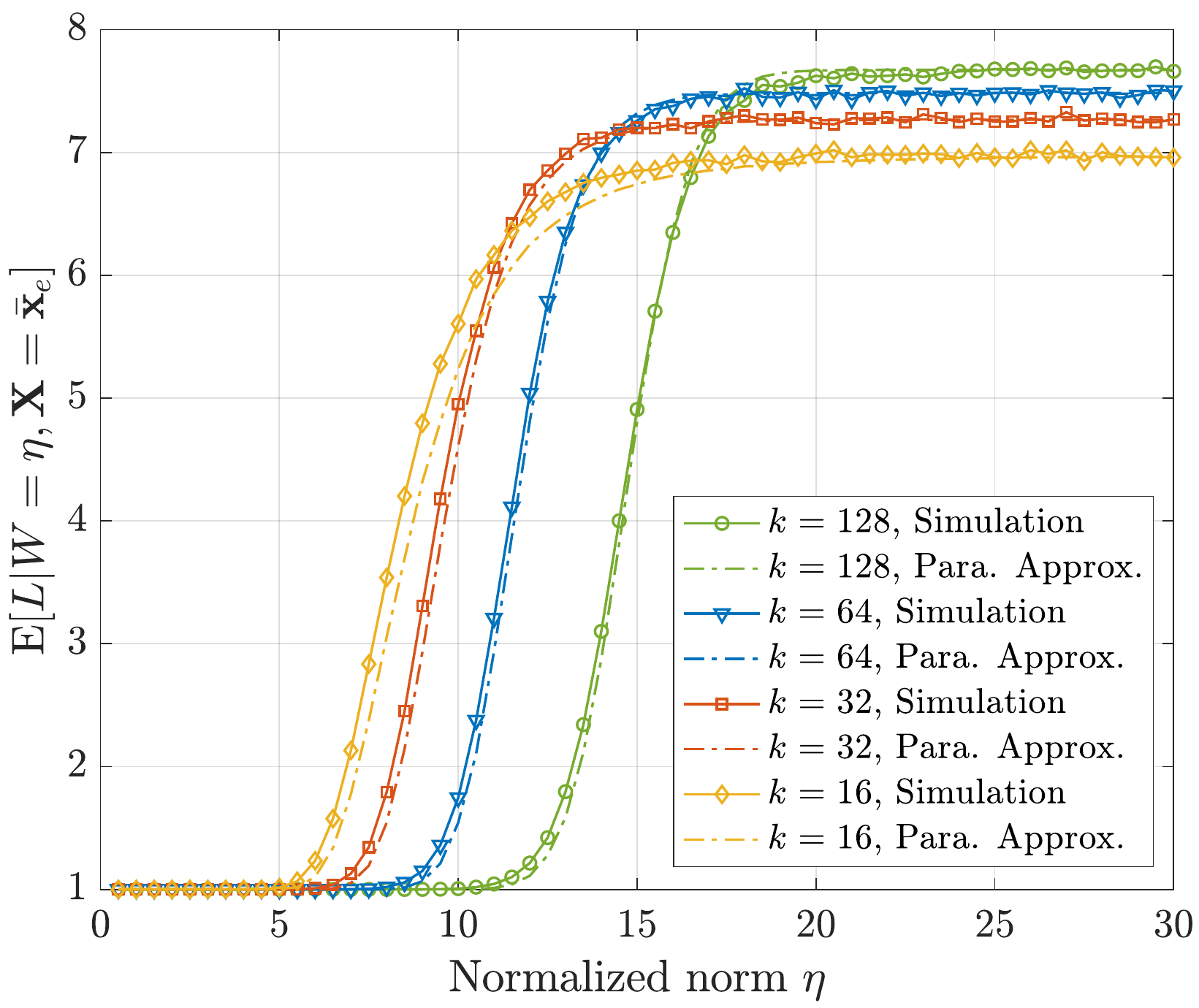}
\caption{The conditional expected list rank $\E[L|W = \eta, \bmX=\bar{\bmx}_e]$ vs. the normalized norm $\eta$ for the CRC-ZTCC generated with the degree-$3$ DSO CRC polynomial 0x9 and ZTCC $(13, 17)$. }
\label{fig: cond_exp_list_rank_vs_eta}
\end{figure}

\begin{figure}[t]
\centering
\includegraphics[width=0.45\textwidth]{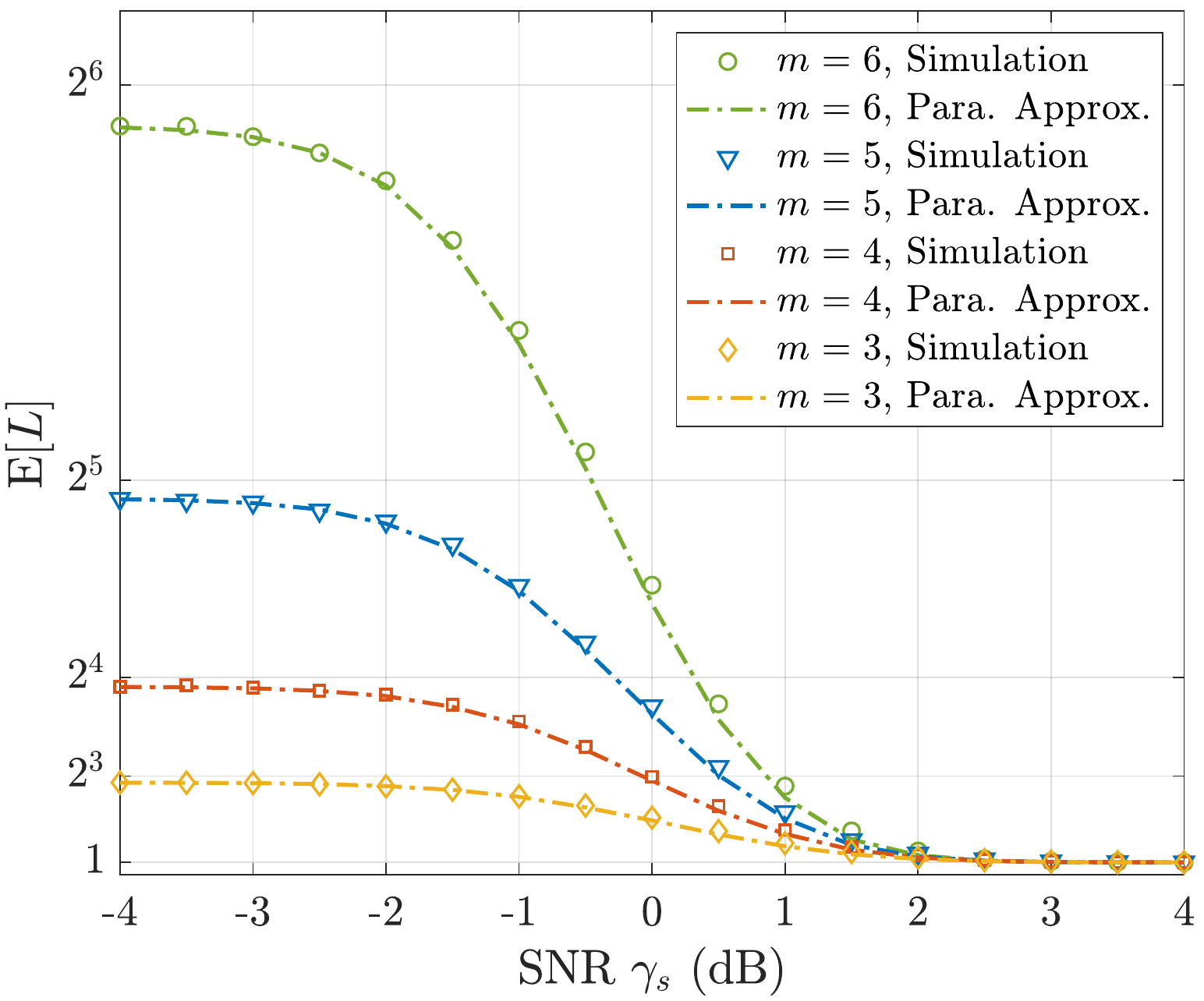}
\caption{The expected list rank $\E[L]$ vs. SNR for various CRC-ZTCCs, where ZTCC is $(13, 17)$  and the DSO CRC polynomials are from Table \ref{table: CRC-ZTCC codes} with degree $m=3,4,\dots, 6$. The information length $k = 64$. }
\label{fig: exp_list_rank_vs_SNR}
\end{figure}

In \eqref{eqL: E[L] decomposition}, it is shown that $\E[L]$ can be fully characterized by its conditional expectation $\E[L|W = w, \bmX=\bar{\bmx}]$. For a given $w$ and $A$, let $\bar{\bmx}_e = \bar{\bmx}/A$ be the transmitted point with unit amplitude per dimension. Then it can be shown that
\begin{align}
\E[L|W = w, \bmX=\bar{\bmx}] = \E[L|W = \eta, \bmX=\bar{\bmx}_e], \label{eq: normalized norm}
\end{align}
where $\eta\triangleq w/A$ is called the \emph{normalized norm}.
Hence, it suffices to compute $\E[L|W = \eta, \bmX=\bar{\bmx}_e]$. The SNR (equivalently, the BPSK amplitude $A$) only exhibits a scaling effect. To evaluate $\E[L|W = \eta, \bmX=\bar{\bmx}_e]$, let $\C_l^-\triangleq\C_l\setminus\{\bar{\bmc}\}$ and define the conditional probability of UE conditioned on the sphere $\B(\bar{\bmx}_e, \eta)$ as
\begin{align}
  P_{e, \lambda}(\eta)\triangleq \sum_{\bmc\in\C_l^- } \Prob(\Y(\bmc)|W =\eta, \bmX=\bar{\bmx}_e).
\end{align}
In general, it is difficult to know the conditional probability of UE $P_{e,\lambda}(\eta)$. Assuming the knowledge of parametric information $P_{e, \lambda}(\eta)$, we first show an approximation that represents $\E[L|W = \eta, \bmX=\bar{\bmx}_e]$ as a linear combination between $L=1$ and $L=\bar{L}$ with coefficient given by $P_{e, \lambda}(\eta)$.

\begin{approximation}[Parametric approximation]\label{approx: parametric approx}
For a CRC-aided convolutional code with corresponding parameters of $\bar{L}$ and $P_{e, \lambda}(\eta)$, where  $\bar{L}\triangleq \E[L|\bmX=\bm{O}]$, 
\begin{align}
\E[L|W = \eta, \bmX=\bar{\bmx}_e]\approx 1-P_{e, \lambda}(\eta)+P_{e, \lambda}(\eta)\bar{L}. \label{eq: parametric approx}
\end{align}
Furthermore, averaging over $W = \eta$ on both sides of \eqref{eq: parametric approx} yields the approximation of $\E[L]$, i.e.,
\begin{align}
  \E[L]\approx 1 - P_{e,\lambda}+ P_{e,\lambda}\E[L|\bmX=\bm{O}].\label{eq: overall genie approximation}
\end{align}
\end{approximation}

\begin{IEEEproof}[Justification]
For ease of discussion, we use the shorthand notation $\Prob(\cdot|\eta, \bar{\bmx}_e) \triangleq \Prob(\cdot|W = \eta, \bmX=\bar{\bmx}_e)$ and $\Prob(\cdot|\bm{O})=\Prob(\cdot|\bmX=\bm{O})$. Let us consider $\eta$ for which $P_{e,\lambda}(\eta)>0$. Hence,
\begin{align}
\phantom{=}&\E[L|W = \eta, \bmX=\bar{\bmx}_e]\notag\\
=& \sum_{s=1}^{\lambda}s\Prob(L=s|\eta, \bar{\bmx}_e)\notag\\
=& \Prob(\Y(\bar{\bmc})|\eta, \bar{\bmx}_e)+\sum_{s=1}^{\lambda}s\Prob(L=s|\eta, \bar{\bmx}_e)-\sum_{s=1}^{\lambda}\Prob(\Z_s(\bar{\bmc})|\eta, \bar{\bmx}_e )\notag\\
\ge& 1- P_{e, \lambda}(\eta)+\sum_{s=1}^{\lambda}s\big(\Prob(L=s|\eta, \bar{\bmx}_e)-\Prob(\Z_s(\bar{\bmc})|\eta, \bar{\bmx}_e )\big)\notag\\
=& 1- P_{e, \lambda}(\eta)+P_{e, \lambda}(\eta)\left(\sum_{s=1}^{\lambda}s\frac{\sum_{\bmc\in\C_l^-}\Prob(\Z_s(\bmc)|\eta, \bar{\bmx}_e)}{\sum_{\bmc\in\C_l^-}\Prob(\Y(\bmc)|\eta, \bar{\bmx}_e)} \right)\label{eq: 98}\\
\approx& 1- P_{e, \lambda}(\eta)+P_{e, \lambda}(\eta)\left(\sum_{s=1}^{\lambda}s\Prob(L=s|\bm{O}) \right)\label{eq: 99}\\
=& 1- P_{e, \lambda}(\eta)+P_{e, \lambda}(\eta)\bar{L},\notag
\end{align}
where \eqref{eq: 99} follows from the substitution below. Consider the conditional list rank distribution
\begin{align}
\bmP_{\eta}=\Big(\frac{\sum_{\bmc\in\C_l^-}\Prob(\Z_1(\bmc)|\eta, \bar{\bmx}_e)}{\sum_{\bmc\in\C_l^- }\Prob(\Y(\bmc)|\eta, \bar{\bmx}_e)},\dots, \frac{\sum_{\bmc\in\C_l^-}\Prob(\Z_{\lambda}(\bmc)|\eta, \bar{\bmx}_e)}{\sum_{\bmc\in\C_l^- }\Prob(\Y(\bmc)|\eta, \bar{\bmx}_e)} \Big).
\end{align}
Using the fact that $\lim_{\eta\to\infty}g_{\eta}(\bmy_p)=1/S_{n-1}(\sqrt{n})$, the conditional list rank distribution $\bmP_{\eta}$ will converge to
\begin{align}
\bmP_{\infty}&=\left(\frac{\Prob(\Z_1(\bmc)|\bm{O})}{\Prob(\Y(\bmc)|\bm{O})},\dots, \frac{\Prob(\Z_{\lambda}(\bmc)|\bm{O})}{\Prob(\Y(\bmc)|\bm{O})} \right)\label{eq: 101}\\
  &=\left(\frac{\sum_{\bmc\in\C_l} \Prob(\Z_1(\bmc)|\bm{O})}{\sum_{\bmc\in\C_l} \Prob(\Y(\bmc)|\bm{O})},\dots, \frac{\sum_{\bmc\in\C_l}\Prob(\Z_{\lambda}(\bmc)|\bm{O})}{\sum_{\bmc\in\C_l}\Prob(\Y(\bmc)|\bm{O})}\right)\notag\\
  &=(\Prob(L=1|\bm{O}), \dots, \Prob(L=\lambda|\bm{O})), \label{eq: 102}
\end{align}
where $\bmc$ is any lower-rate codeword in \eqref{eq: 101}. Hence, we directly replace $\bmP_{\eta}$ with the limit distribution $\bmP_{\infty}$ in \eqref{eq: 98}. Finally, averaging over $W = \eta$ on both sides of \eqref{eq: parametric approx} yields \eqref{eq: overall genie approximation}.
\end{IEEEproof}

Fig. \ref{fig: cond_exp_list_rank_vs_eta} shows the simulation results of the conditional expected list rank $\E[L|W = \eta, \bmX=\bar{\bmx}_e]$ vs. the normalized norm $\eta$ for CRC-ZTCCs with various information lengths. The corresponding parametric approximation is also provided. We see that the parametric approximation exhibits a remarkable accuracy that improves as $k$ increases. Observe that for large values of $k$, the convergent $\E[L|W = \eta, \bmX=\bar{\bmx}_e]$ is close to $2^m$.

Using \eqref{eqL: E[L] decomposition} and \eqref{eq: normalized norm}, we can produce $\E[L]$ as a function of SNR $\gamma_s$. Fig. \ref{fig: exp_list_rank_vs_SNR} shows $\E[L]$ vs. SNR along with its parametric approximations for ZTCC $(13, 17)$ and various DSO CRC polynomials of degree $m=3,4,\dots, 6$. We see that the parametric approximation on $\E[L]$ remains extremely tight.

The parametric approximation provides a practically useful quantitative connection between performance and complexity. Specifically, for CRC-ZTCCs with a target probability of UE $P_{e,\lambda}^*$ and $\bar{L}\approx 2^m$ for CRC degree $m$, \eqref{eq: overall genie approximation} implies that a CRC with degree $m\le -\log(P_{e,\lambda}^*)$ is sufficient to maintain $\E[L]\le 2$, which ensures that the average complexity for SLVD to achieve $P_{e, \lambda}^*$ is at most one more traceback than the standard Viterbi decoding.

As an alternative to Approximation \ref{approx: parametric approx},  we provide a higher-order approximation formula for a good CRC-aided convolutional code that only requires knowledge of $\E[L|\bmX=\bm{O}]$. This alternative approximation is motivated by Shannon's observation \cite{Shannon1959} that an optimal $(n, M)$ code places its codewords on the surface of a sphere such that the total solid angle $\Omega_0$ is evenly divided among the $M$ Voronoi regions, one for each codeword, and that each Voronoi region is a circular cone. Hence, if the CRC-aided convolutional code is good enough, the union of order-$1$ to order-$\mu$ decoding regions $\Z_s(\bmc)$ for a lower-rate codeword $\bmc\in\C_l$ should resemble circular cones, where $\mu$ is a parameter to be optimized. From this perspective, we propose the \emph{onion model} for the order-$1$ decoding region to the order-$\mu$ decoding region based on the following assumptions. 
\begin{enumerate}
  \item The union $\bigcup_{i=1}^s\Z_i(\bmc)$ of order-$1$ to order-$s$ decoding regions, $1\le s\le\mu$, is a circular cone with half-angle $\alpha_s$. This implies that each order-$s$ decoding region, $2\le s\le \mu$ is an \emph{annulus} in between two circular cones.
  \item The solid angle $\Omega(\alpha_s)$ of $\bigcup_{i=1}^s\Z_i(\bmc)$ is equal to $\frac{s}{2^{k+m}}\Omega_0$, $1\le s\le \mu$, where $\Omega_0$ is the total solid angle (i.e., the area of a unit sphere in $\R^n$).
  \item The conditional expected list rank beyond $\bigcup_{i=1}^{\mu}\Z_i(\bar{\bmc})$ is equal to $\bar{L}$ (i.e., $\E[L|\bmX=\bm{O}]$).
\end{enumerate}

\begin{approximation}[Higher-order approximation]\label{approx: higher-order approx}
For a given CRC-aided convolutional code, let $\bar{L}=\E[L|\bmX=\bm{O}]$. With the onion model assumptions and parameter $\mu$, $\mu\in\NN$, $\E[L|W = \eta, \bmX=\bar{\bmx}_e]$ is approximated by
\begin{align}
  &\E[L|W = \eta, \bmX=\bar{\bmx}_e]\notag\\
  &\approx\begin{cases}
      1,\quad \text{if }\eta<\sqrt{n}\sin\alpha_1\\
      \dots   \\
      s - \sum_{i=1}^{s-1}F_{\bar{\bmx}_e}(i),\  \text{if }\sqrt{n}\sin\alpha_{s-1}\le\eta<\sqrt{n}\sin\alpha_s\\
      \dots \\
      \bar{L}-(\bar{L}-\mu)F_{\bar{\bmx}_e}(\mu)-\sum_{i=1}^{\mu-1}F_{\bar{\bmx}_e}(i),\ \text{if }\eta\ge \sqrt{n}\sin\alpha_{\mu},
  \end{cases}
\end{align}
where assuming $\eta \ge \sqrt{n}\sin\alpha_s$,
\begin{align}
  &F_{\bar{\bmx}_e}(s) = \frac{\Gamma\big(\frac{n}{2}\big)}{\sqrt{\pi}\Gamma\big(\frac{n-1}{2}\big)}\notag\\
  &\phantom{F_{\bar{\bmx}_e}(s)=}\cdot\Bigg(\int_0^{\beta_{s,1}}\sin^{n-2}\theta\diff\theta+\int_0^{\beta_{s,2}}\sin^{n-2}\theta\diff\theta\Bigg),\\
    &\beta_{s,1} = \frac{\pi}{2}+\alpha_s-\arcsin\Bigg(\frac{\sqrt{\eta^2-n\sin^2\alpha_s}}{\eta}\Bigg),\\
    &\beta_{s,2} =\left(\frac{\pi}{2}-\alpha_s-\arcsin\Bigg(\frac{\sqrt{\eta^2-n\sin^2\alpha_s}}{\eta}\Bigg)\right)\indicator_{\{\eta\le \sqrt{n}\}},
\end{align}
and $\alpha_s$ is the half-angle for which
\begin{align}
\frac{\Omega(\alpha_s)}{\Omega_0}=\frac{\Gamma\big(\frac{n}{2}\big)}{\sqrt{\pi}\Gamma\big(\frac{n-1}{2}\big)}\int_0^{\alpha_s}\sin^{n-2}\theta\diff\theta=\frac{s}{2^{k+m}}.
\end{align}
\end{approximation}

\begin{IEEEproof}[Justification]
The onion model assumptions implies that each higher order decoding region $\Z_s(\bmc)$, $2\le s\le\mu$, is an annulus in between two circular cones. Hence, $\Prob(L=s|W=\eta, \bmX=\bar{\bmx}_e)$ is simply the spherical area of $\B(\bar{\bmx}_e, \eta)$ cut out by the annulus. To evaluate this quantity, consider the cumulative probability function of $L=s$,
\begin{align}
  F_{\bar{\bmx}_e}(s)&\triangleq\Prob(L\le s, \bmX=\bar{\bmx}_e). \label{eq: p103}
\end{align}
Thus,
\begin{align}
\Prob(L=s|W=\eta, \bmX=\bar{\bmx}_e)=F_{\bar{\bmx}_e}(s)-F_{\bar{\bmx}_e}(s-1).
\end{align}
By the onion model assumptions, for $\eta\ge \sqrt{n}\sin\alpha_{\mu}$,
\begin{align}
&\E[L|W = \eta, \bmX=\bar{\bmx}_e]\\
\approx&\sum_{i=1}^{\mu}i(F_{\bar{\bmx}_e}(i)-F_{\bar{\bmx}_e}(i-1))+\bar{L}(1-F_{\bar{\bmx}_e}(\mu))\\
=&\bar{L} - (\bar{L}-\mu)F_{\bar{\bmx}_e}(\mu)-\sum_{i=1}^{\mu-1}F_{\bar{\bmx}_e}(i).
\end{align}
In the similar fashion, for $\sqrt{n}\sin\alpha_{s-1}\le\eta<\sqrt{n}\sin\alpha_s$, $1\le s\le \mu$,
\begin{align}
\E[L|W = \eta, \bmX=\bar{\bmx}_e]\approx s-\sum_{i=1}^{s-1}F_{\bar{\bmx}_e}(i).
\end{align}
Next, we derive the cumulative probability function $F_{\bar{\bmx}_e}(s)$. Geometrically, $F_{\bar{\bmx}_e}(s)$ is the fraction of the spherical area of $\B(\bar{\bmx}_e, \eta)$ cut out by the circular cone $\bigcup_{i=1}^s\Z_s(\bar{\bmc})$ with half-angle $\alpha_s$ to the total noise spherical area. Assume that $\sqrt{n}\sin\alpha_s\le \eta\le \sqrt{n}$. Fig. \ref{fig: geometry of CPF} shows the side view of this scenario in $\R^3$, in which the blue arc represents the spherical area contained in $\bigcup_{i=1}^s\Z_s(\bar{\bmc})$. It can be seen that $\alpha_s$ will induce two possible half-angles $\beta_{s, 1}$ and $\beta_{s, 2}$. By law of cosines,
\begin{figure}[t]
\centering
\includegraphics[width=0.3\textwidth]{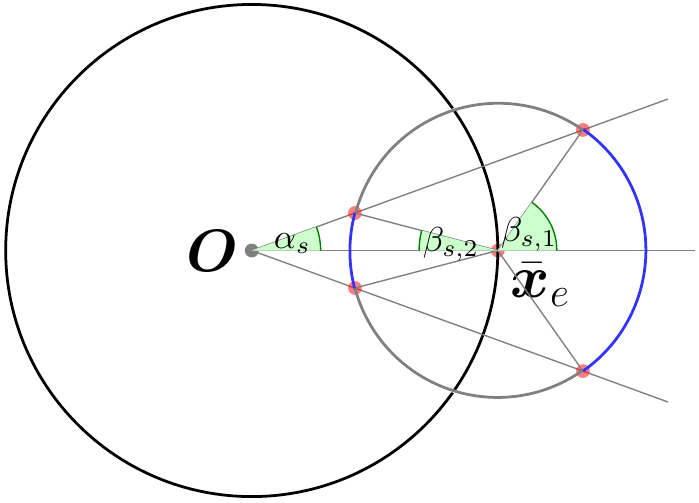}
\caption{The geometry of the cumulative probability function $F_{\bar{\bmx}_e}(s)$, assuming that $\sqrt{n}\sin\alpha_s\le \eta\le \sqrt{n}$. }
\label{fig: geometry of CPF}
\end{figure}

\begin{align}
  \beta &= \frac{\pi}{2} \pm\alpha_s-\arcsin\left(\frac{r_2-r_1}{2\eta}\right) \\
    &= \frac{\pi}{2} \pm\alpha_s-\arcsin\left(\frac{\sqrt{\eta^2-n\sin^2\alpha_s}}{\eta}\right),
\end{align}
where $r_1, r_2$ are solutions to
\begin{align}
r^2 - (2\sqrt{n}\cos\alpha_s)r + (n - \eta^2) = 0.
\end{align}
The induced half-angle $\beta$ becomes unique once $\eta>\sqrt{n}$. 

\begin{figure}[t]
\centering
\includegraphics[width=0.45\textwidth]{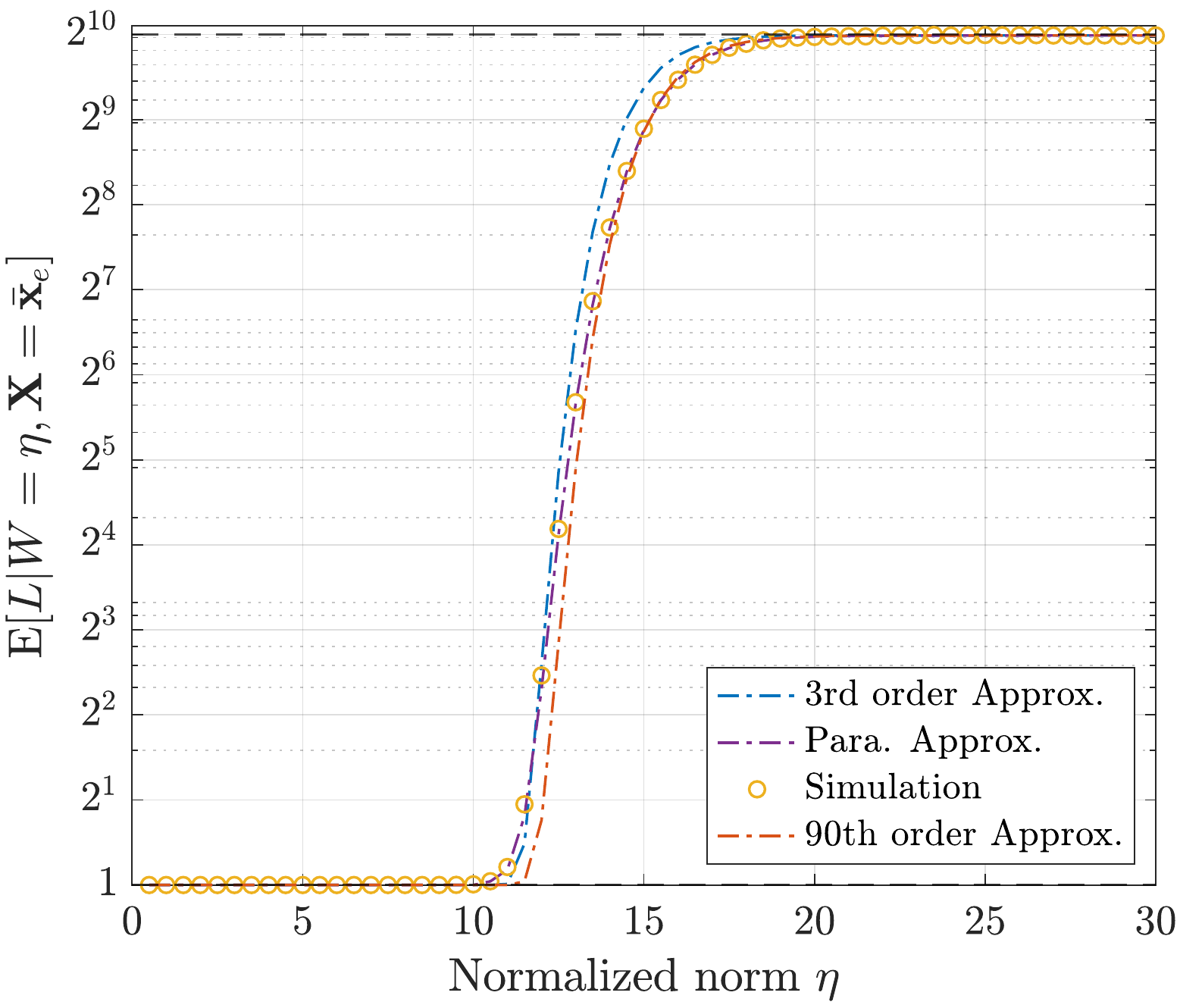}
\caption{The parametric and higher-order approximations of $\E[L|W = \eta, \bmX=\bar{\bmx}_e]$ for ZTCC $(561, 753)$ used with the degree-$10$ DSO CRC polynomial 0x4CF at $k = 64$. Both higher-order approximations assume the knowledge of $\bar{L}=1017$. }
\label{fig: higher order approximation}
\end{figure}

\begin{figure}[t]
\centering
\includegraphics[width=0.45\textwidth]{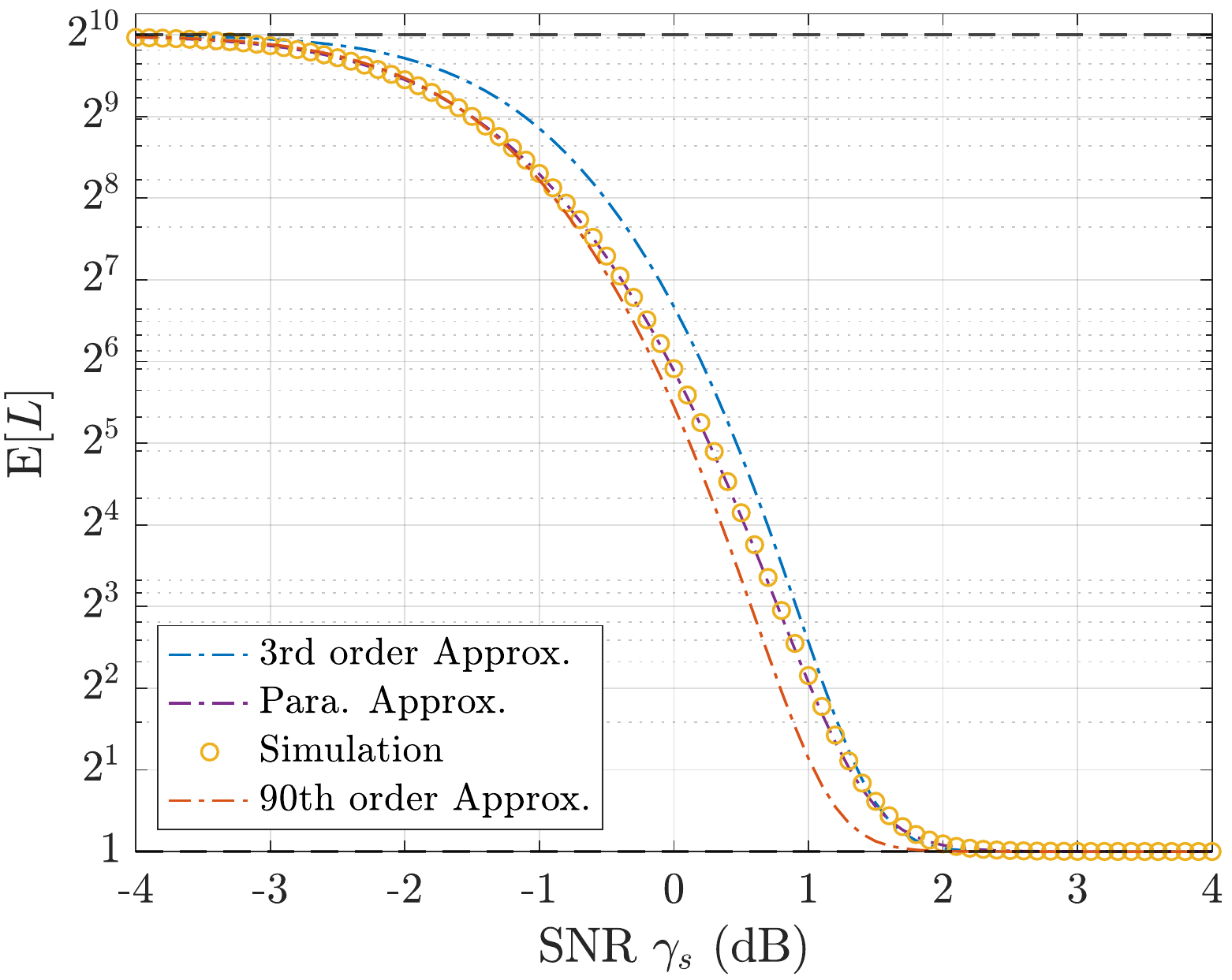}
\caption{The expected list rank $\E[L]$ vs. SNR via \eqref{eqL: E[L] decomposition} and \eqref{eq: normalized norm} for ZTCC $(561, 753)$, degree-$10$ DSO CRC polynomial 0x4CF at $k=64$. }
\label{fig: overall approximation}
\end{figure}

From \cite[Eq. (21)]{Shannon1959}, the solid angle $\Omega(\alpha)$ of a circular cone with center $\bm{O}$ and half-angle $\alpha$ in $n$-dimensional Euclidean space is given by
\begin{align}
    \Omega(\alpha) = \frac{2\pi^{\frac{n-1}{2}} }{\Gamma\big({\frac{n-1}{2} }\big)}\int_0^{\alpha}\sin^{n-2}\theta \diff \theta. \label{eq: p112}
\end{align}
The total solid angle $\Omega_0$ in $n$-dimensional Euclidean space is given by
\begin{align}
    \Omega_0 = \frac{2\pi^{\frac{n}{2}} }{\Gamma\big(\frac{n}{2} \big) }. \label{eq: p113}
\end{align}
Thus, using \eqref{eq: p112}, \eqref{eq: p113}, we can solve $\alpha_s$ from assumption 2 of the onion model. Namely, $\alpha_s$ is the solution to
\begin{align}
    \frac{\Omega(\alpha)}{\Omega_0} = \frac{\Gamma\big({\frac{n}{2} }\big)}{\sqrt{\pi}\Gamma\big({\frac{n-1}{2} }\big)}\int_0^{\alpha}\sin^{n-2}\theta \diff \theta =  \frac{s}{2^{k+m}}.
\end{align}
By geometry in Fig. \ref{fig: geometry of CPF}, $F_{\bar{\bmx}_e}(s)$ in \eqref{eq: p103} is given by
\begin{align}
    F_{\bar{\bmx}_e}(s)&= \frac{\Omega(\beta_{s,1}) + \Omega(\beta_{s,2})}{\Omega_0}.
\end{align}
This concludes the justification of Approximation \ref{approx: higher-order approx}.
\end{IEEEproof}

To demonstrate the tightness of the proposed approximation for good enough CRC-aided convolutional codes, Fig. \ref{fig: higher order approximation} shows the approximations of $\E[L|W = \eta, \bmX=\bar{\bmx}_e]$ for ZTCC $(561, 753)$ used with the degree-$10$ DSO CRC polynomial 0x4CF at $k = 64$ with $\mu=3$ and $90$. This concatenated code has a minimum distance $d_{\min}^l=20$ and thus can be deemed as good enough. When $\mu=3$, our approximation accurately gives the smaller values of the actual conditional expected list rank. As $\mu$ increases, the accuracy of the approximation will shift towards large values of conditional expected list rank. Fig. \ref{fig: overall approximation} illustrates the approximation of $\E[L]$ vs. SNR via \eqref{eqL: E[L] decomposition} and \eqref{eq: normalized norm}. The $3$rd-order and $90$-th order approximations still behave in the similar fashion as in Fig. \ref{fig: higher order approximation}.

\subsection{Complexity Analysis}
There are a variety of implementations of list decoding of convolutional codes as described in, e.g., \cite{Bai2004,Lijofi2004,Roder2006,Kim2018,Shirvanimoghaddam2019}. In this paper, the SLVD implementation maintains a list of path metric differences by using a red-black tree as described in \cite{Roder2006}, which provides the fastest runtime we found among the data structures that support full floating-point precision. The literature mentioned above also analyzed the number of bit operations or the asymptotic complexity of the algorithms presented, but those complexity metrics are not directly connected with actual runtime. To explore how the additional complexity of SLVD of CRC-ZTCCs relative to the standard soft Viterbi (SSV) decoding, we develop an average complexity expression that closely approximates our empirical runtimes.

For our specific implementation, three components comprise the average complexity of SLVD, given by
\begin{align}
C_{\SLVD} = C_{\SSV} + C_{\trace} + C_{\tlist}. \label{eq: SLVD complexity formula}
\end{align}

The first component $C_{\SSV}$ is the complexity required to perform the add-compare-select (ACS) operations on the trellis of the given convolutional code and perform the initial traceback associated with SSV. Specifically, for CRC-ZTCCs, this quantity is given by
\begin{align}
C_{\SSV}&=(2^{\nu+1}-2)+1.5(2^{\nu+1}-2)+1.5(k+m-\nu)2^{\nu+1}\notag\\
  &\phantom{=}+c_1[2(k+m+\nu)+1.5(k+m)]. \label{eq: C_SSV}
\end{align}
For CRC-TBCCs, this quantity is given by
\begin{align}
C_{\SSV}&=1.5(k+m)2^{\nu+1}+2^{\nu}+3.5c_1(k+m). \label{eq: C_SSV for CRC-TBCC}
\end{align}
In order to measure the decoding complexity, define $1$ unit of complexity as the complexity required by performing one addition. In \eqref{eq: C_SSV} and \eqref{eq: C_SSV for CRC-TBCC}, we assign $1$ unit of complexity to each addition per branch and $0.5$ units of complexity to each compare-select operation per branch. In the first and second terms of \eqref{eq: C_SSV}, $(2^{\nu+1}-2)$ counts the number of edges in the initial $\nu$ sections and the final $\nu$ termination sections of a ZT trellis. In the third term of \eqref{eq: C_SSV}, $(k+m-\nu)2^{\nu+1}$ counts the number of edges in the middle $(k+m-\nu)$ sections of a ZT trellis. The fourth term in \eqref{eq: C_SSV} approximates the complexity of the traceback operation, assigning $2$ units of complexity for accessing the parent node per trellis stage and $1.5$ units of complexity per codeword symbol for the CRC verification on the decoded sequence $\hat{\bmv}$. In \eqref{eq: C_SSV for CRC-TBCC}, the second term is because it takes $2^{\nu}$ operations to identify the optimal termination state with minimum metric before the first traceback.

The second component $C_{\trace}$ represents the complexity of the \emph{additional} traceback operations required by SLVD. Specifically, for a given CRC-ZTCC,
\begin{align}
C_{\trace} = c_1(\E[L]-1)[2(k+m+\nu)+1.5(k+m)]. \label{eq: C_trace}
\end{align}
For CRC-TBCCs,
\begin{align}
C_{\trace} = 3.5c_1(\E[L]-1)(k+m). \label{eq: C_trace for CRC-TBCC}
\end{align}

The third component $C_{\tlist}$ represents the average complexity of inserting new elements to maintain an ordered list of path metric differences. For both CRC-ZTCCs and CRC-TBCCs,
\begin{align}
C_{\tlist} = c_2\E[I]\log(\E[I]), \label{eq: C_list}
\end{align}
where $\E[I]$ is the expected number of insertions to maintain the sorted list of path metric differences. According to the mechanism of insertion, for CRC-ZTCCs,
\begin{align}
  \E[I]&\le (k+m)\E[L],
\end{align}
and for CRC-TBCCs,
\begin{align}
  \E[I]&\le (k+m)\E[L]+2^{\nu}-1, \label{eq:EI for CRC-TBCC}
\end{align}
where $2^{\nu}-1$ denotes the number of path metric differences between the optimal terminating state and any other of the $2^{\nu}-1$ terminating states.

In \eqref{eq: C_SSV}, \eqref{eq: C_SSV for CRC-TBCC}, \eqref{eq: C_trace}, \eqref{eq: C_trace for CRC-TBCC}, and \eqref{eq: C_list} the constants $c_1$ and $c_2$ characterize implementation-specific differences in the implemented complexity of traceback and list insertion, respectively, as compared to the ACS operations of Viterbi decoding. For our implementation, we found $c_1 = 1.5$ and $c_2=2.2$.

\begin{figure}[t]
\centering
\includegraphics[width=0.45\textwidth]{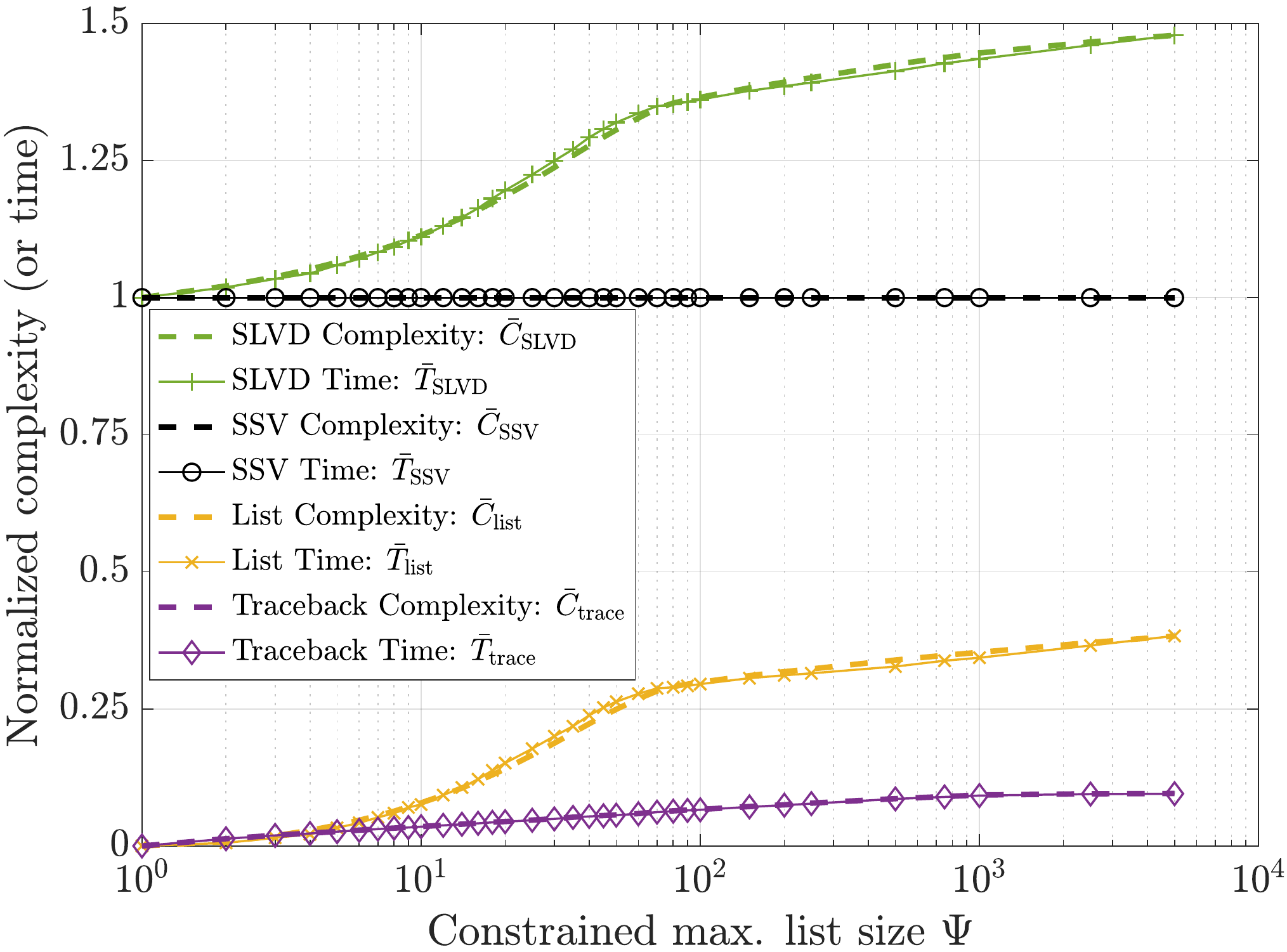}
\caption{The complexity of SLVD with different constrained maximum list sizes for ZTCC $(27,31)$, and degree-$10$ DSO CRC polynomial 0x709, with $k=64$ at SNR $\gamma_s=2$ dB. All variables are normalized by the time or complexity of the SSV algorithm. In the simulation, $c_1=1.5$ and $c_2=2.2$. }
\label{fig: 27_31_complexity_vs_listsize}
\end{figure}

The additional complexity of the SLVD over SSV decoding is completely characterized by the additional tracebacks along the trellis and the maintenance of an ordered list of path metric differences.  We define the \emph{normalized complexity} $\bar{C}_{\SLVD}$ as the complexity of SLVD divided by the complexity of SSV decoding, i.e.,
\begin{align}
\bar{C}_{\SLVD}= \frac{C_{\SLVD}}{C_{\SSV}} = 1+ \bar{C}_{\trace} +\bar{C}_{\tlist}. \label{eq:NormalizedC}
\end{align}
The normalized complexity provides a measure for the additional complexity of operations associated with the SLVD relative to the complexity of the SSV algorithm.

We recorded the runtime $T_{\SLVD}$, $T_{\SSV}$, $T_{\trace}$, and $T_{\tlist}$ on an Intel i7-4720HQ using Visual C{}\verb!++!. We then divided all of these terms by $T_{\SSV}$ to compute a normalized runtime $\bar{T}$. Fig. \ref{fig: 27_31_complexity_vs_listsize} shows normalized complexity based on equation \eqref{eq:NormalizedC} and normalized runtime. In both cases, the normalization is computed by dividing by the complexity or run-time associated with SSV, i.e., performing all ACS operations on the trellis and a traceback from the state with the best metric.  The normalized complexity and normalized runtime curves are indistinguishable. Fig. \ref{fig: 27_31_complexity_vs_listsize} also shows that the additional complexity of SLVD is primarily from maintaining an ordered list of path metric differences.

\begin{figure}[t]
\centering
\includegraphics[width=0.45\textwidth]{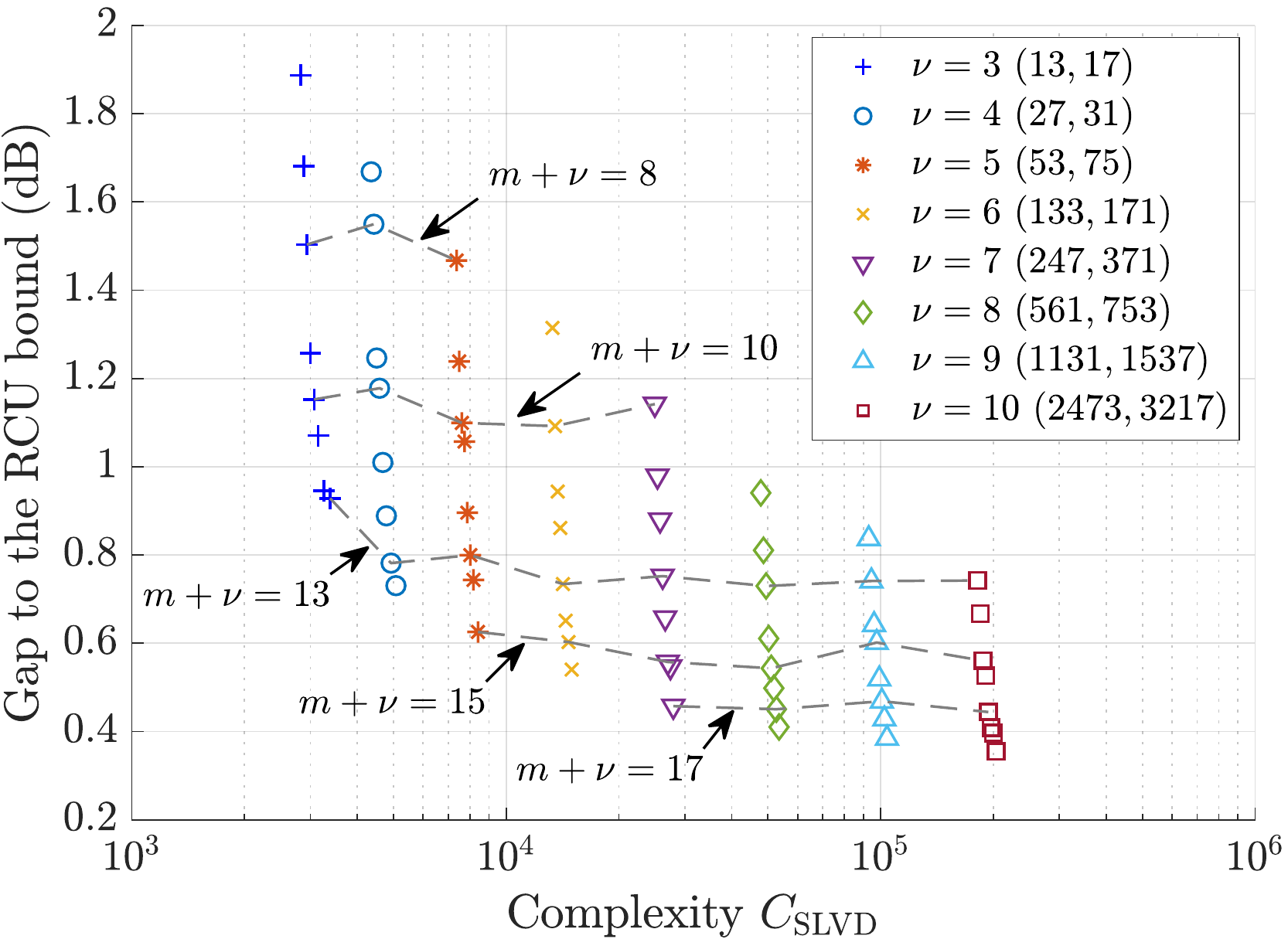}
\caption{The SNR gap to the RCU bound vs. the average complexity of SLVD for the family of CRC-ZTCCs in Table \ref{table: CRC-ZTCC codes} at target $P_{e, \lambda}=10^{-4}$. Each color represents a specific ZTCC shown in parenthesis. Markers from top to bottom with the same color correspond to the DSO CRC polynomials with $m=3,4,\dots, 10$ in Table \ref{table: CRC-ZTCC codes}. The information length and blocklength are given by $k = 64$ and $n = 2(64+m+\nu)$, respectively.}
\label{fig: tradeoff CRC-ZTCC}
\end{figure}

\begin{figure}[t]
\centering
\includegraphics[width=0.45\textwidth]{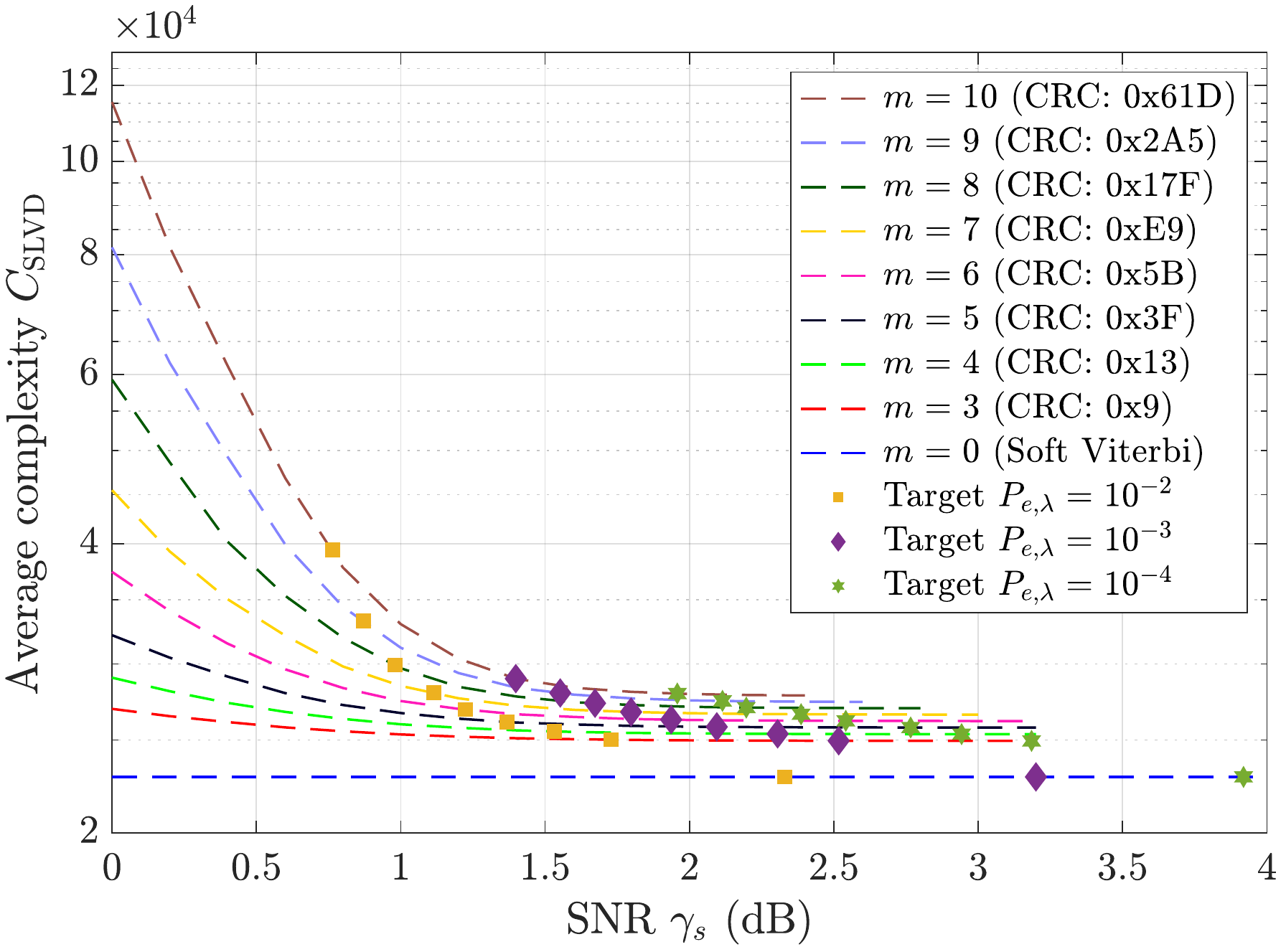}
\caption{The average complexity vs. SNR for ZTCC $(247, 371)$ used with its DSO CRC polynomials. The ZTCC with no CRC using soft Viterbi decoding is also given as a reference. }
\label{fig: complexity_vs_SNR}
\end{figure}

\section{Simulation Results}
\label{sec: simulation results}

In this section, we present our simulation results of CRC-ZTCCs in Table \ref{table: CRC-ZTCC codes} and CRC-TBCCs in Table \ref{table: CRC-TBCC codes} for $k=64$. Finally, we compare the $(128, 64)$ punctured CRC-TBCC designed in our precursor conference paper \cite{Liang2019} with several $(128, 64)$ short blocklength codes presented in \cite{Coskun2019}.

\subsection{Simulation Results for CRC-ZTCCs}
Fig. \ref{fig: tradeoff CRC-ZTCC} shows the trade-off between the SNR gap to the RCU bound and the average decoding complexity computed using \eqref{eq: SLVD complexity formula} for target probability of UE $P_{e, \lambda}=10^{-4}$. It is shown that for a given ZTCC, increasing the degree $m$ of DSO CRC polynomials can significantly diminish the SNR gap to the RCU bound at a relatively small complexity increase. This SNR gap reduction is especially considerable when $\nu$ is small and becomes less significant as $\nu$ becomes large. For all ZTCCs, the complexity cost of increasing $m$ from $3$ to $10$ is within a factor of $2$. This is consistent with Fig. \ref{fig: 27_31_complexity_vs_listsize} in which the complexity increases by a factor less than $1.5$ even for a very large constrained maximum list size $\Psi$. 

A CRC-ZTCC could be decoded using Viterbi alone, without list decoding, on a trellis with $2^{m+\nu}$ states per trellis stage. The dashed lines in Fig. \ref{fig: tradeoff CRC-ZTCC} show that the gap to the RCU bound remains roughly constant for a constant value of $m+\nu$.  However, list decoding with a well chosen $(m,\nu)$ pair achieves this performance with a minimum complexity $C_{\SLVD}$.  Thus, for a given target $P_{e,\lambda}$ and a fixed value of $m+\nu$, the inclusion of CRC-aided list decoding will generally reduce complexity over using Viterbi decoding alone on a convolutional code with $2^{m+\nu}$ states per trellis stage.

Fig. \ref{fig: complexity_vs_SNR} shows the complexity $C_{\SLVD}$ computed using \eqref{eq: SLVD complexity formula} as a function of SNR for ZTCC $(247, 371)$ and its DSO CRC polynomials with degree $m$ from $3$ to $10$ from Table \ref{table: CRC-ZTCC codes}. The ZTCC using soft Viterbi decoding with no CRC is also shown. Here, the target probabilities of UE at $10^{-2}, 10^{-3}, 10^{-4}$ for each CRC-ZTCC are marked by squares, diamonds, and stars, respectively.  For each target probability of UE, the corresponding complexity is within a factor of $2$ compared to the soft Viterbi decoding of ZTCC $(247, 371)$.

The termination overhead associated with ZTCC induces a gap from the RCU bound, which can be closed by using the corresponding TBCC as we will see below.

\subsection{Simulation Results for CRC-TBCCs}

In Section \ref{sec: preliminaries} we use the fact that for a CRC-ZTCC, each SLVD operation yields a valid higher-rate codeword, i.e., a ZT codeword. However,  for a CRC-TBCC, SLVD operations do not always yield a valid higher rate codeword, i.e., a TB codeword, because the TB condition is often not met.  Because of this, we can no longer assume that $\bar{L} \approx 2^m$. Nevertheless, Approximations \ref{approx: parametric approx} and \ref{approx: higher-order approx} still apply for an accurate value of $\bar{L}$ which can be obtained from simulation. 

The increased value of $\bar{L}$ may be understood by considering the higher-rate code $\C_h$ to be the pseudo code represented by all paths on the trellis regardless of whether they meet the TB condition. Due to the additional complexity required to check the TB condition, $\E[I]$ is significantly increased compared to the CRC-ZTCC. While we identified the empirical value of $\E[I]$  for CRC-ZTCCs, in this section we simply assume $\E[I]$ attains the upper bound in \eqref{eq:EI for CRC-TBCC} for CRC-TBCCs. Hence, using \eqref{eq: C_SSV for CRC-TBCC}, \eqref{eq: C_trace for CRC-TBCC} with $c_1=1.5$, \eqref{eq: C_list} with $c_2=2.2$, together with the aforementioned assumption on $\E[I]$, we can compute an estimate of the average complexity $C_{\SLVD}$ of our implementation of SLVD of CRC-TBCCs. 

\begin{figure}[t]
\centering
\includegraphics[width=0.45\textwidth]{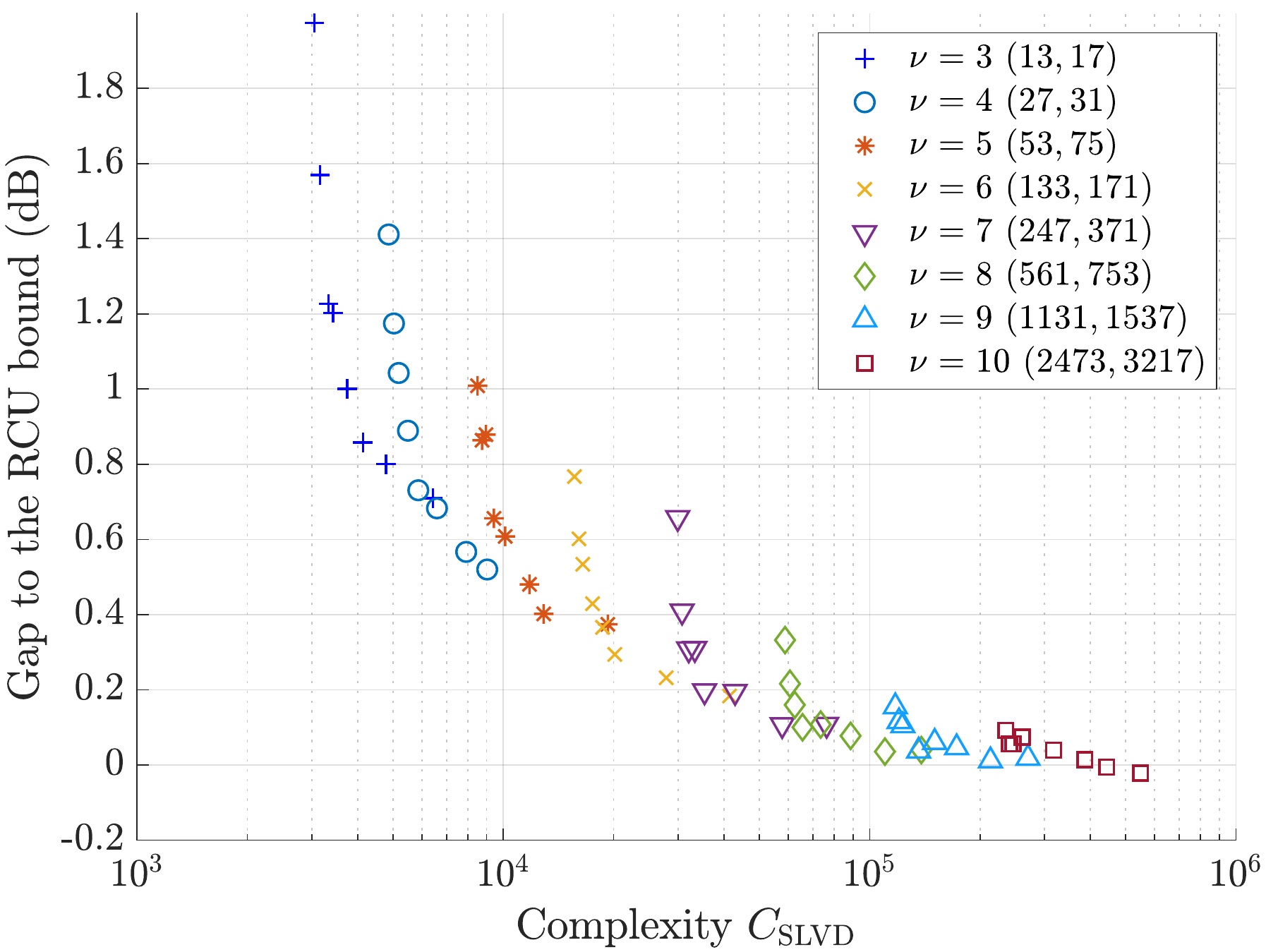}
\caption{The SNR gap to the RCU bound vs. the average complexity of SLVD for the family of CRC-TBCCs in Table \ref{table: CRC-TBCC codes} at target $P_{e, \lambda}=10^{-4}$. Each color represents a specific TBCC shown in parenthesis. Markers from top to bottom with the same color correspond to the DSO CRC polynomials with $m=3,4,\dots, 10$ in Table \ref{table: CRC-TBCC codes}. The information length and blocklength are given by $k = 64$ and $n = 2(64+m)$, respectively.}
\label{fig: tradeoff CRC-TBCC}
\end{figure}

Fig. \ref{fig: tradeoff CRC-TBCC} shows the SNR gap to the RCU bound vs. the average complexity for target probability of UE $P_{e,\lambda}=10^{-4}$ for all CRC-TBCCs designed in Table \ref{table: CRC-TBCC codes}. Compared to Fig. \ref{fig: tradeoff CRC-ZTCC}, TB encoding significantly reduces the SNR gap to the RCU, because the overhead of termination is avoided.  However, this reduction of the gap comes at the expense of a slight increase in average complexity for checking the TB condition. Note the exciting result that some CRC-TBCCs outperform the RCU bound for $\nu=9$ and $10$. Another phenomenon distinct from CRC-ZTCCs is that for TBCCs with large $\nu$, increasing the DSO CRC polynomial degree from $m=3$ to $10$ only provides a small benefit.  Note, however, that the degree-$3$ DSO CRC polynomial does provide a benefit over a TBCC used with no CRC at all.

\begin{figure}[t]
\centering
\includegraphics[width=0.45\textwidth]{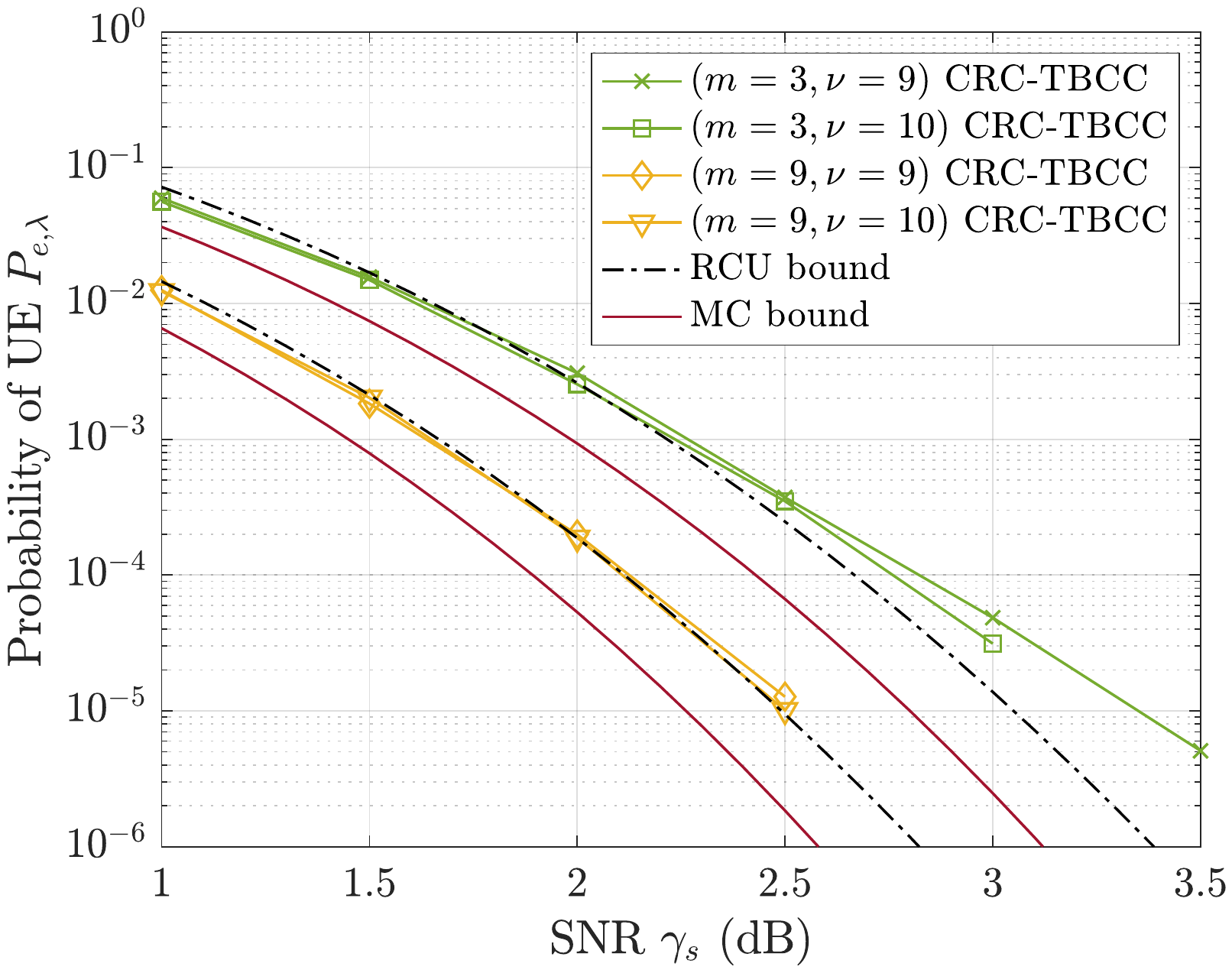}
\caption{Comparison between $P_{e, \lambda}$ and RCU and MC bounds at rates $R = 64/134$ ($m=3$) and $R=64/146$ ($m=9$) for the CRC-TBCCs designed in Table \ref{table: CRC-TBCC codes}. For the sake of clarity, only $\nu=9, 10$ TBCCs are displayed. }
\label{fig: P_UE_vs_RCU_MC}
\end{figure}

\begin{figure}[t]
\centering
\includegraphics[width=0.45\textwidth]{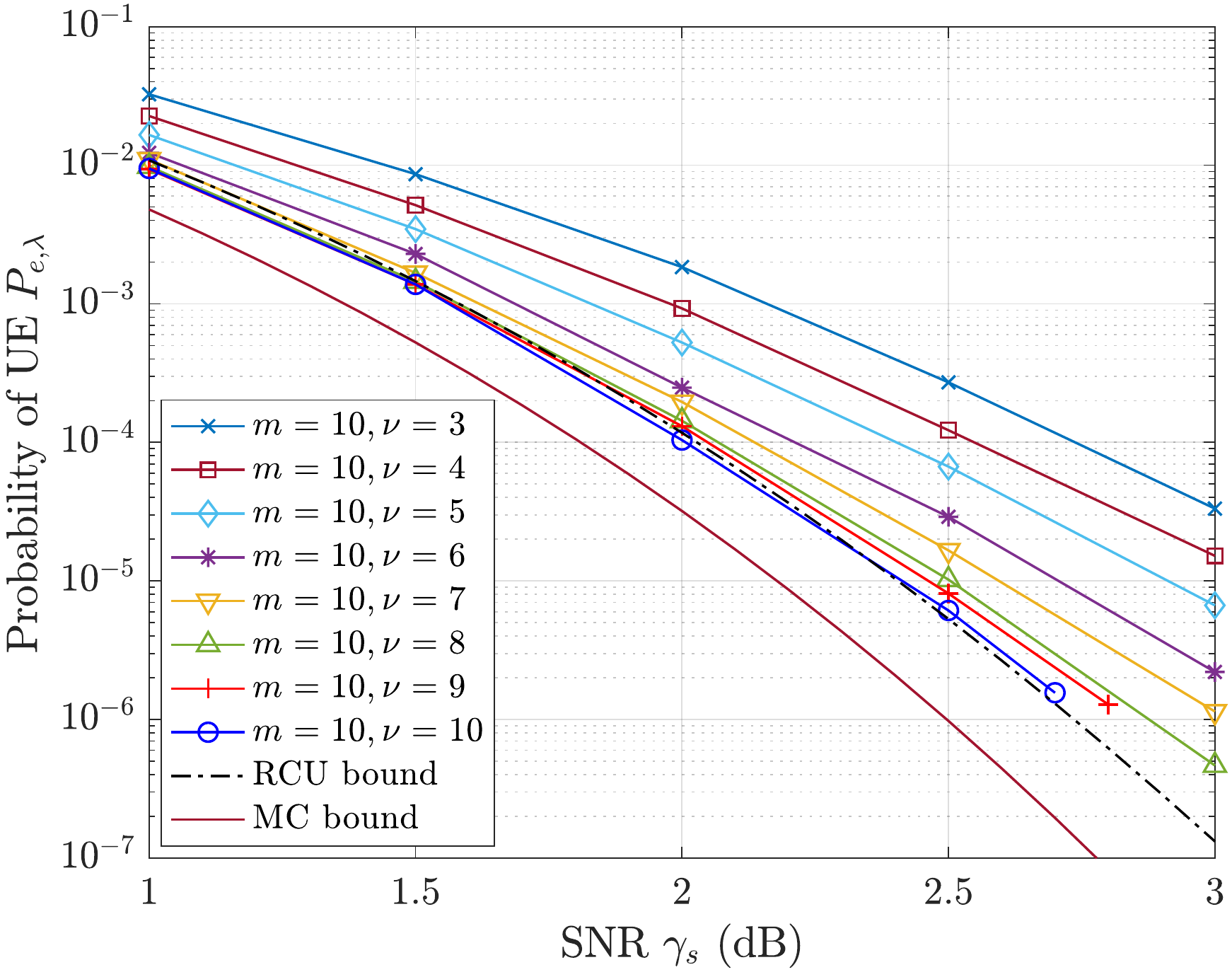}
\caption{Comparison between $P_{e, \lambda}$ and RCU and MC bounds at rate $R=64/148$ (i.e., $m=10$) for the CRC-TBCCs designed in Table \ref{table: CRC-TBCC codes}. }
\label{fig: P_UE_vs_RCU_MC_m_10}
\end{figure}

To illustrate the performance of the best CRC-TBCCs designed in Table \ref{table: CRC-TBCC codes}, we select $\nu=9$ and $\nu=10$ TBCCs as an example. Fig. \ref{fig: P_UE_vs_RCU_MC} shows two cases: $R = 64/134$ corresponding to $m=3$ and $R=64/146$ corresponding to $m=9$. The MC bound and the RCU bounds for these rates are plotted using the saddlepoint approximations provided in Approximations \ref{approx: MC bound} and \ref{approx: RCU bound}, respectively. We see that in these two cases, the CRC-TBCCs in Fig. \ref{fig: P_UE_vs_RCU_MC} beat the RCU bound at low SNR values. However, this superiority gradually fades away as SNR increases, although for $m=9$, the performance is very close to the RCU bound even at $P_{e,\lambda}=10^{-5}$. Simulations also suggest that it is extremely difficult to further improve the code performance once beyond the RCU bound at low probability of UE.

Fig. \ref{fig: P_UE_vs_RCU_MC_m_10} shows the family of CRC-TBCCs with $k = 64$ and $n = 148$ (corresponding to $m=10$). For small $\nu$, we see a visible improvement as $\nu$ increases. However, once performance reaches the RCU bound, further increases in $\nu$ provide little benefit. For example with $m=10$, the CRC-TBCC with $\nu=9$ attains similar performance to that with $\nu=10$.

%\begin{remark}
\subsection{Comparison of $(128, 64)$ Linear Block Codes}
\begin{figure}[t]
\centering
\includegraphics[width=0.45\textwidth]{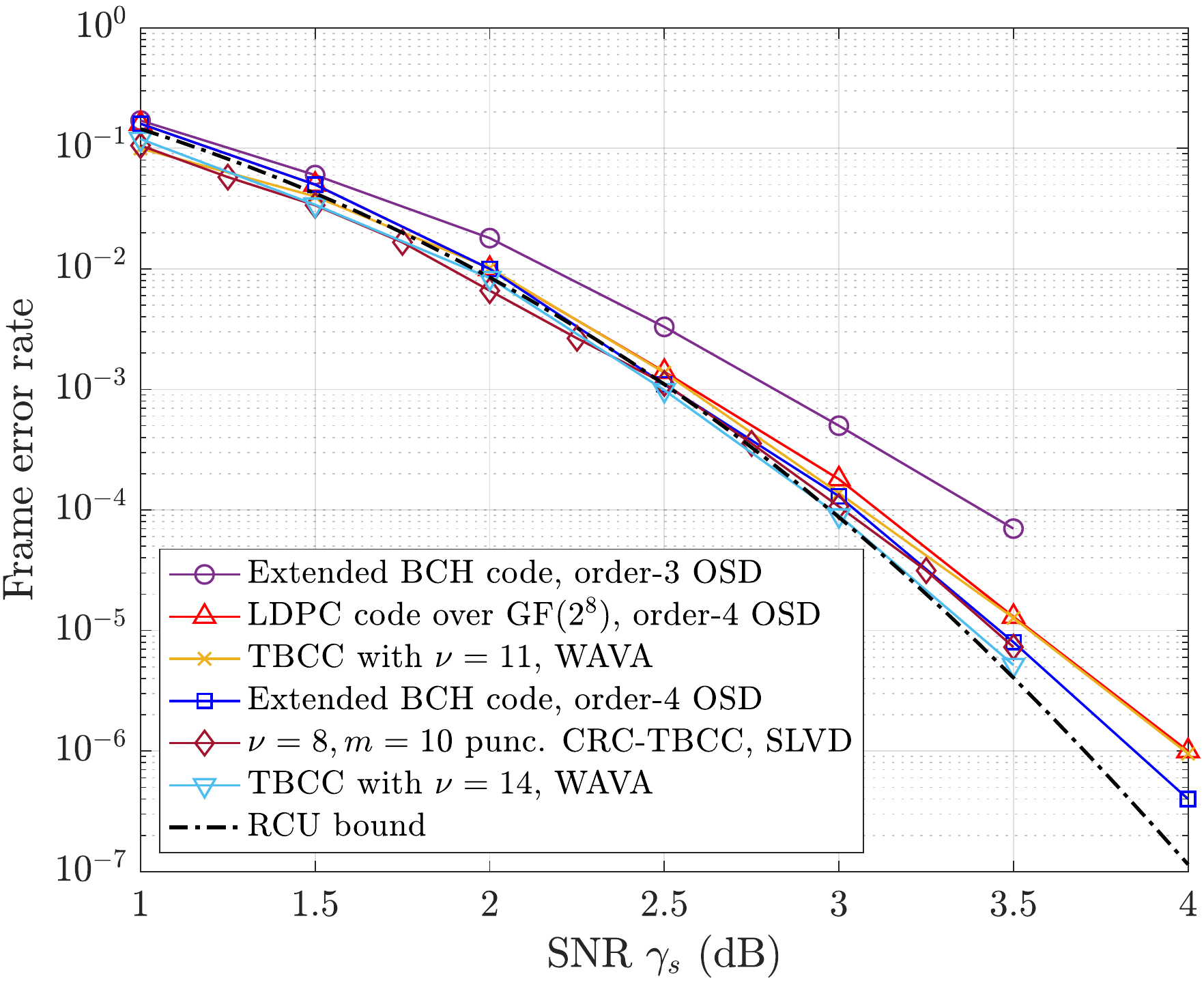}
\caption{Comparison of $(128, 64)$ linear block codes.}
\label{fig: FER_vs_SNR_128_64_codes}
\end{figure}

Direct comparison of CRC-TBCCs with other codes often requires puncturing to match rates. For simplicity, we have excluded puncturing from analysis in this paper. However, our precursor conference paper \cite{Liang2019} designed a  $v=8$, $m=10$ punctured CRC-TBCC with $k=64$ and $n=128$ whose FER performance can be directly compared to the $(128, 64)$ linear block codes presented in \cite{Coskun2019}, as shown in Fig. \ref{fig: FER_vs_SNR_128_64_codes}.

At SNR of $3$ dB, the  $v=8$, $m=10$ punctured CRC-TBCC in \cite{Liang2019} and the best codes studied in \cite{Coskun2019} all perform similarly.  Specifically, the four codes in \cite{Coskun2019} with similar performance at  $3$ dB to the  $v=8$, $m=10$ punctured CRC-TBCC are the following:  the $v=14$ and $v=11$ TBCCs decoded with WAVA, the extended BCH code with order-$4$ OSD, and a  non-binary LDPC code over $\F_{256}$ with order-$4$ OSD.  As shown in Fig. \ref{fig: FER_vs_SNR_128_64_codes}, at higher SNR, the FER performance is more differentiated with the best performance provided by the $v=14$ TBCC, slightly worse performance provided by the  $v=8$, $m=10$ punctured CRC-TBCC and the extended BCH code with order-$4$ OSD and further degraded performance by the $v=11$ TBCC and the non-binary LDPC code over $\F_{256}$ with order-$4$ OSD.

We now consider the decoding complexity of the three best codes described above at $3$ dB, excluding the discussion of the non-binary LDPC code due to its further degraded performance. Actual complexity depends on specific implementation choices, here we consider the total number of computations per codeword as a way to give some flavor of the complexity differences between these approaches. At SNR of $3$ dB, simulation shows that $\E[L]=44.41$ for the $v=8$, $m=10$ punctured CRC-TBCC. Using \eqref{eq: C_SSV for CRC-TBCC}, \eqref{eq: C_trace for CRC-TBCC}, \eqref{eq: C_list} together with \eqref{eq:EI for CRC-TBCC}, we obtain $C_{\SLVD}\le 1.67\times10^5$. 

In terms of WAVA complexity, let $I$ be the number of iterations in WAVA. By assuming $0.5$ units of complexity for compare/select operation per branch and $1$ unit of complexity for one addition, the WAVA complexity for a rate-$1/\omega$ TBCC with $\nu$ memory elements at information length $k$ is given by
\begin{align}
    C_{\WAVA} = kI(0.5\cdot2^{\nu} + 2^{\nu+1}).\label{eq: WAVA complexity}
\end{align}
Using \eqref{eq: WAVA complexity}, the complexity of $3$-round WAVA for $v = 11$ TBCC in \cite{Coskun2019} is $9.83\times10^5$, which is higher than for the $v=8, m = 10$ punctured CRC-TBCC. The best $v = 14$ TBCC in \cite{Coskun2019} under $3$-round WAVA achieves a complexity of $7.86\times10^6$.

A direct complexity comparison of SLVD with OSD is more difficult, but Table V in \cite{Fossorier1995} indicates that at $3$ dB, the order-$3$ OSD of the $(128,64)$ extended BCH code requires $2.83\times10^5$ operations per codeword on average, which indicates that the order-$4$ OSD would likely have a higher complexity than the SLVD of $v=8$, $m=10$ punctured CRC-TBCC.  Based on this analysis, the CRC-TBCC paradigm appears to be competitive with the existing approaches that provide similarly excellent FER performance at short blocklength.
%\end{remark}

\section{Conclusion}
\label{sec: conclusion}

In this paper, we consider the CRC-aided convolutional code as a promising short blocklength code. The concatenated nature permits the use of SLVD that allows the code to attain the ML decoding performance at low complexity. For $k=64$, we identified the DSO CRC polynomial for a family of ZTCCs and TBCCs generated with the optimum rate-$1/2$ convolutional encoders identified by \cite{Lin2004} at sufficiently low target probability of UE. Several CRC-TBCCs beat the RCU bound at practically interesting values of SNR. In a recent work \cite{Schiavone2021}, Schiavone confirmed that the CRC-TBCC is indeed a powerful short blocklength code by showing that its performance matches the expurgated ensemble. 

All CRC-aided convolutional codes considered in this paper are designed based on an optimum convolutional encoder. It would be interesting to investigate whether a suboptimal convolutional code used with the DSO CRC polynomial can also lead to a good concatenated code. Another interesting direction is to explore the performance of CRC-aided convolutional codes in the moderately short blocklength regime, e.g., $256\le n< 1000$. If puncturing is introduced in the code design, it remains open as to how to jointly design the puncturing pattern and the optimal CRC polynomial for a given convolutional code.

%To show the excellent performance of the CRC-aided convolutional code in an absolute sense, this paper provides results with respect to the RCU bound.  In this way, the gap to the RCU bound can be easily compared with both previous and future short-blocklength designs with similar rates and blocklengths.  Fig. \ref{fig: tradeoff CRC-ZTCC} shows that the CRC-ZTCC with $\nu=8$ and $m=10$ performs within $0.4$ dB of the RCU bound for $P_{e,\lambda}=10^{-4}$. Fig. \ref{fig: tradeoff CRC-TBCC} shows that the CRC-TBCC with $\nu=8$ and $m=9$ essentially attains the RCU bound for $P_{e,\lambda}=10^{-4}$.  Thus, this CRC-TBCC achieves a similar proximity to the RCU bound as the $\nu=14$ TBCC under WAVA from \cite{Liang2019,Coskun2019}, but with less decoder complexity than a TBCC with $\nu=9$ as shown in Fig. \ref{fig: tradeoff CRC-TBCC}.

The beauty of SLVD lies in the fact that its average complexity is governed by its expected list rank $\E[L]$, a quantity that is inversely proportional to the SNR value. This allows a huge complexity reduction at interesting operating SNR values that guarantee a low target probability of UE. In particular, the parametric approximation of $\E[L]$ provides an explicit characterization of the performance-complexity trade-off. It shows that for CRC-ZTCCs with a target error probability $P_{e,\lambda}^*$ and $\bar{L}\approx 2^m$, a CRC degree $m\le -\log(P_{e,\lambda}^*)$ is sufficient to maintain $\E[L]\le2$.  However, several problems are still open, for instance, how to upper bound $\E[L|\bmX = \bm{O}]$, and how to upper bound $P_{e, 1}$ using the weight spectrum. In addition, the behavior of the supremum list rank $\lambda$ is also less understood and is worth future investigation.

\appendices
\section{Derivation of the induced density function}
\label{appendix: induced density function}

\begin{figure}[h]
\centering
\includegraphics[width=0.25\textwidth]{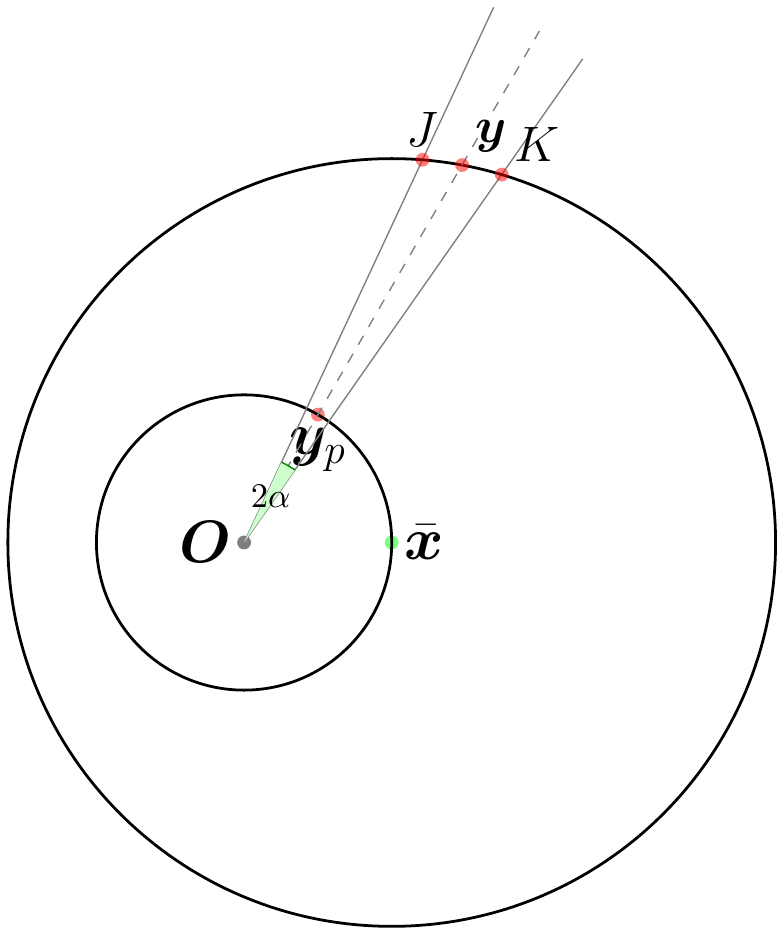}
\caption{Derivation of the induced density function $g_w(\bmy_p)$ in $\R^n$.}
\label{fig: induced density function}
\end{figure}

Let $\B(\bm{a}, r)$ denote the spherical surface of center $\bm{a}\in\R^n$ and radius $r$ in $\R^n$. In this section, we derive the induced density function $g_{w}(\bmy_p)$ incurred when projecting a received point $\bmy$ uniformly distributed on $\B(\bar{\bmx}, w)$ to point $\bmy_p = (r/\norm{\bmy})\bmy$ that lies on the codeword sphere $\B(\bm{O}, A\sqrt{n})$ in $\R^n$. As an illustration, Fig. \ref{fig: induced density function} depicts this scenario in $\R^2$. For our purposes, we assume that $w\ge A\sqrt{n}$ to ensure the bijective relationship between $\bmy$ and $\bmy_p$.

Let us consider a circular cone $Q_{\alpha}$ in $\R^n$ with apex at the origin $\bm{O}$, axis along $\bm{O}\bmy_p$ and half-angle $\alpha$. Algebraically, define the direction vectors
\begin{align}
\bmy_e&\triangleq\frac{\bmy}{\norm{\bmy}},\\
\bmz_e&\triangleq\frac{\bmy-\bar{\bmx}}{\norm{\bmy-\bar{\bmx}}}.
\end{align}
Hence, the circular cone $Q_{\alpha}$ is given by
\begin{align}
Q_{\alpha} &= \left\{\bmr\in\R^n: \frac{\bmr^\top\bmy_e}{\norm{\bmr}}\ge\cos\alpha\right\}\notag\\
  &= \big\{\bmr\in\R^n:  (\bmr-\bm{0})^\top(I-\epsilon^2(\alpha)\bmy_e\bmy_e^\top)(\bmr-\bm{0})\le0\big\},
\end{align}
where $\epsilon(\alpha)\triangleq 1/\cos\alpha$ denotes the eccentricity of the cone. Cone $Q_{\alpha}$ intersects with the noise sphere $\B(\bar{\bmx}, w)$, thus producing a surface area $Q_{\alpha}\cap\B(\bar{\bmx}, w)$ delimited by $J$ and $K$ on Fig. \ref{fig: induced density function}. Thus, the induced density at $\bmy_p$ is given by
\begin{align}
g_w(\bmy_p) &=\lim_{\alpha\to0} \frac{S(Q_{\alpha}\cap\B(\bar{\bmx}, w))/S_{n-1}(w)}{S(Q_{\alpha}\cap\B(\bm{O}, A\sqrt{n}))}, \label{eq: 96}
\end{align}
where $S(\cdot)$ denotes the surface area in $\R^n$. Note that for sufficiently small $\alpha$, the spherical surface around $\bmy$ is equivalent to the tangent hyperplane at $\bmy$, given by
\begin{align}
H &= \big\{\bmr\in\R^n: \bmz_e^\top(\bmr-\bmy) = 0 \big\}\notag\\
  &= \big\{\bmr\in\R^n: \bmz_e^\top(\bmr - \bm{0}) = \hat{h}\big\},
\end{align}
where $\hat{h} \triangleq \bmz_e^\top \bmy$. Define $\rho \triangleq \sqrt{1-(\bmz_e^\top\bmy_e)^2}$. Thus, using the result by Dearing \cite[Eq.~(15)]{Dearing2020}, if $\epsilon(\alpha)\rho<1$, the intersection of hyperplane $H$ and cone $Q_{\alpha}$ is an ellipsoid of dimension $(n-1)$, which, after proper rotation $T$ around $\bm{O}$, can be written as
\begin{align}
&T(Q_{\alpha})\cap T(H)\notag\\
&=\Big\{(r_1, \dots, r_{n-1}, \hat{h}): \frac{(r_1-\hat{c}_1)^2}{\hat{a}^2}+\frac{\sum_{j=2}^{n-1}(r_j-\hat{c}_j)^2}{\tilde{b}} = 1\Big\},\notag
\end{align}
where
\begin{align}
\sigma&= \bmz_e^\top\bmy_e, \\
\hat{c}_1&=\frac{\epsilon^2(\alpha)\rho\sigma\hat{h}}{1-\epsilon^2(\alpha)\rho^2},\quad \hat{c}_j = 0,\ j=2,\dots,n-1,\\
\hat{a}^2&= \frac{(\epsilon^2(\alpha)-1)\hat{h}^2}{(1-\epsilon^2(\alpha)\rho^2)^2}, \\
\tilde{b}&= \hat{a}^2(1-\epsilon^2(\alpha)\rho^2).
\end{align}
Since $\bmz_e$ and $\bmy_e$ are non-orthogonal, $1/\rho > 1$. Hence, for sufficiently small $\alpha$, $\epsilon(\alpha)<1/\rho$ and thus Dearing's result follows. Summarizing the analysis above, we obtain
\begin{align}
&\lim_{\alpha\to0}S\big(Q_{\alpha}\cap\B(\bar{\bmx},w)\big)\notag\\
&=\lim_{\alpha\to0}S\big(T(Q_{\alpha})\cap T(H)\big)\label{eq: 104}\\
&=\lim_{\alpha\to0}\frac{\pi^{\frac{n-1}{2}}}{\Gamma(\frac{n+1}{2})}\hat{a}\Big(\sqrt{\tilde{b}}\Big)^{n-2}\\
&= \lim_{\alpha\to0}\frac{\pi^{\frac{n-1}{2}}}{\Gamma(\frac{n+1}{2})}\Bigg(\frac{(\epsilon^2(\alpha)-1)\hat{h}^2}{(1-\epsilon^2(\alpha)\rho^2)^2} \Bigg)^{\frac{n-1}{2}}\big(1-\epsilon^2(\alpha)\rho^2\big)^{\frac{n-2}{2}}\\
&= \lim_{\alpha\to0}\frac{\pi^{\frac{n-1}{2}}}{\Gamma(\frac{n+1}{2})}2^{\frac{n-1}{2}}(\epsilon(\alpha)-1)^{\frac{n-1}{2}}\hat{h}^{n-1}(\bmz_e^\top\bmy_e)^{-n}\\
&= \lim_{\alpha\to0}\frac{\pi^{\frac{n-1}{2}}}{\Gamma(\frac{n+1}{2})}2^{\frac{n-1}{2}}\Big(\frac{1-\cos\alpha}{\cos\alpha}\Big)^{\frac{n-1}{2}}\Big(\frac{\bmz_e^\top\bmy}{\bmz_e^\top\bmy_e}\Big)^{n-1}\frac{1}{\bmz_e^\top \bmy_e}\\
&= \lim_{\alpha\to0}\frac{\pi^{\frac{n-1}{2}}}{\Gamma(\frac{n+1}{2})}2^{\frac{n-1}{2}}\Big(2\sin^2\Big(\frac{\alpha}{2}\Big)\Big)^{\frac{n-1}{2}}\frac{\norm{\bmy}^{n-1}}{\cos\angle \bar{\bmx}\bmy\bm{O}}\\
&= \lim_{\alpha\to0}\frac{\pi^{\frac{n-1}{2}}}{\Gamma(\frac{n+1}{2})}\alpha^{n-1}\frac{\norm{\bmy}^{n-1}}{\cos\angle \bar{\bmx}\bmy\bm{O}},\label{eq: 110}
\end{align}
where  \eqref{eq: 104} follows since for sufficiently small half-angle, the spherical surface around $\bmy$ is equivalent to that of the tangent hyperplane $H$ at $\bmy$. From \cite[Eq.~(21)]{Shannon1959}, the area of the spherical cap $S(Q_{\alpha}\cap\B(\bm{O}, A\sqrt{n}))$ is given by
\begin{align}
&S(Q_{\alpha}\cap\B(\bm{O}, A\sqrt{n}))\notag\\
&=\frac{(n-1)\pi^{\frac{n-1}{2}}(A\sqrt{n})^{n-1}}{\Gamma\big(\frac{n+1}{2}\big)}\int_0^{\alpha}\sin^{n-2}\theta\diff\theta. \label{eq: 111}
\end{align}
Substituting \eqref{eq: 110}, \eqref{eq: 111} into \eqref{eq: 96}, we obtain
\begin{align}
g_w(\bmy_p)=&\lim_{\alpha\to0}\frac{S(Q_{\alpha}\cap\B(\bar{\bmx}, w))}{S(Q_{\alpha}\cap\B(\bm{O}, A\sqrt{n}))}\frac{S_{n-1}(A\sqrt{n})}{S_{n-1}(w)S_{n-1}(A\sqrt{n})}\notag\\
=&\lim_{\alpha\to0}\frac{\alpha^{n-1}\frac{\norm{\bmy(\bmy_p)}^{n-1}}{\cos\angle \bar{\bmx}\bmy(\bmy_p)\bm{O}}}{(n-1)\int_0^{\alpha}\theta^{n-2}\diff\theta}\frac{1}{w^{n-1}}\frac{1}{S_{n-1}(A\sqrt{n})}\notag\\
=& \Big(\frac{\norm{\bmy(\bmy_p)}}{w}\Big)^{n-1}\frac{1}{\cos\angle \bar{\bmx}\bmy(\bmy_p)\bm{O}}\frac{1}{S_{n-1}(A\sqrt{n})},\label{eq: 112}
\end{align}
where $\bmy(\bmy_p)$ is the pre-image of $\bmy_p$ on the noise sphere $\B(\bar{\bmx}, w)$. Here, \eqref{eq: 112} is the induced density function of $\bmy_p\in\B(\bm{O}, A\sqrt{n})$. Observe that it is rotationally symmetric with respect to axis $\bm{O}\bar{\bmx}$.

Next, we give an alternative expression of $g_w(\bmy_p)$ to derive its upper bound and lower bound. First, we rotate the coordinate system such that axis $\bm{O}\bar{\bmx}$ is the first coordinate and the remaining $(n-1)$ coordinates are orthogonal to $\bm{O}\bar{\bmx}$. In the new coordinate system, let $\bar{\bmx}=(A\sqrt{n}, 0,\dots, 0)\in\R^n$. Hence, for an arbitrary projected point $\bmy_p=(y_1, y_2, \dots, y_n)\in\B(\bm{O}, A\sqrt{n})$, assume that $\rho\triangleq \norm{\bmy(\bmy_p)}$. Thus,
\begin{align}
\bmy(\bmy_p)=\frac{\rho}{A\sqrt{n}}(y_1, y_2, \dots, y_n).
\end{align}
Since $\bmy(\bmy_p)\in\B(\bar{\bmx}, w)$, 
\begin{align}
\left(\frac{\rho}{A\sqrt{n}}y_1 - A\sqrt{n} \right)^2+\left(\frac{\rho}{A\sqrt{n}}\right)^2\sum_{i=2}^ny_i^2 = w^2.
\end{align}
Solving for $\rho$ yields
\begin{align}
\rho = y_1 + \sqrt{y_1^2 + w^2-A^2n}. \label{eq: 115}
\end{align}
By law of cosines, it is shown that
\begin{align}
\cos\angle\bar{\bmx}\bmy\bm{O}=\frac{\rho^2+w^2-A^2n}{2\rho w}=\frac{\sqrt{y_1^2+w^2-A^2n}}{w}. \label{eq: 116}
\end{align}
Hence, substituting \eqref{eq: 115} and \eqref{eq: 116} into \eqref{eq: 112} and expressing $g_w(\bmy_p)$ in terms of $y_1\in[-A\sqrt{n}, A\sqrt{n}]$, we obtain
\begin{align}
g_w(y_1)&=\frac{1}{S_{n-1}(A\sqrt{n})}\frac{(y_1+\sqrt{y_1^2+w^2-A^2n})^{n-2}}{w^{n-2}}\notag\\
&\phantom{=}\cdot\left(1+\frac{y_1}{\sqrt{y_1^2+w^2-A^2n}}\right).
\end{align}
Clearly, $g_w(y_1)$ is monotonically increasing in $y_1$. Hence,
\begin{align}
g_w(y_1)\ge g_w(-A\sqrt{n})&=\frac{1}{S_{n-1}(A\sqrt{n})}\left(1-\frac{A\sqrt{n}}{w}\right)^{n-1}\\
g_w(y_1)\le g_w(A\sqrt{n})&=\frac{1}{S_{n-1}(A\sqrt{n})}\left(1+\frac{A\sqrt{n}}{w}\right)^{n-1}.
\end{align}
Geometrically, this implies that the maximum induced density is attained at the transmitted point $\bar{\bmx}$, whereas the minimum induced density is attained at $-\bar{\bmx}$.

% or



\section*{Acknowledgment}
The authors would like to thank the anonymous reviewers for their careful reading and constructive comments that improved this paper.


\ifCLASSOPTIONcaptionsoff
  \newpage
\fi

\bibliographystyle{IEEEtran}
\bibliography{IEEEabrv,references}

\end{document}